\documentclass[psfig,epsfig]{report}

\newcommand{\beq}{\begin{equation}}
\newcommand{\eeq}{\end{equation}}
\newcommand{\beqa}{\begin{eqnarray}}
\newcommand{\eeqa}{\end{eqnarray}}
\newcommand{\bitm}{\begin{itemize}}
\newcommand{\eitm}{\end{itemize}}
\newcommand{\btab}{\begin{table}}
\newcommand{\etab}{\end{table}}

\newcommand{\pip}{\pi^{+}}

\newcommand{\numu}{\nu_\mu}

\newcommand{\nue}{\nu_e}

\def\numubar{\bar\nu_{\mu}}
\def\nuebar{\bar\nu_e}

\def\dmsq{\Delta m^{2}~}

\def\gtwid{\mathrel{\raise.3ex\hbox{$>$\kern-.75em\lower1ex\hbox{$\sim$}}}}
\def\ltwid{\mathrel{\raise.3ex\hbox{$<$\kern-.75em\lower1ex\hbox{$\sim$}}}}

\def\sinsqtheta{\sin^2 2 \theta~}
\usepackage{epsf}
\usepackage[section] {placeins}
\usepackage{graphicx}
\usepackage{subfigure}
\usepackage{wrapfig}
\usepackage{lscape}
\usepackage{amsmath}
\usepackage{fancyhdr}
\usepackage{enumerate}
\usepackage{indentfirst}
\usepackage{xspace}
\usepackage{array}
\usepackage{rotating}
\usepackage{bbold}
\usepackage[linktocpage=true]{hyperref}

\begin{document}


\begin{titlepage}

\vspace*{-0.75in}

\centerline{\huge The OscSNS White Paper}

\vspace{0.15in}

\centerline{\large \today}

\vspace{0.3in}


\vspace{0.1in}

\centerline{ \large M. Elnimr, I. Stancu}

\centerline{\it University of Alabama, Tuscaloosa, AL 35487}

\vspace{0.1in}

\centerline{ \large M. Yeh}

\centerline{\it Brookhaven National Laboratory, Upton, NY 11973}

\vspace{0.1in}

\centerline{ \large R. Svoboda}

\centerline{\it University of California, Davis, CA 95616}

\vspace{0.1in}

\centerline{ \large M. J. Wetstein}

\centerline{\it University of Chicago, Chicago, IL 60637}

\vspace{0.1in}

\centerline{ \large F. G. Garcia}

\centerline{\it Fermi National Accelerator Laboratory, Batavia, IL 60510}

\vspace{0.1in}

\centerline{ \large B. Osmanov, H. Ray}

\centerline{\it University of Florida, Gainesville, FL 32611}

\vspace{0.1in}

\centerline{ \large R. Tayloe}

\centerline{\it Indiana University, Bloomington, IN 47405}

\vspace{0.1in}

\centerline{ \large J. Boissevain, G. T. Garvey, W. Huelsnitz, W. C. Louis, G. B. Mills, } 
\smallskip
\centerline{\large Z. Pavlovic, R. Van de Water, D. H. White}

\centerline{\it Los Alamos National Laboratory, Los Alamos, NM 87545}

\vspace{0.1in}

\centerline{ \large R. Imlay}

\centerline{\it Louisiana State University, Baton Rouge, LA 70803}

\vspace{0.1in}

\centerline{ \large B. P. Roe}

\centerline{\it University of Michigan, Ann Arbor, MI 48109}

\vspace{0.1in}

\centerline{ \large R. Allen}

\centerline{\it Oak Ridge National Laboratory, Oak Ridge, TN 37831}

\vspace{0.1in}

\centerline{ \large Y. Efremenko, T. Gabriel, T. Handler, Y. Kamyshkov}

\centerline{\it University of Tennessee, Knoxville, TN 37996}

\vspace{0.1in}

\centerline{ \large F. T. Avignone, S. R. Mishra, C. Rosenfeld}

\centerline{\it University of South Carolina, Columbia, SC 29208}

\vspace{0.1in}

\centerline{ \large J. M. Link}

\centerline{\it Virginia Tech, Blacksburg, VA 24060}

\vspace{0.35in}

\centerline{~~~\hrulefill~~}
\centerline{\it Contact Persons: H. Ray and W. C. Louis}


\end{titlepage}

\newpage

\tableofcontents

\newpage

\listoftables

\newpage

\listoffigures

\newpage

\chapter{Executive Summary}

There exists a need to address and resolve the growing evidence 
for short-baseline neutrino oscillations and the possible existence of sterile neutrinos.  
Such non-standard particles require a mass of $\sim 1$ eV/c$^2$, 
far above the mass scale associated with active neutrinos, and were first invoked to
explain the LSND $\bar \nu_\mu \rightarrow \bar \nu_e$
appearance signal~\cite{lsnd}. More recently, the 
MiniBooNE experiment has reported a $2.8 \sigma$ excess of events 
in antineutrino mode consistent with neutrino oscillations and 
with the LSND antineutrino appearance signal~\cite{mb_osc_anti}. MiniBooNE also observed a 
$3.4 \sigma$ excess of events in their neutrino mode data.  Lower than 
expected neutrino-induced event rates using calibrated radioactive sources~\cite{radioactive} and 
nuclear reactors~\cite{reactor} can also be explained by the existence of sterile neutrinos. 
Fits to the world's neutrino and antineutrino data are consistent with 
sterile neutrinos at this $\sim 1$ eV/c$^2$ mass scale, although there is some 
tension between measurements from disappearance and appearance experiments~\cite{3+N,kopp}.  
In addition to resolving this potential major extension of the Standard Model, the existence of 
sterile neutrinos will impact design and planning for all future neutrino experiments. It should be 
an extremely high priority to conclusively establish if such unexpected light sterile neutrinos exist. 
{\bf The Spallation Neutron Source (SNS) at Oak Ridge National Laboratory~\cite{ornl}, 
built to usher in a new era in neutron research, provides a unique opportunity 
for US science to perform a definitive world-class search for sterile neutrinos.}

The 1.4 MW beam power of the SNS is a prodigious source of neutrinos from the decay of $\pi^+$ 
and $\mu^+$ at rest. These decays produce a well specified flux of neutrinos via 
$\pi^+ \rightarrow \mu^+$ $\nu_\mu$, $\tau_\pi = 2.7 \times 10^{-8}$ s, and $\mu^+ \rightarrow e^+$
 $\nu_e$ $\bar \nu_\mu$, $\tau_\mu = 2.2 \times 10^{-6}$ s.  
The low duty factor of the SNS ($\sim 695$ ns beam pulses at 60 Hz, $DF=4.2 \times 10^{-5}$) is more than 
1000 times less than LAMPF.  This much smaller duty factor provides a reduction in backgrounds 
due to cosmic rays, and allows the $\nu_\mu$ induced events from $\pi^+$ 
decay to be separated from the $\nu_e$ and $\bar \nu_\mu$ induced events from $\mu^+$ decay.

The OscSNS experiment will make use of this prodigious source of 
neutrinos.  The OscSNS detector will be centered 60 meters from the SNS target, in the 
backward direction.  The cylindrical detector design is based upon the LSND and MiniBooNE detectors 
and will consist of an 800-ton tank of mineral oil with a small concentration of b-PBD 
scintillator dissolved in the oil, 
that is covered by approximately 3500 8-inch phototubes for a photocathode coverage of 25\%.  The 
cylindrical design will allow us to map the event rates as a function of L/E, to look for any 
sinusoidal dependence indicative of oscillations.

This experiment will use the mono-energetic 29.8 MeV 
$\nu_\mu$ to investigate the existence of light sterile neutrinos via the neutral-current 
reaction $\nu_\mu C \rightarrow \nu_\mu C^*(15.11 MeV) \rightarrow \nu_\mu C \gamma$.  This reaction 
has the same cross section for all active 
neutrinos, but is zero for sterile neutrinos. An observed oscillation in this reaction is 
direct evidence for sterile neutrinos.  OscSNS can also carry out an unique and decisive test of the 
LSND $\bar \nu_\mu \rightarrow \bar \nu_e$ appearance signal.  
In addition, OscSNS can make a sensitive search for $\nu_e$ disappearance by searching for 
oscillations in the reaction 
$\nu_e C \rightarrow e^- N_{gs}$, where the $N_{gs}$ is identified by its beta decay. It is 
important to note that all cross sections involved are known to two percent or better.

The SNS represents a unique opportunity to pursue a critical neutrino physics program in a cost-effective manner, 
as an intense flux of neutrinos from stopped $\pi^+$ and $\mu^+$ decay are produced during normal SNS operations.  
The existence of light sterile neutrinos would be the first major extension of the Standard Model.  
Sterile neutrino properties are central to dark matter, cosmology, astrophysics, and future neutrino research.  
{\bf The OscSNS experiment would be able to prove whether 
sterile neutrinos can explain these existing short-baseline anomalies.}  

\chapter{Physics Goals}

Observations of neutrino oscillations, and therefore neutrino mass, have been made by solar-neutrino 
experiments at a $\Delta m^2 \sim 8 \times 10^{-5}$ eV$^2$, and by atmospheric-neutrino experiments at 
a $\Delta m^2 \sim 3 \times 10^{-3}$ eV$^2$, where $\Delta m^2$ is the difference in mass squared of the 
two dominant mass eigenstates contributing to the oscillations \cite{fogli}. In addition to these observations, the LSND 
experiment, which took data at Los Alamos National Laboratory (LANSCE) for six years from 1993 to 1998, 
obtained evidence for $\bar \nu_\mu \rightarrow \bar \nu_e$ oscillations at a $\Delta m^2 \sim 1$ eV$^2$ \cite{lsnd}. 
Oscillations at the mass splittings seen by LSND do not fit with well-established oscillation observations 
from solar and atmospheric experiments. The Standard Model, with only three flavors of neutrinos, cannot 
accommodate all three observations. Confirmation of LSND-style oscillations would require further non-trivial 
extensions to the Standard Model.

\begin{figure}
\vspace{5mm}
\centering
\includegraphics[width=10cm,angle=0,clip=true]{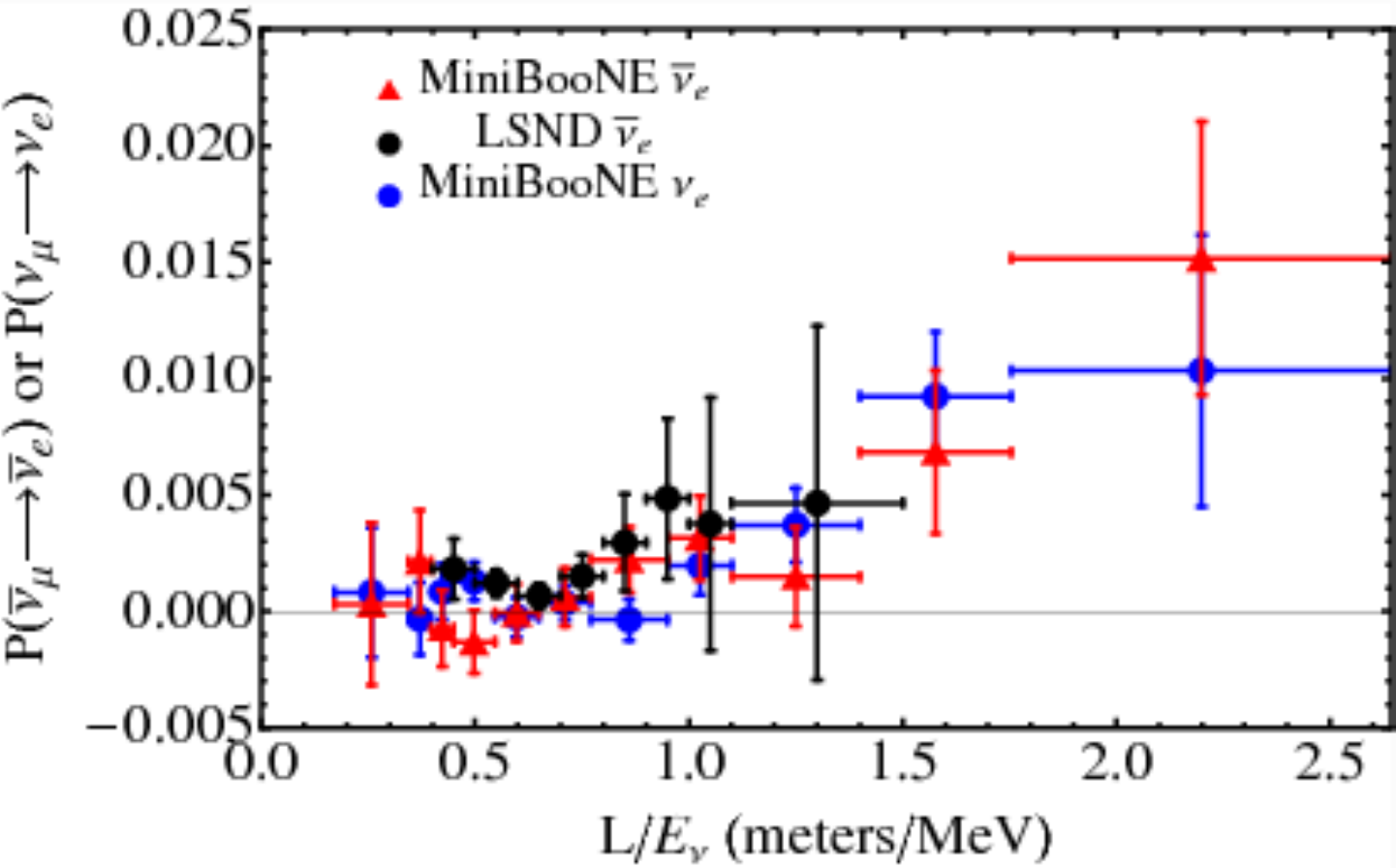}
\vspace{1mm}
\caption{
The probability of $\bar \nu_\mu \rightarrow \bar \nu_e$ from LSND and
$\bar \nu_\mu \rightarrow \bar \nu_e$ and $\nu_\mu \rightarrow \nu_e$ from MiniBooNE as a function of the 
neutrino proper time.
}
\label{L_E}
\end{figure}

\begin{figure}
\vspace{5mm}
\centering
\includegraphics[width=14cm,angle=90,clip=true]{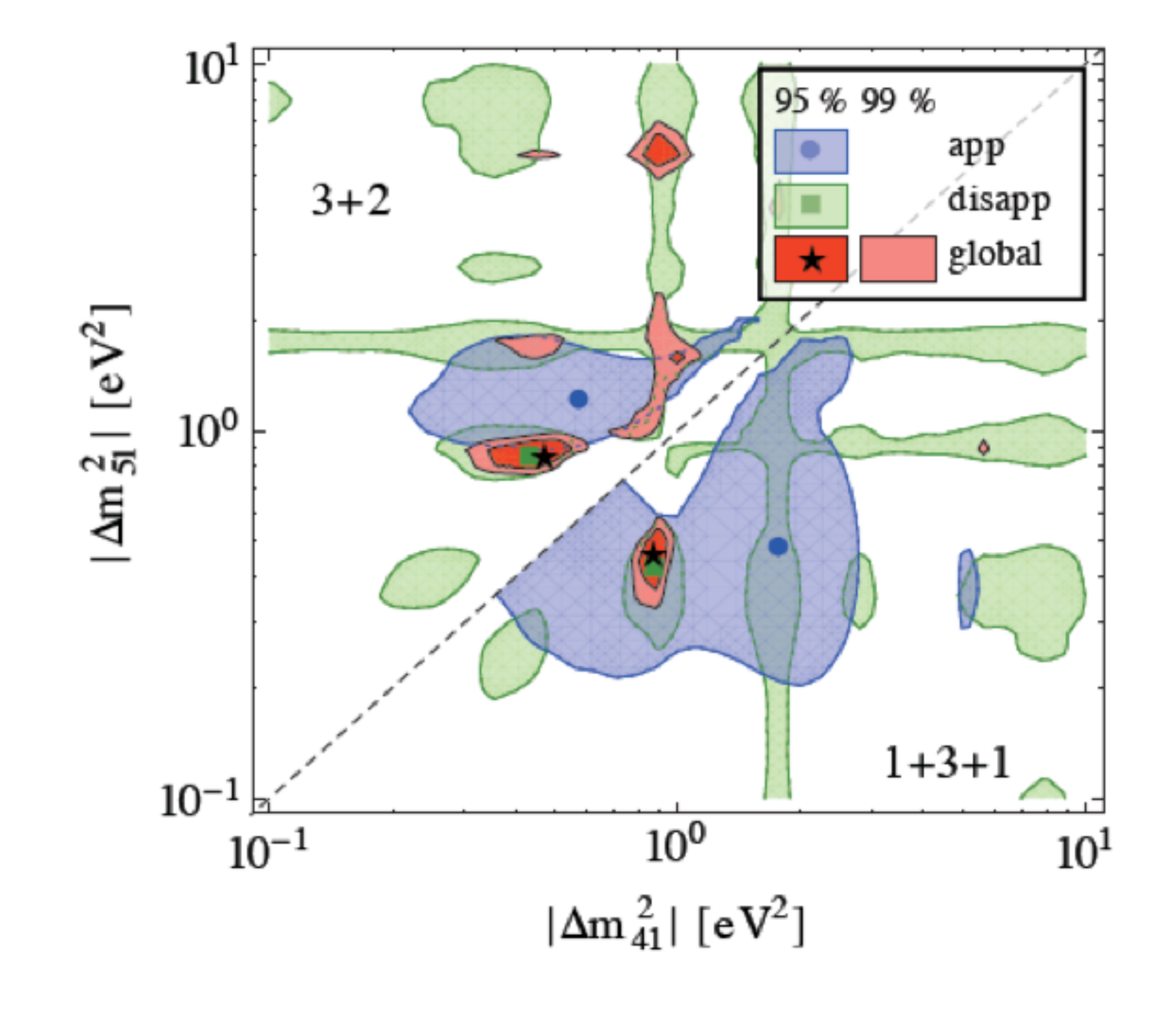}
\vspace{1mm}
\caption{
A global fit to the world
neutrino plus antineutrino data indicates that the world data fit reasonably well to a 3+2 model with
three active neutrinos plus two sterile neutrinos \cite{kopp}.
}
\label{global}
\end{figure}

The MiniBooNE experiment at Fermilab, designed to search for $\nu_\mu \rightarrow \nu_e$ and
$\bar \nu_\mu \rightarrow \bar \nu_e$ oscillations and to further explore the LSND neutrino oscillation 
evidence, has presented separate neutrino and antineutrino oscillation results. 
Combining these results, MiniBooNE 
observes a 3.8$\sigma$ excess of events in the 200-1250 MeV oscillation energy range that is consistent with
the LSND signal~\cite{mb_lowe,mb_osc_anti}. Many of the beyond 
the Standard Model explanations of this excess involve sterile neutrinos, which would have a huge impact 
on astrophysics, supernovae neutrino bursts, dark matter, and the creation of the heaviest elements. Fig. \ref{L_E} 
shows the 
L/E (neutrino proper time) dependence of $\bar \nu_\mu \rightarrow \bar \nu_e$ from LSND and 
$\bar \nu_\mu \rightarrow \bar \nu_e$ and $\nu_\mu \rightarrow \nu_e$ from MiniBooNE. 
The correspondence between the two experiments is striking. Furthermore, Fig. \ref{global} shows fits to the world 
neutrino plus antineutrino data that indicate that the world data fit reasonably well to a 3+2 model with
three active neutrinos and two sterile neutrinos \cite{kopp}.

The SNS \cite{ornl} offers many advantages for neutrino oscillation physics, including known neutrino spectra, well 
understood neutrino cross sections (uncertainties less than a few percent), low duty factor for cosmic ray 
background rejection, low beam-induced neutrino background, and a very high neutrino rate of greater than $10^{15}$/s 
from the decay of stopped pions and muons in the Hg beam dump. Stopped pions produce 29.8 MeV mono-energetic 
$\nu_\mu$ from $\pi^+ \rightarrow \mu^+ \nu_\mu$ decay, while stopped muons produce $\bar \nu_\mu$ and 
$\nu_e$ up to the 52.8 MeV endpoint from $\mu^+ \rightarrow e^+ \nu_e \bar \nu_\mu$ decay. Note that greater than 
99\% of $\pi^-$ and $\mu^-$ capture in Hg before they have a chance to decay, so that hardly any neutrinos 
are produced from either $\pi^- \rightarrow \mu^- \bar \nu_\mu$ or $\mu^- \rightarrow e^- \bar \nu_e \nu_\mu$
decay. 

The SNS neutrino flux is good for probing $\bar \nu_\mu \rightarrow \bar \nu_e$ and
$\nu_\mu \rightarrow \nu_e$ appearance,
as well as $\nu_\mu$ and $\nu_e$ disappearance into sterile neutrinos. The appearance searches
both have a two-fold coincidence for the rejection of background. For $\bar \nu_\mu \rightarrow \bar \nu_e$
appearance, the signal is an $e^+$ in coincidence with a 2.2 MeV $\gamma$: $\bar \nu_e p \rightarrow e^+ n$,
followed by $n p \rightarrow D \gamma$. For $\nu_\mu \rightarrow \nu_e$ appearance, the signal is a mono-energetic
12.5 MeV $e^-$ 
in coincidence with an $e^+$ from the $\beta$ decay of the ground state of $^{12}N$: $\nu_e~^{12}C \rightarrow
e^-~^{12}N_{gs}$, followed by $^{12}N_{gs} \rightarrow ~^{12}C e^+ \nu_e$. The $\nu_\mu$ disappearance 
search will detect 
the prompt 15.11 MeV $\gamma$ from the neutral-current reaction $\nu_\mu C \rightarrow \nu_\mu C^*$(15.11). 
This reaction has been measured by the KARMEN experiment, which has determined a cross section that is 
consistent with theoretical expectations \cite{karmen}. However, the KARMEN result was measured in a sample of 86 events, 
and carries a 20\% total error. OscSNS will be able to greatly improve upon the statistical and systematic 
uncertainties of this measurement. If OscSNS observes an event rate from this neutral-current reaction that 
is less than expected, or if the event rate displays a sinusoidal dependence with distance ($L/E$ can be measured
with a resolution of $\sim 1\%$), then this will be 
evidence for $\nu_\mu$ oscillations into sterile neutrinos. The $\nu_e$ disappearance search will measure the
reaction $\nu_e C \rightarrow e^- N_{gs}$ followed by $N_{gs}$ beta decay. This reaction is very clean with a very
low background due to the two-fold signature. Furthermore, the neutrino energy is approximately equal to the 
electron energy, so that $L/E$ can be measured with a resolution of $<5\%$, which allows for a sensitive test
for oscillations in the detector.

In addition to the neutrino oscillation searches, OscSNS will also make precision cross section measurements of 
$\nu_e C \rightarrow e^- N_{gs}$ scattering and $\nu e^- \rightarrow \nu e^-$ elastic scattering. The former reaction 
has a well-understood cross section and can be used to normalize the total neutrino flux, while the latter 
reaction, involving $\nu_\mu$, $\nu_e$, and $\bar \nu_\mu$, will allow a precision measurement of 
$\sin^2 \theta_W$.

\begin{figure}
\vspace{5mm}
\centering
\includegraphics[width=5.5cm,angle=90,clip=true]{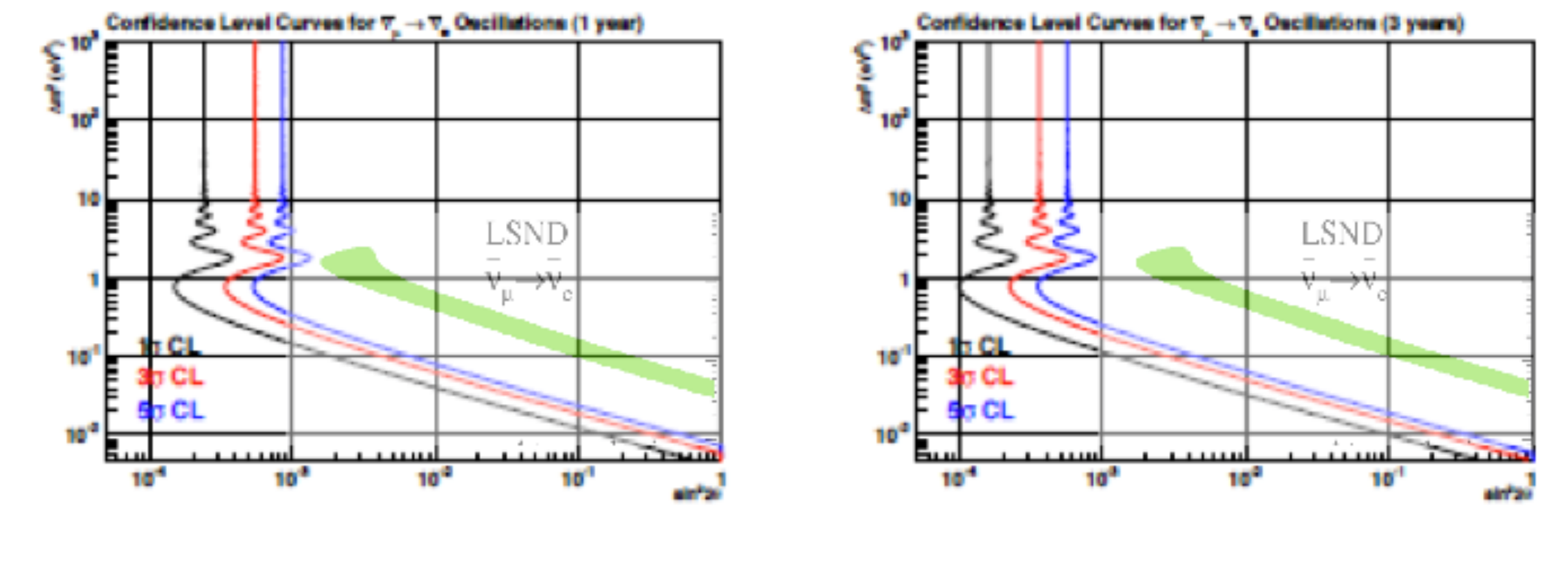}
\vspace{1mm}
\caption{
The OscSNS sensitivity curves for the simulated sensitivity to $\bar \nu_\mu \rightarrow \bar \nu_e$  
oscillations after two (left) and six (right) {\it calendar} years of operation, assuming a 50\% beam on-time (one and three years of running at 100\% beam-on). Note that it has more than 5$\sigma$
sensitivity to the LSND result in 2 years.
}
\label{app}
\end{figure}

\begin{figure}
\vspace{5mm}
\centering
\includegraphics[width=5cm,angle=90,clip=true]{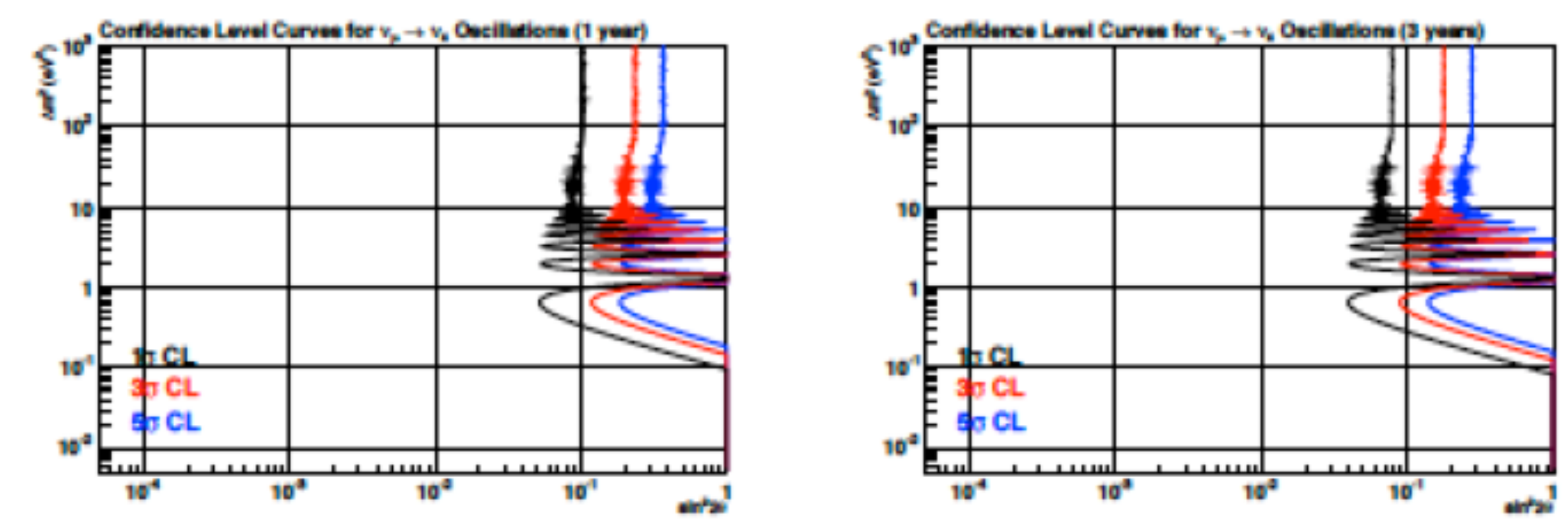}
\vspace{1mm}
\caption{
The OscSNS sensitivity curves for $\nu_\mu$ disappearance for two and six {\it calendar} years, respectively.  
}
\label{disap}
\end{figure}

\begin{figure}
\centering
\includegraphics[width=8cm,angle=90]{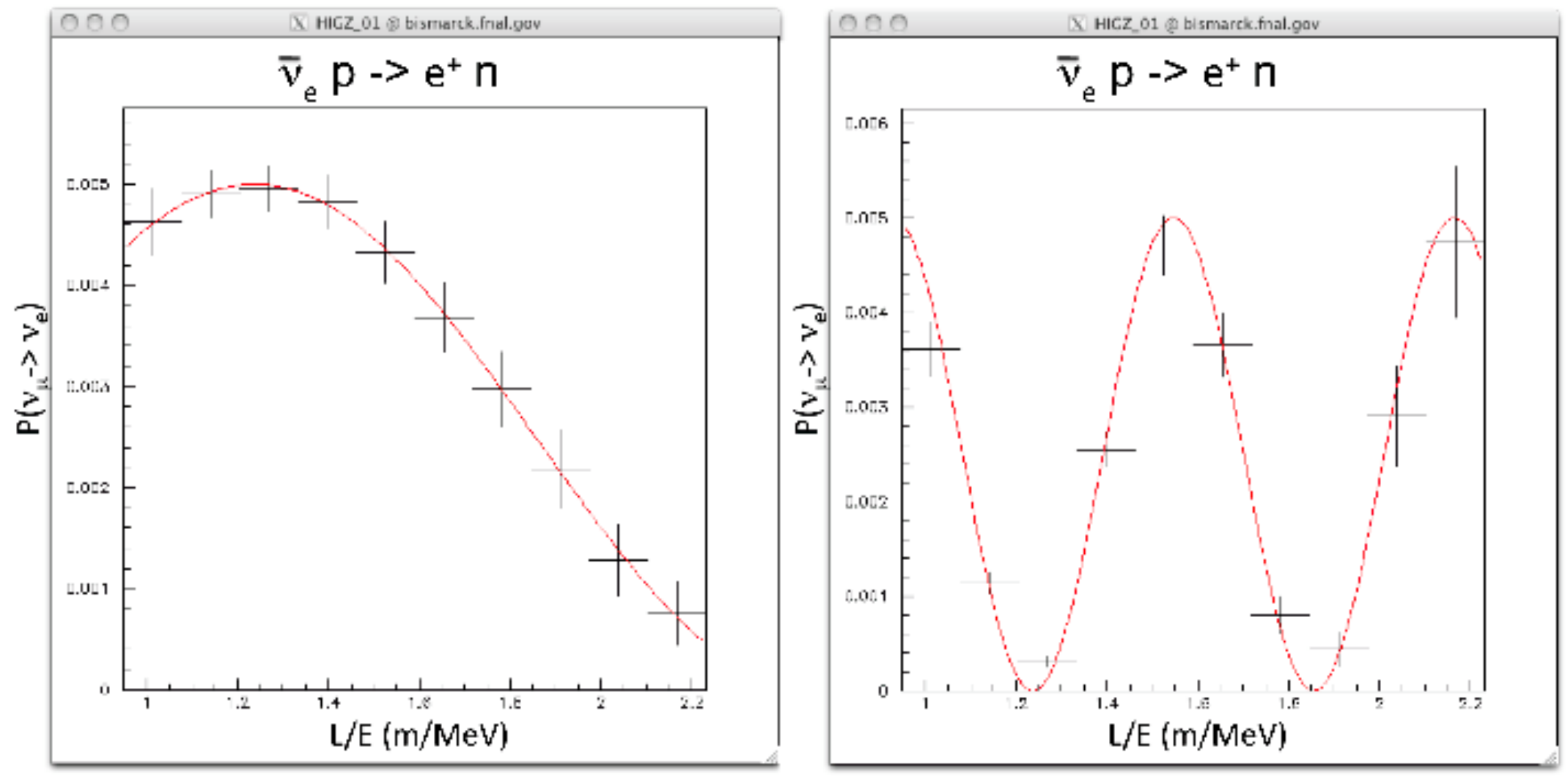}
\caption{The expected oscillation probability from $\bar \nu_e$ appearance as a
function of $L/E$ for
$\sin^22\theta = 0.005$ and $\Delta m^2 = 1$ eV$^2$ (left plot) and $\Delta m^2 = 4$ eV$^2$ (right plot).
The plot assumes ten calendar years of data collection at 50\% beam live-time.}
\label{loe_nuebar}
\end{figure}

Table~\ref{table:nums_hr} summarizes the expected event sample sizes for the disappearance and appearance
oscillation searches, per calendar year, at the OscSNS.  Figure~\ref{app} shows the expected sensitivity for 
$\bar \nu_e$ appearance after two and six calendar years of run time, while Figure \ref{disap} presents 
the sensitivity for $\nu_\mu$ disappearance.  The LSND allowed region is fully
covered by more than 5$\sigma$.  The cylindrical design of the OscSNS detector allows for detection of 
oscillations as a function of $L/E$, as shown in Figure~\ref{loe_nuebar}.  Such an observation would
prove that any observed excess is due to short-baseline neutrino oscillations, and not due to a misunderstood 
background. The oscillation sensitivities are further improved by the construction of a near detector 
and by the planned construction of a second target station that is located at a longer neutrino baseline.

\vspace{2cm}
\begin{table}[hb]
\begin{center}
\begin{tabular}{|c|c|c|} \hline
 \multicolumn{1}{c}{Channel} &  \multicolumn{1}{c}{Background} & \multicolumn{1}{c}{Signal}  \\ \hline \hline
 \multicolumn{3}{c}{Disappearance Search}  \\ \hline
$\nu_\mu \ ^{12}C \rightarrow \nu_\mu \ ^{12}C^*$    &      & \\
$\nu_e \ ^{12}C \rightarrow \nu_e \ ^{12}C^*$    &      & \\
$\bar{\nu_\mu} \ ^{12}C \rightarrow \bar \nu_\mu \ ^{12}C^*$ & 1060 $\pm$ 36 & 3535 $\pm$ 182  \\ \hline
$\nu_\mu \ ^{12}C \rightarrow \nu_\mu \ ^{12}C^*$  & 224 $\pm$ 75 & 745 $\pm$ 42  \\ \hline
$\nu_e \ ^{12}C \rightarrow e^- \ ^{12}N_{gs}$  & 24 $\pm$ 13 & 2353 $\pm$ 123  \\ \hline
\multicolumn{3}{c}{Appearance Search} \\ \hline
$\bar \nu_\mu \rightarrow \bar \nu_e$: $\bar \nu_e \ ^{12}C \rightarrow e^+  \ ^{11}B \ n$  &  & \\
$\bar \nu_\mu \rightarrow \bar \nu_e$: $\bar \nu_e \ p \rightarrow e^+  \ n$          & 42 $\pm$ 5 & 120 $\pm$ 10 \\ \hline
$\nu_\mu \rightarrow \nue$: $\nue \ ^{12}C \rightarrow e^- \ ^{12}N_{gs}$  & 12 $\pm$ 3 & 3.5 $\pm$ 1.5 \\
\hline
\hline
\end{tabular}
\caption{Summary of per calendar year event rate predictions for a detector located at the SNS, centered at a distance of 60 meters from the interaction point, at $\sim$150
degrees in the backward direction from the proton beam.  The first column is the 
oscillation channel, the second column is the
expected intrinsic background, and the third column is the expected signal for appearance searches and the total
number of events for disappearance searches.  All event rates account for a 50\% detector efficiency, a 50\%
beam-on efficiency, a fiducial volume of 523 m$^3$, and are in units of expected events per calendar
year.  Appearance signal estimates assume a 0.26\% oscillation probability.}
\label{table:nums_hr}
\end{center}
\end{table}

\chapter{Path to Funding}

The OscSNS experiment is proposing to operate at the SNS, a laboratory currently funded under Basic Energy 
Sciences (BES).  There isn't a strong history of funding at Oak Ridge by the High Energy Physics (HEP) program, and 
that makes our path forward for requesting funding less straight forward.  This chapter describes our justification for 
running at Oak Ridge, and the path we propose to take to achieve full funding.

\section{Why the SNS?}

There are currently 8 national laboratories that have or are seeking funding from the 
Department of Energy High Energy Physics Intensity Frontier (DOE HEP IF) program.  
Due to funding challenges, it is valid to question why 
we are seeking to open another laboratory for DOE HEP funding, instead of proposing our 
experiment at a lab that is already being funded under this program. The simple answer to this question is two-fold.  
The neutrinos at the SNS come for free, as they are created during the process that produces the spallation neutrons.  
This dramatically increases the cost effectiveness of the experiment 
and provides ``more bang for your buck'' than can be had at any 
other facility.  Also, the energy spectrum of the neutrino beam provides a golden opportunity for precision 
short-baseline neutrino experiments and the possible discovery of sterile neutrinos.
Due to the low-energy neutrinos from stopped $\pi^+$ and $\mu^+$ decays, the OscSNS experiment has 
the potential of actually observing oscillations in the detector from both charged-current and
neutral-current interactions, definitively proving the existence of sterile neutrinos.  This is not possible at a 
higher energy neutrino source, where the parent pions and muons decay in flight.

\subsection{SNS vs FNAL facilities}

FNAL provides high-energy neutrinos from $\pi^+$ and $K^+$ decay in flight (DIF) that are ideal for long-baseline
neutrino experiments at the atmospheric neutrino mass scale of $\Delta m^2 \sim 2\times 10^{-3}$ eV$^2$.
However, for short-baseline neutrino experiments at a mass scale of $\Delta m^2 \sim 1$ eV$^2$, there is a
clear advantage in using neutrinos from $\pi^+$ and $\mu^+$ decay at rest (DAR).  For DAR neutrino sources, 
the neutrino energy spectra are known perfectly, the neutrino 
cross section uncertainties are $<2\%$, and the low neutrino energies allow
the possibility of actually observing an oscillation pattern of event rates in the detector. 

FNAL has two running neutrino beamlines: the NuMI beamline, and the Booster beamline (BNB).  Both of these 
are DIF neutrino sources with a peak energy of 700 MeV (BNB) and 4 GeV (NuMI).  In the BNB beam stop 
there are also neutrinos produced from $\pi^+$ and $\mu^+$ decay at rest.  There are several factors that make this a 
far less attractive DAR source.  The proton beam power at FNAL is ~30 kW, compared to the ~1400 kW 
of the SNS.  As pion production, and therefore neutrino production, is roughly proportional to beam power, this results in a far 
less intense neutrino beam for a detector placed 60 meters away from the beam stop.  The identification of an oscillatory 
pattern in event rates is crucial for definitively claiming discovery of sterile neutrinos.  The oscillation pattern will 
only be apparent in the detector if the neutrino production region is small.  At FNAL, the 
pions and muons flying through the air come to rest in the beam stop located 50 meters from the proton-Be target 
region, in addition to coming to rest near the Be target and along the walls of the 50 meter decay pipe.  
This results in the spreading out of the DAR neutrino beam.  
Use of the Booster DAR neutrino source will tend to wash out any oscillation pattern and will result in an experiment that 
will observe an excess or deficit of events, but that will be unable to conclusively observe oscillations 
to sterile neutrinos.  Compare this to the SNS, where pions produced in the proton-Hg interaction are immediately brought 
to rest inside of the target (at $>$99\%).  Thus, the uncertainty of the neutrino production location at the SNS is 
limited to the size of the Hg target, or $\sim 50$ cm.

However, the \underline{results found by OscSNS will directly and positively impact the Intensity Frontier program}
\underline{at Fermilab}.  
The resolution to the hints of physics beyond the Standard Model in the neutrino sector will dramatically impact the program 
plan at Fermilab, allowing them to best choose next generation experiments to either elucidate further the properties of the 
sterile neutrino, or to pursue other Intensity Frontier goals should the existence of sterile neutrinos at this energy scale be 
disproven.

\subsection{SNS vs All Other World Facilities}

The only other world facility that is comparable to the SNS is the J-PARC spallation source in Japan, which
has similar proton beam power (1.0 MW for J-PARC vs. 1.4 MW for SNS).  
J-PARC has the advantage of a lower duty factor ($5 \times 10^{-6}$
for J-PARC vs. $42 \times 10^{-6}$ for SNS), while the SNS has the advantage of a lower proton beam energy
(3 GeV for J-PARC vs. 1 GeV for SNS). The lower proton energy is desirable due to the lack of backgrounds 
from produced kaons
and a lower energy neutron background.  The lower energy proton beam at the SNS will cause more of the produced pions to come to 
rest inside the target, resulting in a smaller size of the neutrino source. As mentioned above, the smaller the neutrino 
source the more clearly the pattern of oscillations to sterile neutrinos can be distinguished in the detector.  


\section{Why OscSNS is Unique}

OscSNS is a unique experiment that will make full use of the capabilities of the SNS facility. With
OscSNS, the neutrino energy spectra are known perfectly, the neutrino oscillation channel cross section uncertainties
are $<2\%$, there is no bias from nuclear effects in the determination of neutrino energy, the
intrinsic $\bar \nu_e$ background is very low at the level of 0.1\%, and the potential exists 
to observe short-baseline oscillations in the detector for both charged-current and neutral-current 
interactions to prove (or disprove) the existence of sterile neutrinos. OscSNS will have excellent sensitivity for 
$\bar \nu_\mu \rightarrow \bar \nu_e$ appearance, as well as $\nu_e$ and $\nu_\mu$ disappearance.
Overall, OscSNS will be able to make a definitive test of the current evidence for short-baseline
neutrino oscillations, which imply the existence of physics beyond the Standard Model.

\subsection{Complementariness to Other Experiments}

OscSNS is complementary to the MicroBooNE and NuSTORM experiments at FNAL 
and to other short-baseline experiments
around the world (IsoDAR). If light sterile neutrinos exist in nature, then it will be important to
confirm the oscillation signal in a variety of experiments at different energy scales, especially 
if there are two or more types of sterile neutrinos and short-baseline CP violation. 
MicroBooNE will play an important role in determining whether the excess of events observed by MiniBooNE is due to
electron events or photon events; however, MicroBooNE by itself will not be able to prove 
whether there are short-baseline neutrino oscillations and sterile neutrinos.  Additionally, IsoDAR will only be able to 
observe oscillations in the CC channel; they will be unable to measure NC scattering, the smoking gun for sterile neutrinos 
vs some other anomalous oscillation theory.

The OscSNS experiment is an order of magnitude less expensive and can
be built more quickly than NuSTORM~\cite{nustorm}. Indeed, if OscSNS proves the existence of sterile
neutrinos, then these results will provide stronger motivation for building NuSTORM (as well as Project X).  When comparing 
OscSNS and NuSTORM, it is important to compare the energy spectrum of the neutrino beam as well as the detector 
technology.  NuSTORM is proposing to use a DIF neutrino source, with a far detector composed of steel and scintillator.  
NuSTORM will not be able to observe oscillations of a neutral-current interaction, the golden channel for claiming 
observation of oscillations to sterile neutrinos. Furhermore, OscSNS will not be affected
by nuclear effects in the reconstruction of neutrino energy, which is not the case for NuSTORM,
and the OscSNS neutrino cross sections are all known to within 1-2\%. 
Finally, the strongest argument for OscSNS when compared to NuSTORM, 
is that OscSNS will be a direct test of LSND. If the LSND signal is due to some exotic physics other than neutrino 
oscillations, such as Lorentz violation, then OscSNS will be able to discover this new physics.  
This may not possible with the NuSTORM experiment.

\subsection{DOE Mission Need}

The question of short-baseline oscillations and sterile neutrinos is one of the most pressing
issues in neutrino physics today. Light, sterile neutrinos would have a big impact on high-energy physics, 
nuclear physics, and astrophysics, and would contribute to the dark matter of the universe. Furthermore,
short-baseline oscillations would affect present and future long-baseline neutrino experiments.
Therefore,
OscSNS represents a unique opportunity to provide a definitive test of short-baseline oscillations and
to prove the existence of sterile neutrinos in a timely manner.


\section{Path to Funding}

The SNS currently does not have facilities in place for acceptance of an LOI, or for presentation 
to a PAC, for non-neutron experiments. Therefore, we propose the following course of action:  

\begin{itemize}
\item Visit SNS, present physics plan.  (done: April 12, 2013)
\item Attend Snowmass, garner support from the community. (done)
\item Obtain letter of support from SNS management and have it sent to DOE. (done)
\item Submit R\&D proposals to DOE for the following ground work (Fall, 2013):
        \begin{itemize}
        \item design new electronics
        \item test oil and scintillators from various sources
        \item develop simulations for the main detector, including reconstruction and particle ID algorithms
        \item develop improved neutrino flux simulations
        \end{itemize}
\item Submit white paper to DOE.
\end{itemize}

Following the DOE Critical Decision Approval process, we believe this program 
will allow us to solidly argue for CD-0 and CD-1 approval.

Upon completion of the R\&D program, we will include results from the R\&D program into 
the white paper, producing a Technical Design Report.  The TDR will be submitted to the DOE for 
consideration in the MIE funding process.

\chapter{The SNS Neutrino Source}

\begin{figure}
\centering
\includegraphics[scale=0.65,angle=0]{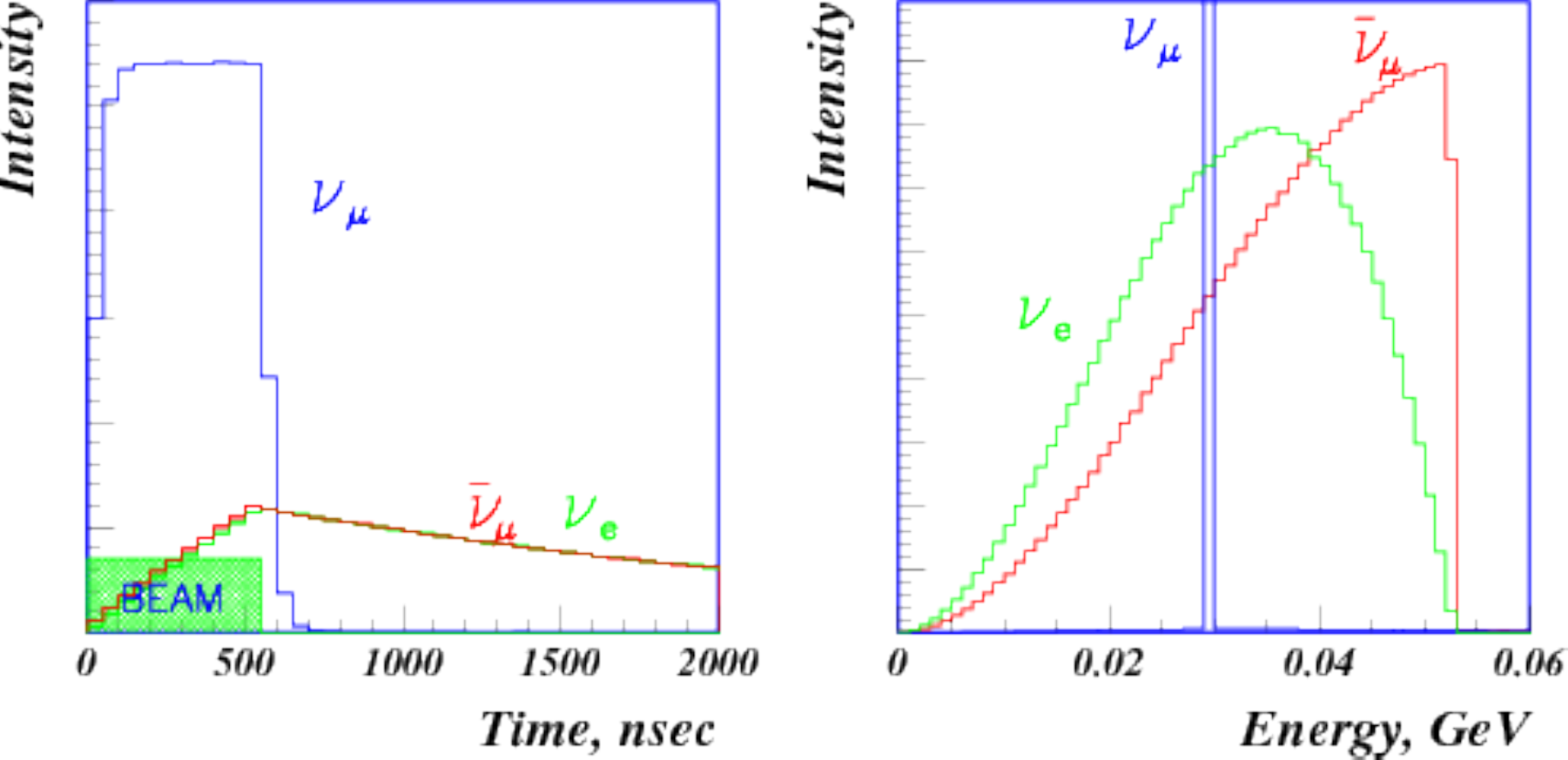}
\caption{The neutrino time and energy spectra 
of the different neutrino species produced isotropically
from a stopped pion source \cite{ornl}.}
\label{sns_flux}
\end{figure}

To search for neutrino oscillations in the mass range $\Delta m^2 > 0.1$ eV$^2$
requires an intense source of well characterized neutrinos. 
The decay of stopped pions from the 1.4 MW, 1.3 GeV, short duty-cycle SNS \cite{ornl}
proton beam 
provides such a source. Neutrinos from stopped pion decay
have a well defined flux, well defined energy spectrum, and measurements of their interactions tend to have low background rates.
The dominant decay scheme that produces neutrinos from a stopped pion
source is 
\begin{equation}
 \pi^+ \rightarrow \mu^{+} \numu, \hspace{0.5cm} \tau = 26 \hspace{0.1cm} \mbox{
ns}
\end{equation}
followed by
\begin{equation}
\mu^+ \rightarrow e^+ \numubar \nu_e, \hspace{0.5cm} \tau = 2.2 \hspace{0.1cm} \mu \mbox{s}.
\end{equation} 
The neutrinos from stopped $\pi^-$'s are highly suppressed because the
negative pions are absorbed in the surrounding target
material.  Thus, neutrinos from the $\pi^-$ decay chain are
significantly depleted and can be estimated from the measured $\nu_\mu$,
$\bar \nu_\mu$, and $\nu_e$ flux. Fig. \ref{sns_flux} shows the neutrino
time and energy spectra from the SNS stopped pion source.  As shown in the
right hand plot, the $\nu_\mu$ energy is mono-energetic ($E_{\nu_\mu} =
29.8$\,MeV), while the
$\bar \nu_\mu$ and $\nu_e$ have known Michel energy
distributions with an end-point energy of
52.8\,MeV.  Furthermore, the left hand plot shows the SNS beam timing
and the time distributions for the three neutrino species.  
With a simple beam-on timing cut, one can obtain a fairly pure
$\nu_\mu$ sample with only a 14\% contamination of $\bar \nu_\mu$ and 
$\nu_e$ each.  
This remaining background is easily measured from the time distribution
and subtracted.  

The expected proton rate from the SNS of $2.2\times 10^{23}$ protons/yr,
coupled with a yield of 0.12 neutrinos per proton, produces
$2.8\times 10^{22}$ $\nu$/yr for 100\% operation efficiency.
Furthermore, the SNS is planning a future upgrade 
that will deliver MW beams to two sources, providing an interesting
experimental environment of multiple baselines with a single detector.

A key component of the neutrino oscillation measurement
is the physical size  of the stopped pion source, which adds an uncertainty
to the neutrino path length. For the SNS, the compact liquid mercury
target will contribute approximately 25\,cm (FWHM), 
or $\sim 0.4\%$ to the neutrino path
length uncertainty.  

\section{The Accelerator and the Target Station}
 
\medskip
 
The Spallation Neutron Source (SNS), now operating at the Oak Ridge
National
Laboratory, produces an intense 1.4 MW pulsed-proton beam.
It
consists of a 493 meter-long linac, an accumulator ring with a radius of just over 35 meters, and a
target station. A second target station is planned in the future.
The SNS is used primarily
for material science research, however, it also is the world's most
intense accelerator neutrino source.
The high intensity and the pulsed nature of the SNS
provide an ideal laboratory for neutrino physics research at medium energies.
The short-pulsed beam (695 ns, 60 Hz) provides a virtually
cosmic-ray-free
measurement of various neutrino interactions. It also provides the ability to
distinguish between neutrinos from pion decay and muon decay. The only neutrino
detector with this capability was the much smaller KARMEN experiment at the Rutherford
Laboratory with significantly lower beam intensity ($200 \mu$A).
 
\begin{figure}
\centering
\includegraphics[width=10cm,angle=90]{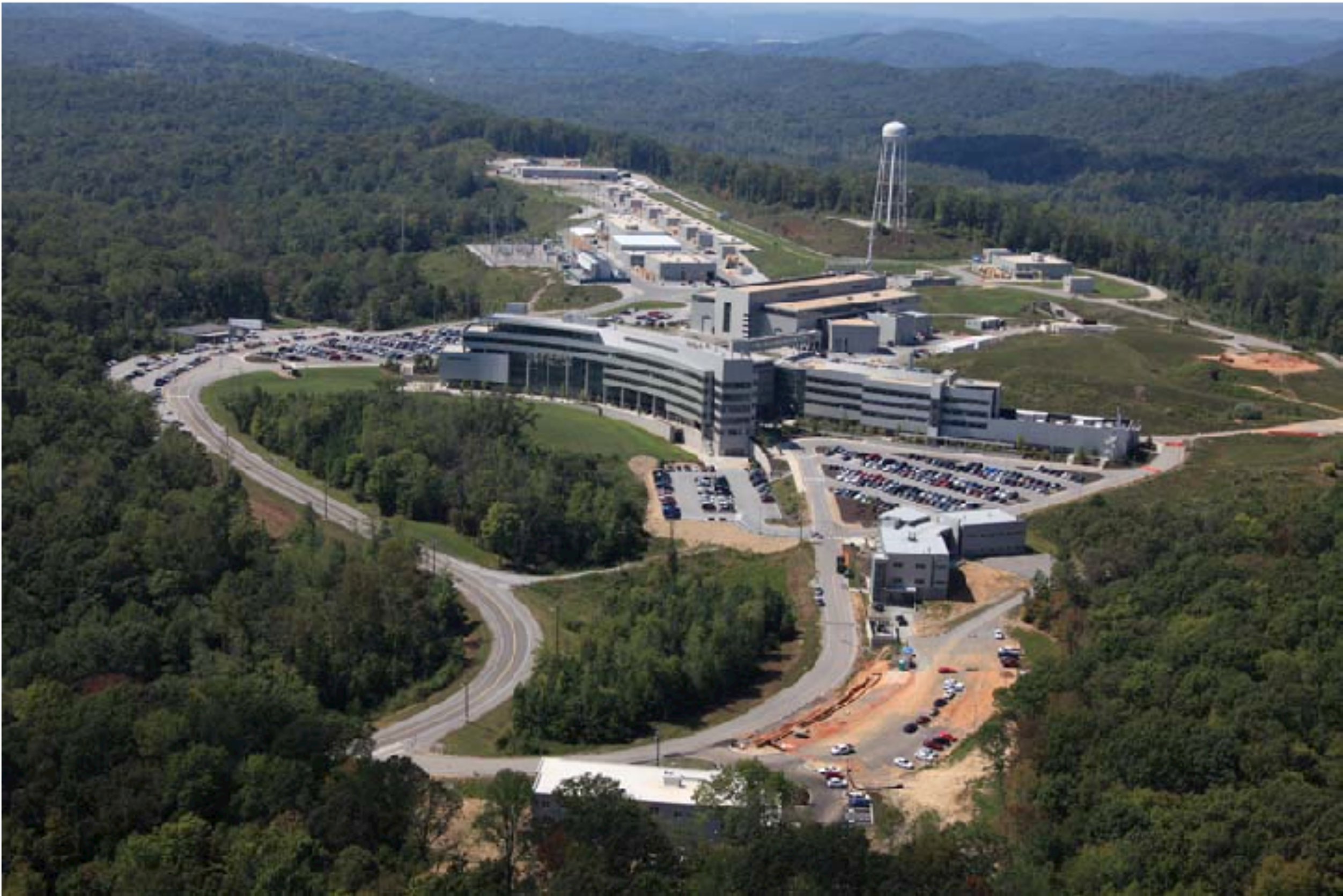}
\caption{A photograph of the completed SNS facility.}
\label{SNS_photo}
\end{figure}

\begin{figure}
\centering
\includegraphics[scale=0.65]{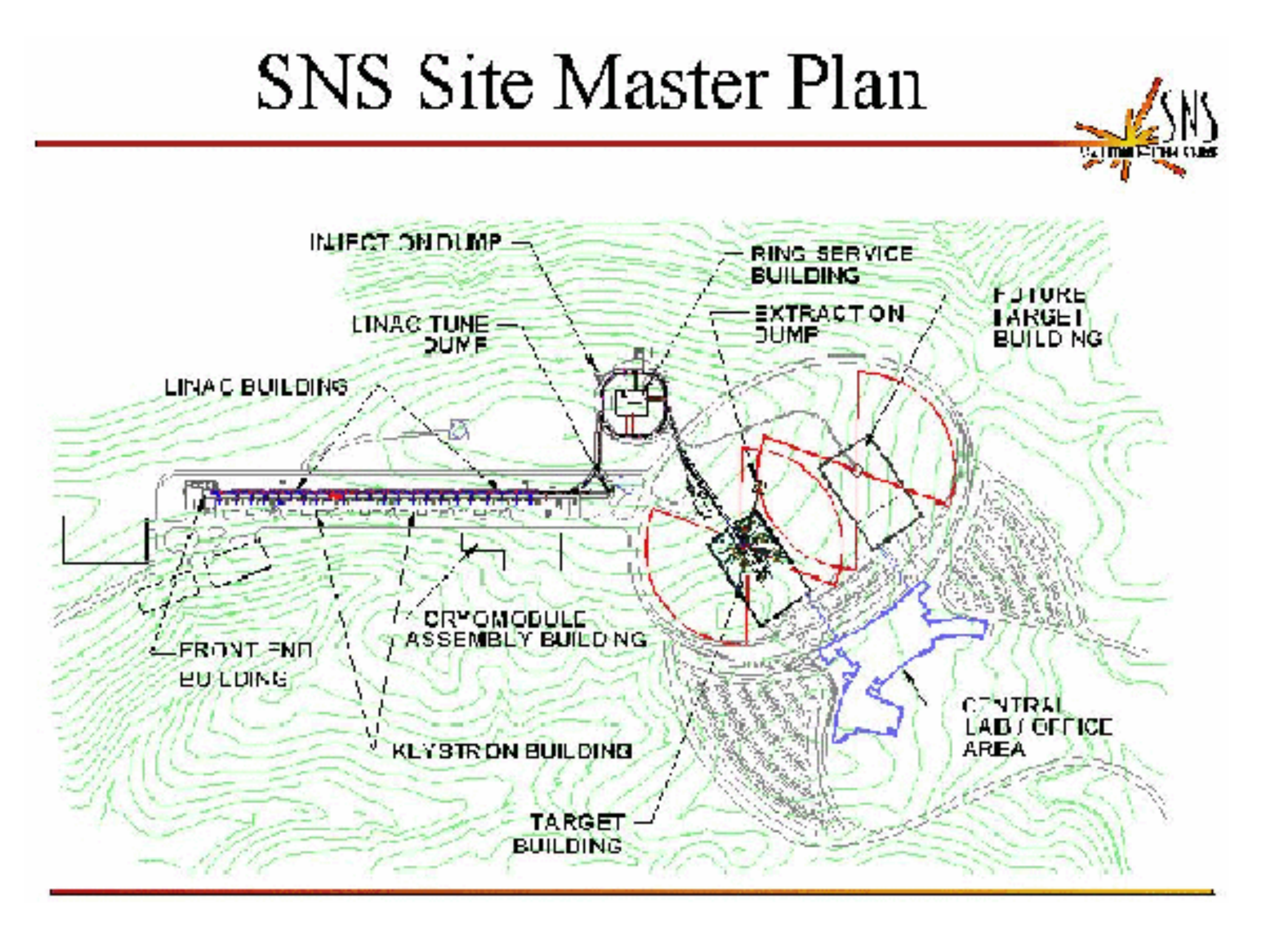}
\caption{Plan view of the SNS facility.  The proposed second target hall is located in the right portion of the view.}
\label{fig5}
\end{figure}

A photograph of the completed SNS facility is shown in Fig.~\ref{SNS_photo},
while Fig.~\ref{fig5} shows a plan view of the facility.
Note that in the plan view the future second target station is shown.
It will have  an intensity of 0.6 mA and repetition rate of 10 Hz.
The two target stations will receive beams at different time slots.
Table~\ref{snspar} shows the design parameters of the SNS.

\begin{figure}
\centering
\includegraphics[scale=0.65]{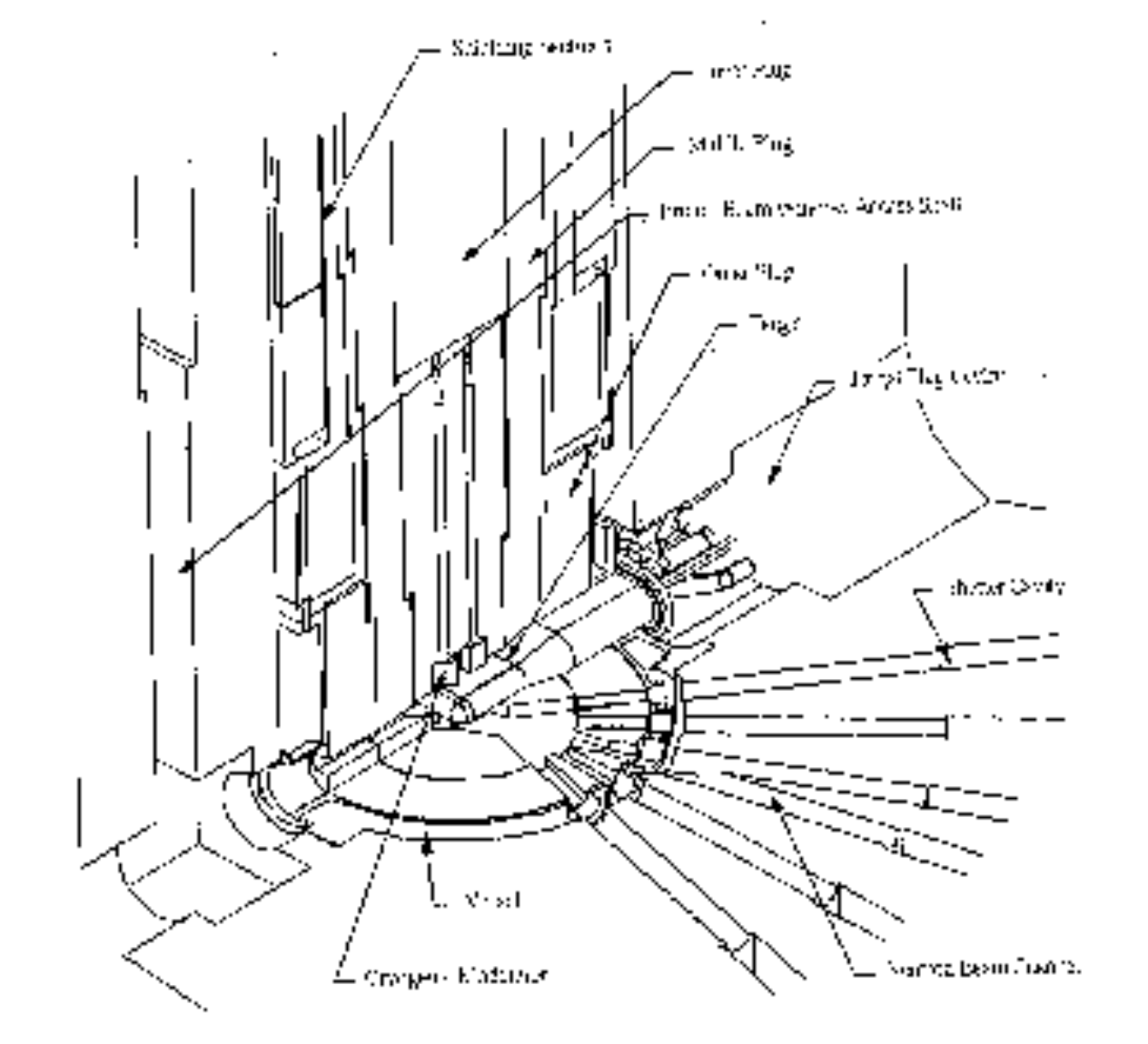}
\caption{Perspective view of the target station.}
\label{fig6}
\end{figure}

\begin{figure}
\centering
\includegraphics[scale=0.65]{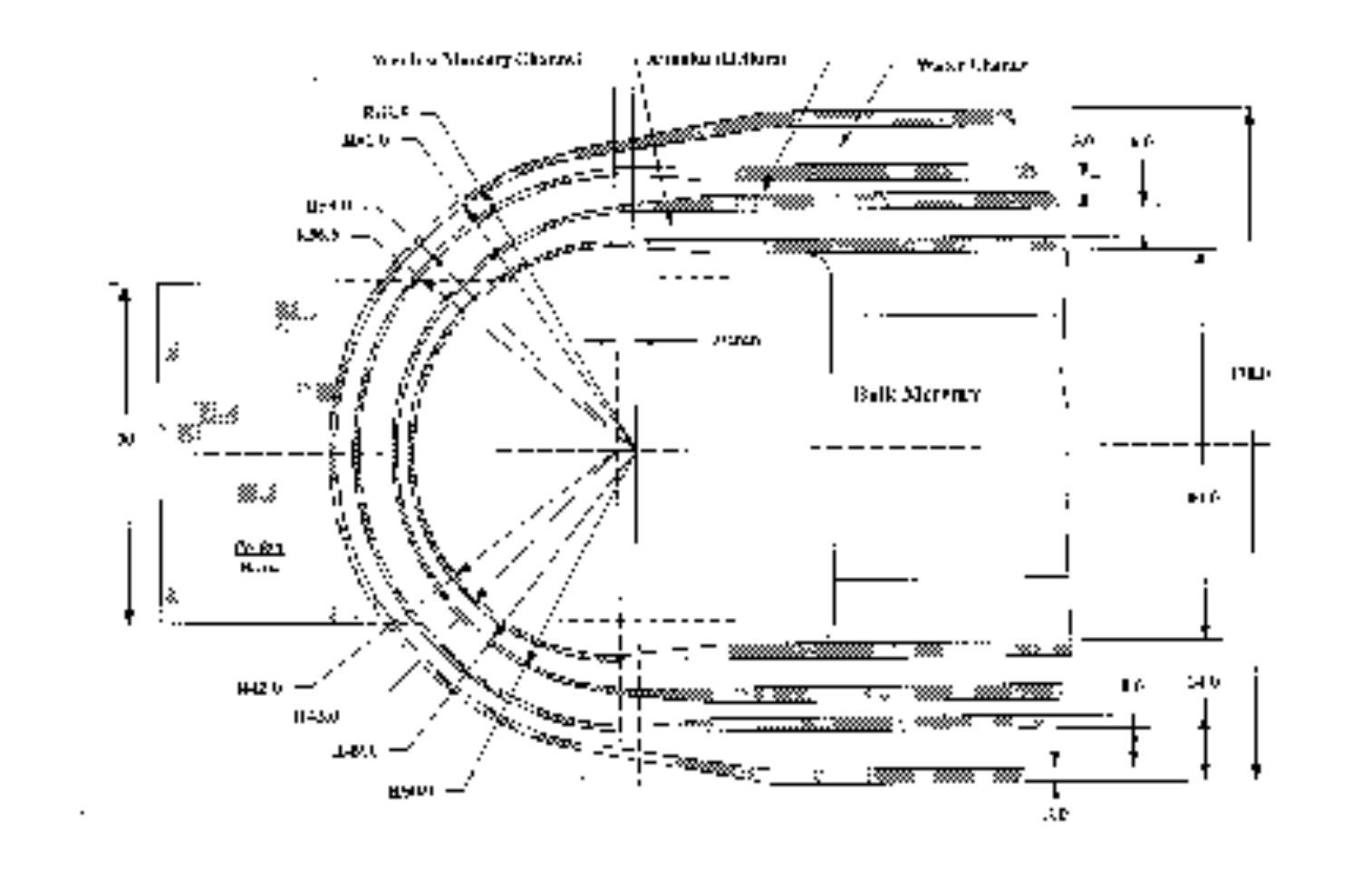}
\caption{An enlargement of the mercury target.}
\label{fig7}
\end{figure}

Fig.~\ref{fig6} shows a cross section of the target station. A beam of
protons enters from the lower left and strikes the target. The beam has
a 7-cm by 20-cm spot size in order to reduce the local heat load in
the target. The target, shown in Fig.~\ref{fig7}, has dimensions of 40 cm
by 10.4 cm by 50 cm. The mercury is contained within a multiple wall structure
made of 316-type stainless steel. To remove heat, the mercury of the target
is constantly circulated at the rate of 140 kg/sec. Room temperature
moderators, filled with water, are placed under the target. Cryogenic
liquid hydrogen moderators are located at the top. Target and moderators are
surrounded by a lead reflector which  extends at least to a radius of 1 meter
around the target. All of this structure is encapsulated inside a thick steel vessel,
to prevent
neutrons from escaping into the experimental hall. There are 18 neutron channels
looking at the moderators, rather than at the target. Shutters are provided
on each channel.

\begin{figure}
\centering
\includegraphics[scale=0.5,angle=0]{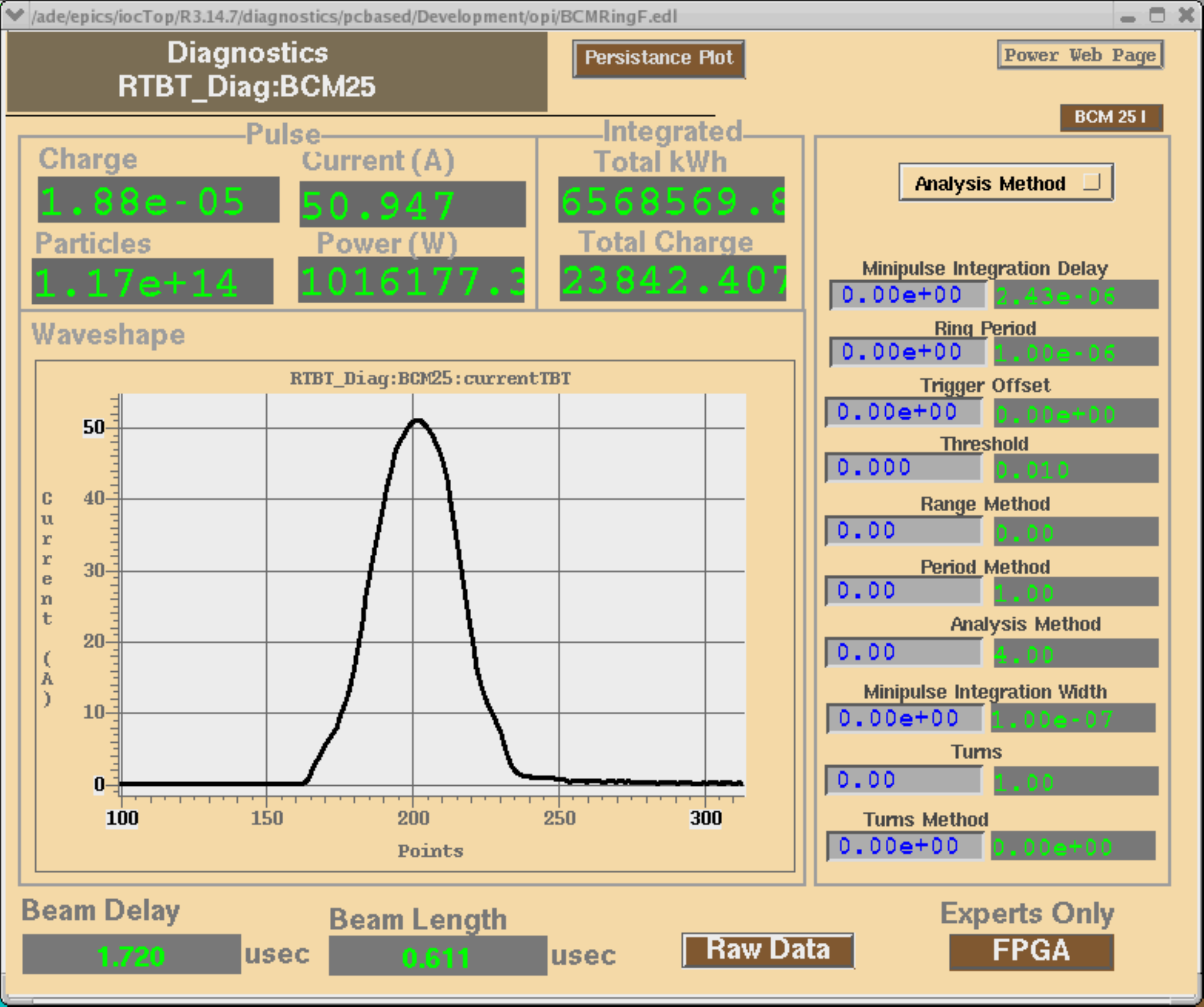}
\caption{The time distribution of a typical beam pulse, where the
vertical axis is the current in Amperes and the horizontal axis is the
time in units of 10 ns.}
\label{SNS_pulse}
\end{figure}

Fig. \ref{SNS_pulse} shows the time distribution of a typical beam pulse, where the
vertical axis is the current in Amperes and the horizontal axis is the 
time in units of 10 ns. The beam pulse has a total width of approximately 695 ns.

\begin{table}[htbp]
\centering
\begin{tabular}{|c|c|}
\hline
   &{\bf baseline }  \\
\hline
linac length & 493 m  \\
accumulator ring circumference &221 m  \\
beam power on the target &1.4 MW \\
beam energy on the target &1.3 GeV  \\
average beam current &1.54 mA \\
repetition rate &60 Hz  \\
ion type, source-linac &$H^{-}$  \\
linac-beam duty factor &$6.2\%$  \\
number of injected turns &1225  \\
particles stored in ring &$1.5 \times 10^{14}$ \\
extracted pulse length &695 nsec  \\
peak current on target &45 A \\
target & mercury  \\
beam spot on target &$7 \times 20$ cm \\
moderators ambient &2 (water) \\
moderators cryogenic &2 ($LH_{2}$) \\
neutron beam ports &18 \\
\hline
\end{tabular}
\caption{SNS design parameters.}
\vspace{0.2in}
\label{snspar}
\end{table}
 
\chapter{The OscSNS Detector}

\section{Overview}

\begin{figure}
\vspace{5mm}
\centering
\includegraphics[scale=0.5,angle=90]{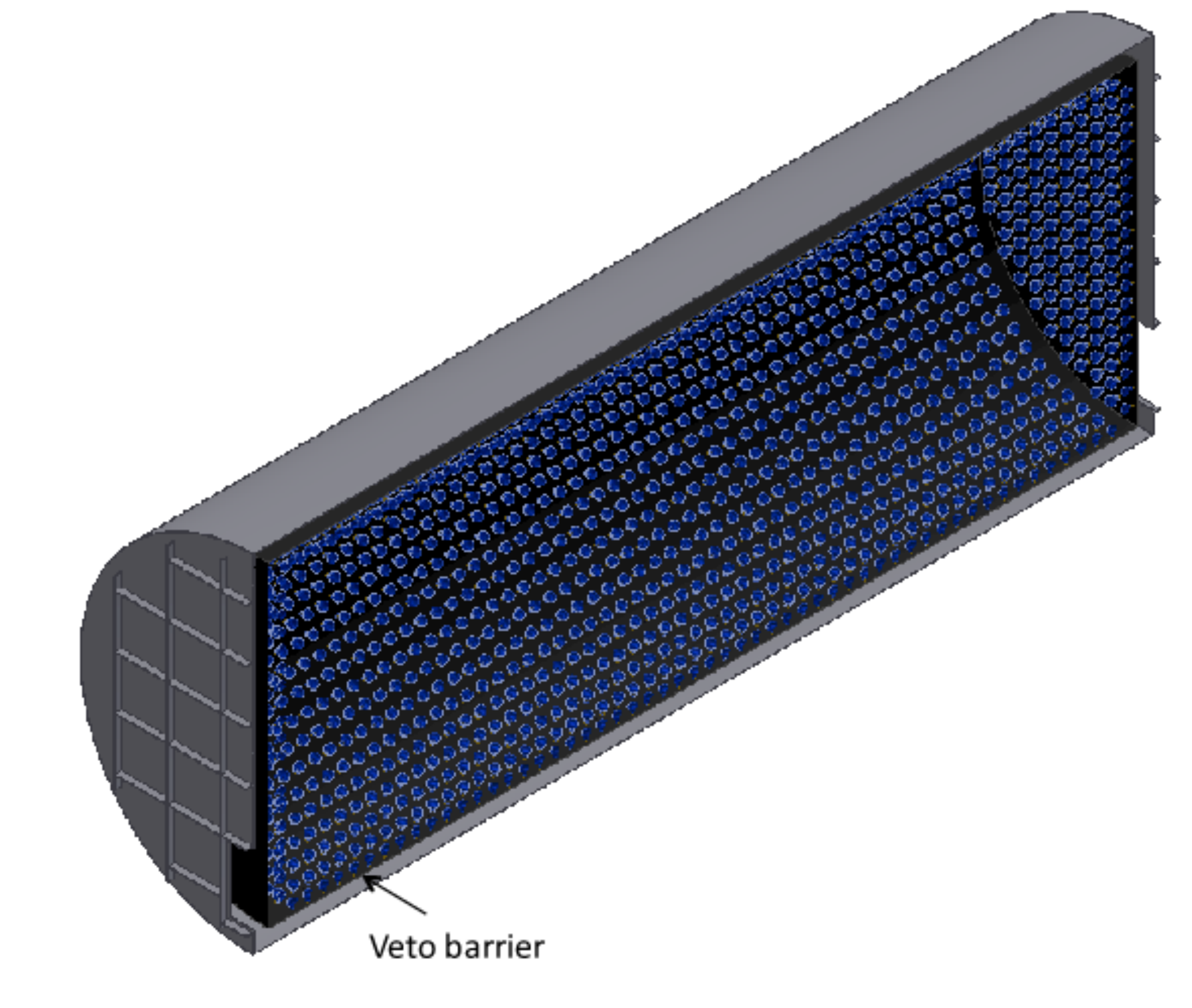}
\vspace{1mm}
\caption{
A cut-away schematic drawing of the OscSNS cylindrical detector tank.
}
\label{detector_schematic}
\end{figure}

The OscSNS detector is a cylindrical detector, filled with mineral oil and lined with photomultiplier tubes, 
similar to the MiniBooNE detector at Fermilab.  OscSNS differs from MiniBooNE in that it will have 
a higher phototube coverage and the addition of butyl-PBD scintillator to the mineral oil, ala LSND.
Fig. \ref{detector_schematic} is
a cut-away schematic drawing of the proposed detector, which consists of a 20.5 m long 
by 8 m diameter cylindrical tank
lined with 4290 8-in phototubes at a radius of 3.5 m (3900 detector phototubes, corresponding to 25\% coverage,
and 390 veto phototubes, corresponding to 2.5\% coverage).  The detector is filled with 886 tonnes
of mineral oil (density = 0.86), corresponding to a fiducial volume of 450 tonnes. 
Also, as in the LSND experiment \cite{lsnd}, 
approximately 0.031 g/l ($\sim 30$ kg)
of butyl-PBD scintillator will be added to the mineral oil in order to 
increase the light output of low-energy particles and provide neutron 
detection. 

\begin{figure*}[htbp]
\centering
\includegraphics[scale=0.50,angle=90]{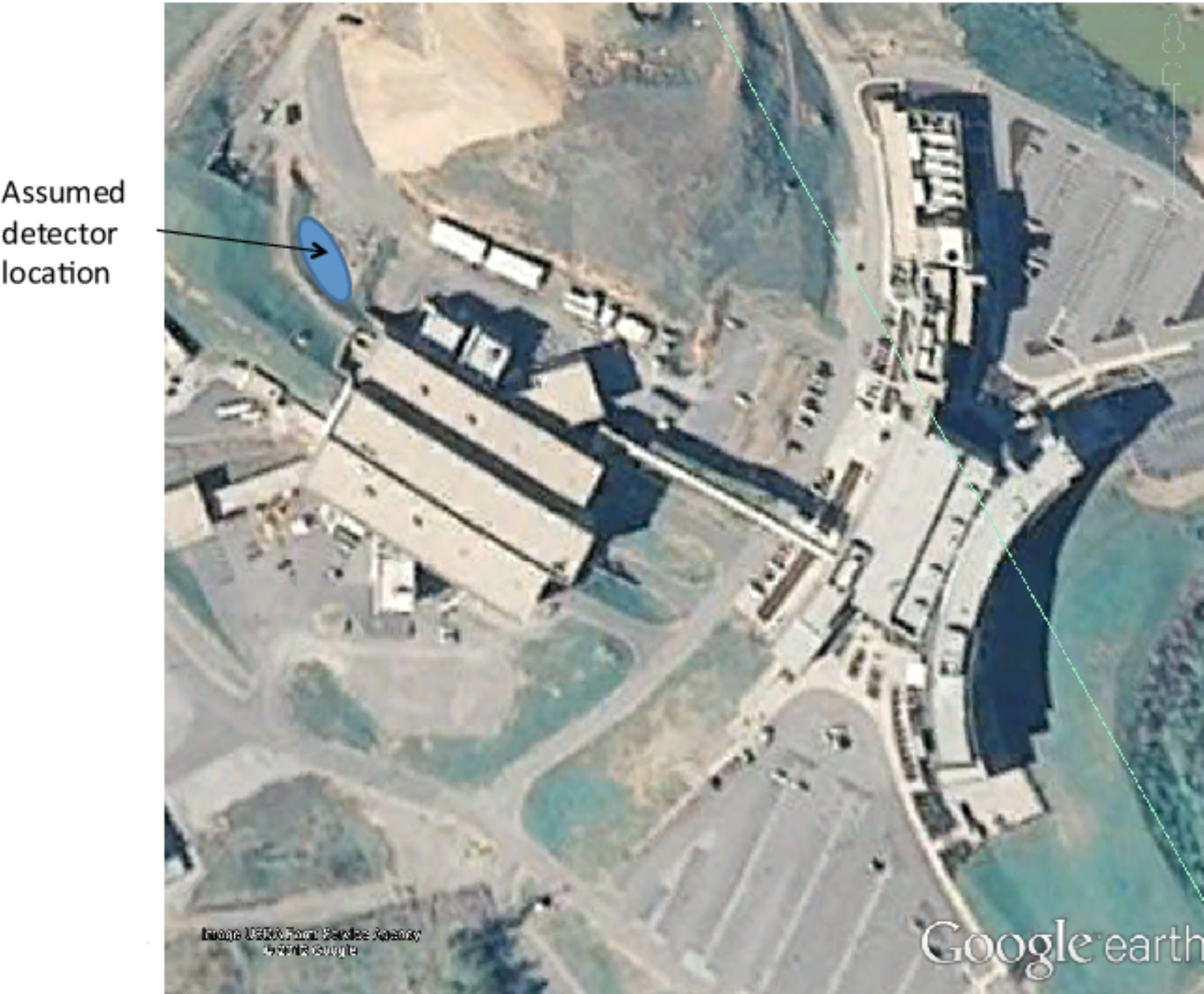}
   \caption{A photograph of the suggested detector location in relation to the SNS target hall.}
\label{sns_detector_photo}
\end{figure*}

\begin{figure*}[htbp]
\centering
\includegraphics[scale=0.60]{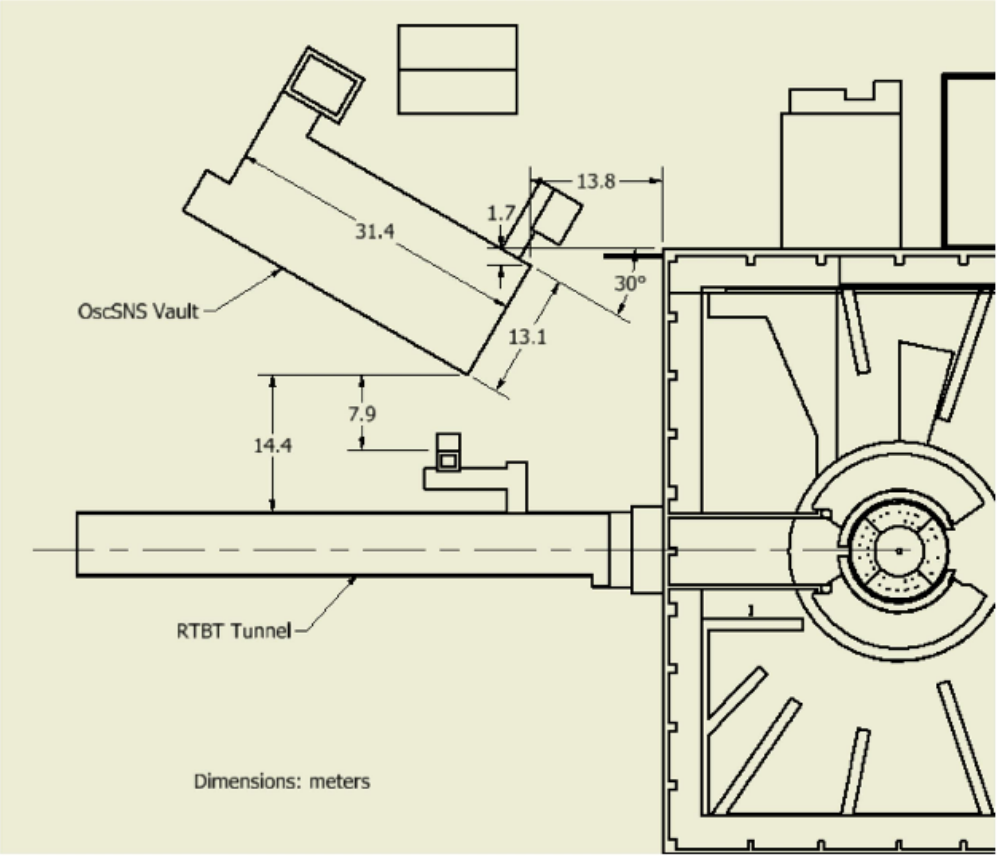}
   \caption{A schematic drawing of the suggested detector location in relation 
to the SNS target hall.}
\label{detector_target_schematic2}
\end{figure*}

\begin{figure*}[htbp]
\centering
\includegraphics[scale=0.60]{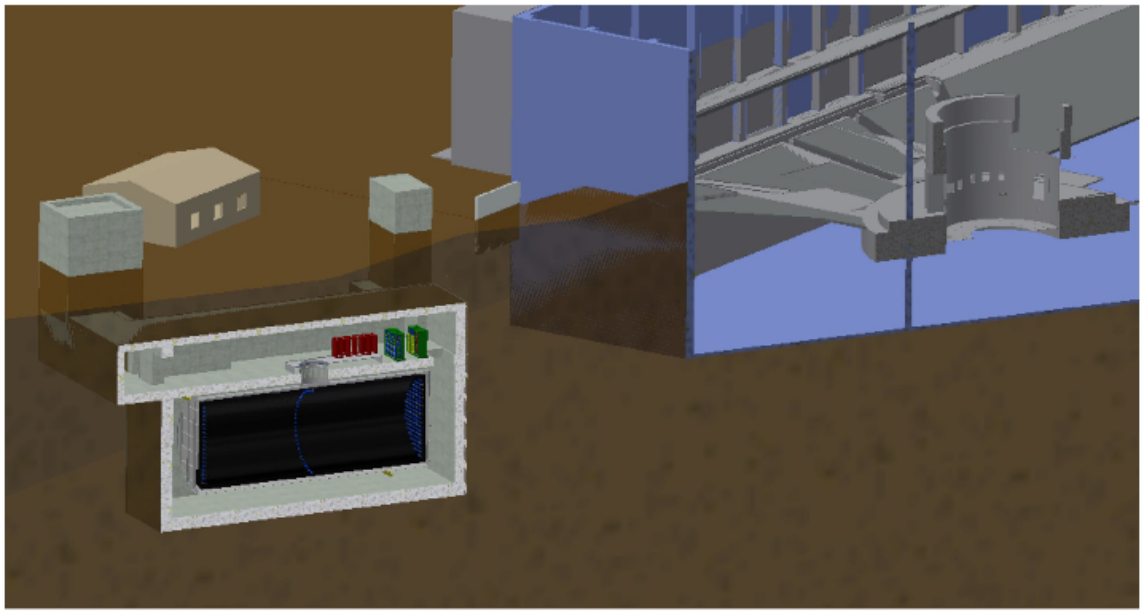}
   \caption{A schematic drawing of the suggested detector location in relation 
to the SNS target hall. Dirt provides the overburden for this design.}
\label{detector_target_schematic1}
\end{figure*}

\begin{figure*}[htbp]
\centering
\includegraphics[scale=0.60]{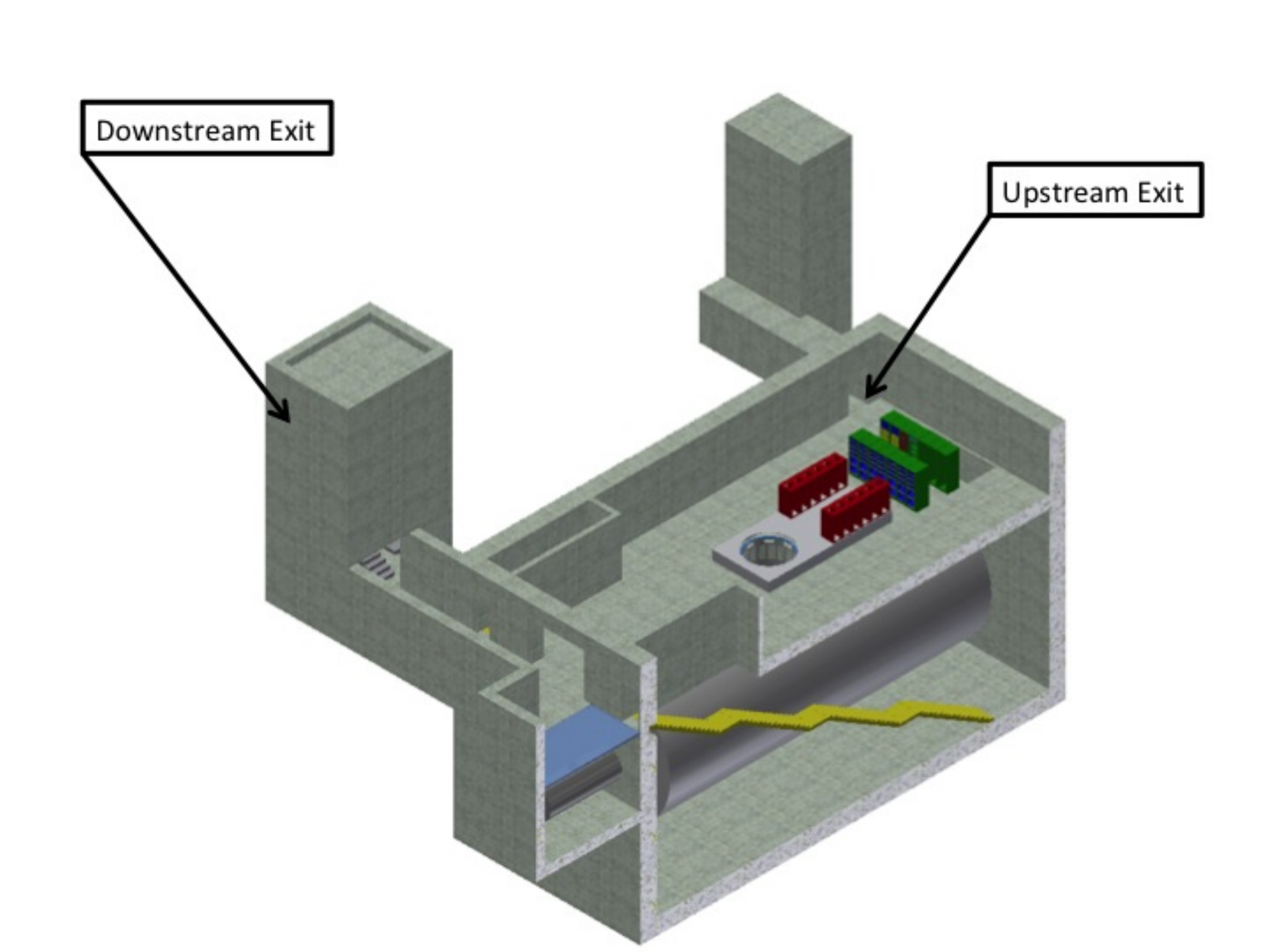}
   \caption{A cut-away schematic drawing of the detector hall.}
\label{schematic3}
\end{figure*}

\begin{figure*}[htbp]
\centering
\includegraphics[scale=0.60]{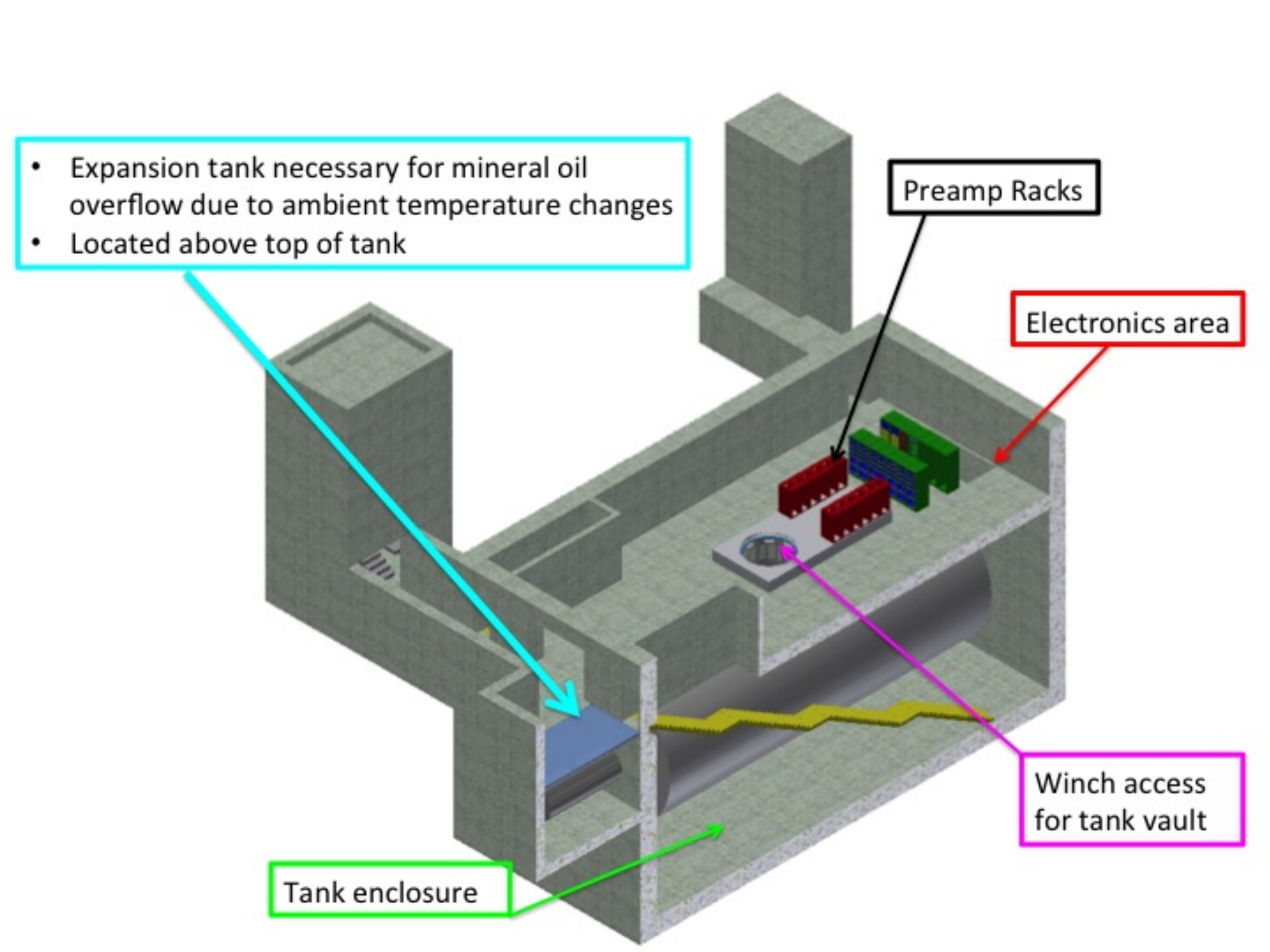}
   \caption{A cut-away schematic drawing of the electronics area above the
detector tank.}
\label{schematic4}
\end{figure*}

The suggested location for the OscSNS detector 
is shown in Figure \ref{sns_detector_photo}, where the center of the detector is located 60 m 
from the Hg beam dump at the SNS.
By placing OscSNS somewhat upstream of the beam dump, 
the decay-in-flight neutrino background to the neutrino measurements will be negligible. The
OscSNS detector will be buried under 6 m of dirt (or 2 m of steel) 
overburden, to
provide shielding from cosmic rays and beam-induced neutrons. New
electronics are being designed for OscSNS that will use ADCs with a
faster clock speed than
were used in LSND and MiniBooNE (200 MHz versus 10 MHz). Figures \ref{detector_target_schematic1}
and \ref{detector_target_schematic2} show schematic drawings of the detector location in relation
to the SNS target hall, while Figures \ref{schematic3}
and \ref{schematic4} show schematic drawings of the electronics area above
the detector tank.

The Detector Plant consists of two main elements:
the Detector Containment and the Support Plant. The Detector Containment
includes a cylindrical tank with access portal that sits in a rectangular
vault. The Support Plant lies above the Detector Containment and
includes area for tank access, utilities,
and electronics.
 
\section{Rectangular Vault}
 
The rectangular vault will house the detector tank and will serve as the
secondary containment for the oil.
The vault
will also provide the means for personnel access during the installation
of phototubes. The excess
soil from the excavation of the vault
can be stored and reused as overburden for the
detector enclosures to provide shielding from cosmic rays.
 
\section{Cylindrical Tank}
 
The cylindrical tank will have a diameter of 8 m, a length of 20.5 m,
and will be constructed
out of steel panels inside the rectangular vault. Bosses that will be used
to support the phototube support structure and cable plant will be welded
on the inside walls of the tank.
An 8-foot diameter access portal
is attached to the top of the tank and
provides the principal means for equipment access.
The diameter of the access portal is sufficient for all expected
penetrations of the tank, including plumbing, cabling, and calibration
lines. There will also be a personnel access port near the bottom
of the tank.
After the tank is filled with mineral oil, the oil level in the
access portal will be stabilized below the penetrations
by an overflow pipe that runs to the overflow tank. A small inflow of oil,
when required, will make the oil level stable at the level of the overflow
runoff.
 
\section{Cylindrical Tank Internals}
 
The phototube support structure will consist of an inner shell consisting of
650 aluminum plates on each which will be mounted five phototubes. The
plates themselves will be mounted on a series of latitudinal rings
supported from the bosses on the tank wall.
The volume inside the
aluminum plates will form the detector volume, while the outside volume will
form the veto shield region. Therefore, the inside surface of
the aluminum plates will be painted black, while the outside surface will
be painted white. There will be a total of 3900 8-inch detector phototubes,
providing 25\% coverage of the surface area, and
an additional 390 8-inch veto phototubes mounted directly on the tank wall.
Each phototube will have one RG58 cable carrying both the high voltage
to the phototube and the signal from the phototube. The cables will
penetrate the side of the access portal through close fitting holes
in the penetration panel, which will be above the oil
level.
 
Light from a laser located in the utility enclosure will be piped
along fiber optic cable through an access portal
penetration to four 10 cm diameter
glass bulbs
that are located at various positions in the detector
volume of the tank. The bulbs
will be filled with LUDOX and will disperse the laser light isotropically,
so that the phototubes can be timed accurately relative to each other.
In addition, there will be six small scintillator cubes (each 5 cm on a side)
that will be used for tracking cosmic-ray muons.
 
The tank will be filled with oil through a 3-inch diameter pipe that connects
to the bottom of the tank. An overflow pipe in the access portal
connects to a storage tank and defines the oil level. Nitrogen gas
will be bubbled into the tank at various levels and will help maintain
a small nitrogen overpressure in the access portal. The nitrogen
will purge
oxygen and water from the oil.
 
\section{Detector Enclosure}
 
The Support Plant is located above the detector tank but below existing grade and
consists of areas for electronics, utilities,
and tank access.
The utilities area includes a reservoir tank to
accommodate the variations in oil volume due to small
variations in temperature.
It also includes
the active plumbing elements necessary for oil and nitrogen circulation
and the laser used for phototube calibration. The electronics area includes
fast electronics, a data acquisition system, a farm of workstations,
and a taping system, while the tank access area contains the phototube
preamplifier electronics and will provide for personnel and equipment
access to the detector tank.
An HVAC system, located in and adjacent
to the utilities area, will be provided
for the enclosure and detector vault. The main entrance
and exit will be near the reservoir tank. The
electrical power required for operation of the OscSNS experiment
is estimated to be $\sim 250$ kW. It is necessary to separate the
electronics power source from the power used in the utility building
for pumps, plumbing, and laser calibration. No water cooling is required.

\section{Gadolinium-loaded Liquid Scintillator Option}

The detection of electron anti-neutrinos via the inverse beta-decay reaction 
is one of the main signals for OscSNS.  To maximize observation of this channel, 
we will need to dope our mineral oil with a liquid scintillator (LS) that will 
produce large numbers of photons at these lower energies of a few MeV.  
Previous neutrino experiments have used butyl-PBD and various organic liquid scintillators as the dopant.  
We are exploring the use of both dopants for OscSNS.

The antineutrino signal is a delayed coincidence between the prompt positron and 
the capture of the neutron in an $(n,\gamma)$ reaction after it has been 
thermalized in the LS. This delayed coincidence tag serves as a powerful tool 
to reduce random backgrounds. 

The neutron capture can occur on the hydrogen in the organic LS, but the 
cross section is small at 0.332 barns, and only one $\gamma$-ray is emitted 
with the energy of 2.2 MeV. There are several important advantages that come 
from adding a metallic ion, such as Gadolinium, Gd, to the LS: 

(a)     The $(n,\gamma)$ cross-section for natural Gd is high, 49,000 barns 
(with major contributions from the following Gd isotopes, Gd-152, 1100 barns, 
0.2\%; Gd-155, 61,000 barns, 14.9\%; Gd-157, 254,000 barns, 15.7\%). Because 
of this high cross section, only a small concentration of Gd, $\sim 0.1-0.2\%$ 
by weight, is needed in the LS.  

(b)     The neutron-capture reaction by Gd releases a sum of 8-MeV energy in 
a cascade of 3-4 $\gamma$-rays. The higher total energy release of the 
$\gamma$-rays and their enhanced isotropy help to exclude low-energy 
backgrounds from other sources, such as radioactive decay in the surrounding 
environment and materials. In addition, the higher energy release of 8 MeV, 
occurs only in the Gd-loaded region of the detector; thus, its fiducial volume 
and target mass are better defined. 

(c)     The time delay for the neutron-capture is also significantly shortened 
to $\sim 27 \mu$s in 0.1\% Gd, as compared to $\sim 200 \mu$s in H. This 
shortened delay time reduces the accidental background rate, by a factor of 
$\sim 7$.

A multi-ton scintillation detector for an antineutrino oscillation experiment 
must satisfy a number of stringent requirements:
\begin{itemize}
\item The Gd-LS must be chemically 
stable for the life of the experiment; periods of $>3$ years are nominally 
quoted for today's liquid scintillator experiments. 
\item The Gd-LS must be 
optically transparent, have high light output, and have ultra-low 
concentrations of radioactive and chemical contaminants. ``Chemical 
stability'' means any formation over time of any components in the liquid 
that will absorb or scatter light, or change the concentration. For example, 
development of color, or gels, or precipitates, or colloids, or hydrolysis, 
etc., cannot be tolerated. 
\item The Gd-LS must also be chemically compatible with 
the containment vessel; for example, it is known that various organic 
solvents can attack acrylic plastic. 
\end{itemize}

Several organic scintillation solvents were studied to test their 
feasibilities for the above-mentioned criteria. 1,2,4-trimethylbenzene 
(pseudocumene, PC) is a clear, fluorescent LS that has been used in previous 
neutrino experiments at nuclear reactors, because of its high light yield. 
However, pseudocumene is also known for low flash point, which imposes extra 
safety concerns, and high chemical reactivity, which causes the problem of 
chemical incompatibility with the detector vessel that contains the LS. 
Phenyl cyclohexane (PCH), a colorless benzene-based liquid, has a lower 
reactivity, but only half of the light yield compared to PC. Both 
di-isopropylnaphthalene (DIN) and 1-phenyl-1-xylyl ethane (PXE) have 
absorption bands in the UV region below 450-nm that are difficult to be 
removed by conventional purification processes, neither by Al2O3 
exchange-column nor by vacuum distillation (since they have high boiling 
points). 
Linear alkylbenzene, LAB, first identified as a scintillation liquid from 
the SNO+ R\&D, is composed of a linear alkyl chain of 10-13 carbons attached 
to a benzene ring; it is commercially used primarily for the industrial 
production of biodegradable synthetic detergents. LAB has a light yield 
comparable to that of PC and a high flash point, which significantly reduces 
the safety concerns. These notable characteristics make it suitable for a 
large-scale neutrino experiment. Current ongoing or proposed experiments for 
reactor electron anti-neutrinos, Daya Bay and RENO, double-beta decay, SNO+, 
and solar neutrinos, LENS, unanimously select LAB as their primary 
scintillation liquid. Similarly, OscSNS will use LAB as the baseline solvent 
for the Gd-loaded option; which has advantage of stability, high light-yield, 
and optical transmission over a binary solvent system (i.e. PC or LAB in 
dodecane or mineral oil).

It is commonly known in the community that the quality of Gd-LS is the key 
to the success of the LS experiments. There is heightened concern for a new 
long-duration oscillation experiment such as we propose here. The BNL 
neutrino and nuclear chemistry group has been involved in R\&D of chemical 
techniques for synthesizing metal-loaded organic liquid scintillators since 
2000 and is currently a member of several liquid scintillator experiments: 
Daya Bay, SNO+ and LENS. A highly stable 0.1\% Gd-LS 
with attenuation length of $\sim 20$m and $\sim 10,000$ optical 
photons/MeV has been developed by the BNL group for the reactor antineutrino 
experiments. Indeed, Daya Bay has published its first observation of 
non-zero $\theta_{13}$ in 2012, based on the successful detection of the IBD 
reaction by Gd-loaded liquid scintillator. 

Even though the synthesis procedures already give satisfactory, consistent 
results and has been demonstrated by modern reactor experiments, once the 
Gd-loaded option is selected, we intend to further perfect the methods of 
synthesis and purification (both chemical and radioactive) to advance their 
reliability and reproducibility, especially to further enhance the optical 
transmission of Gd-LS to better than 25m for large LS detectors ($>$kton) used 
for neutrino oscillation experiments. A list of the tasks that we will 
undertake for OscSNS includes: 

(1)     Continue the R\&D of Gd-loaded in singular system (LAB).

(2)     Start the R\&D of Gd-loaded in binary system (LAB in dodecane or 
mineral oil). 

(3)     Design the mass-production scheme and develop the QA/QC protocols 
for raw material and liquid quality control. 

(4)     Test a broad selection of candidate organic scintillators, in addition 
to LAB and their mixtures of any inert solvents.

(5)     Study the chemical compatibility of these organic LS with the 
materials that will be used to construct the OscSNS detector vessel. 

(6)     Develop methods to remove and assay radioactive contaminants, mainly 
from the naturally occurring U-238 and Th-232 decay chains.

\section{Electronics and Data Acquisition}
 
The phototubes, preamplifier boards, and digitization electronics for
OscSNS will be similar to the system used for LSND and MiniBooNE.
This system has a clock rate of 10MHz,
sufficient memory available for 200 $\mu$s of data before being overwritten,
and is described in detail elsewhere~\cite{bigpaper1}. Improved electronics will have 
a clock rate of 200MHz.  A precursor signal from the accelerator
will be used to record data for a period of about $20 \mu$s
around each 695 ns beam spill. In addition, there will be other
triggers used for calibration.
 
The OscSNS data acquisition system will be based on the LSND/MiniBooNE system.
The 14 crates of electronics will be read out by monoboards located
in each crate, and the data will be sent to the multi-processor computer
that will build and reconstruct the events in real time. The reconstructed
events will then be written to disk and logged to tape. A high-speed
link will connect the control room to the
multi-processor computer, so that the experiment can be monitored
and analyzed in real time from the OscSNS control room.
 
\section{Calibration}

The accurate measurement of event interactions and neutrino oscillations requires 
a well-calibrated detector in the range from 1 MeV to 50 MeV.

The calibration system is designed to
(1) provide information on the PMT response that is needed
as input for the event reconstruction and particle identification calculations,
(2) calibrate the position, energy and direction determination of
the reconstruction program using stopping cosmic ray muons and
Michel electrons from muon decay (end-point energy of 52.8 MeV) and
(3) measure the attenuation of light by the oil in the detector tank.
 
The cosmic-ray muons can be reconstructed by a high-resolution muon 
tracker mounted above the detector tank, and scintillator cubes inside the detector.  
The tracker is made of scintillator strips and
will also help with the rejection of cosmic-ray background and
improve the efficiency of the veto system.
The muon tracker will give the direction of cosmic-ray muons
while the cubes will give a precise position measurement of stopping muons
and decay electrons.

In addition to the cosmic-ray calibration, 
radioactive calibration sources producing both gamma-rays
and neutrons are desirable.  A partial list of potential sources are:

\begin{itemize}

\item $^{16}N$ source: [$^{16}O(n,p)^{16}N^*$] producing a $\beta$-tagged 6.1 
MeV gamma-ray.
\item $^{8}Li$ source: electron energy spectrum up to 15 MeV.
\item pT source: [$^3(p,\gamma)^4He$] producing a 19.8 MeV gamma-ray.
\item $^{252}Cf$ source: producing fission neutrons.
\end{itemize}

A flexible deployment system similar to that of SNO is envisioned,
whereby  the above sources can be moved to any position within
the detector volume \cite{sno}.  This will allow a precise determination of
optical properties and detector response throughout the entire
detector volume.  This is crucial for the precision neutrino measurements
envisioned for OscSNS.

The laser calibration system will provide short pulses of light from a
tunable dye laser to 4 flasks at various locations in the detector.
The system is very similar to the system used successfully in LSND and MiniBooNE.
It is used to determine phototube time-offsets and gains and to determine
time-slewing corrections.
 
The oil monitoring system will measure attenuation of light
as a function of wavelength in the detector.


\chapter{SNS Neutrino Flux Simulation }

\label{chap:flux}

The OscSNS experiment uses Geant4 \cite{geant4} to simulate the interaction of the proton beam
 with the liquid mercury target.  
Approximately 17\% of the incident protons interact to produce a charged pion ($\pi^+$, $\pi^-$).  
These charged pions are brought to rest on a short time scale ($<$ 3 ns), due to
 their relatively low energy at creation ($\sim$ 200 MeV), and to the high stopping power of 
the target material.

Over 99\% of the produced $\pi^+$ decay at rest.  The pion has spin 0, which causes
 it's decay products, $\mu^+$ and $\nu_{\mu}$, to be emitted isotropically. 
The subsequent $\mu^+$ decay occurs within 0.2 g/cm$^2$ of the point at which the $\pi^+$ stopped. 
The $\bar{\nu}_{\mu}$ and $\nu_e$ fluxes are also isotropic. 

The bulk of the $\pi^-$ produced are absorbed by the target before they are able to decay.  
Less than 1\% of the produced $\pi^-$ decay, all of which decay in flight.

The majority of the pions and muons decay in the immediate vicinity of 
their production point. A neutron spallation facility, therefore, produces an 
extremely intense, approximately point-like, source of neutrinos for nuclear and 
particle physics studies.

The complete simulation chain consists of the following stages:
\begin{itemize}
\item Simulate p+Hg interaction
        \begin{itemize}
        \item This interaction produces the mesons ($\pi$, $\mu$), and allows the particles 
        to decay to produce the full neutrino flux coming from the SNS.  We then select only those 
        neutrinos whose direction is 150 degrees, with respect to the proton direction.
        \end{itemize}
\item The generated fluxes are used to produce the interactions inside detector volume.  These 
        interactions produce outgoing charged leptons.
\item The resulting leptons are fed into another Geant4 package that models the OscSNS
    detector.  
The photons from Cerenkov radiation and scintillation produced by 
    the leptons are recorded in the PMTs. Coordinates of the activated PMTs, together 
             with the number of photons and timing information, are saved as histograms.  This 
        information is then used as input for the particle identification and reconstruction algorithms.
\end{itemize}

\section{Generation of Neutrino Fluxes}

A preliminary Monte Carlo simulation for the interaction of the proton beam with the
mercury target was done with Geant4 \cite{geant4}. This software package is currently 
the most widespread simulation tool in the nuclear and high energy physics community.  This is  
due to its versatility and preciseness in the description of incident particle beams, 
involved physics models, and geometry visualization.  Geant4 is open source code that is maintained 
by a core group of personnel, and as such is constantly improving.

 There are four major components to our neutrino flux simulation: 
generation of primary particles, simulation of the target hall geometry, physics processes available to the incident protons and produced mesons, and particle counters and histograms.

\subsection{Generation of Primary Particles}

        Geant4 provides a convenient tool for configuration 
of the incident beam (General Particle Source \cite{gps}). The 1 GeV proton beam was randomly 
distributed within a rectangle with 200 mm x 70 mm cross-sectional area (similar to the SNS 
beam spot size), and was perpendicularly incident on the target surface. 
Beam timing information (695 ns pulse duration with 60 Hz repetition rate) 
was also included to better approximate the experimental conditions.

\subsection{Description of Geometry} 

\begin{figure*}[htbp]
\centering
\includegraphics[scale=0.60]{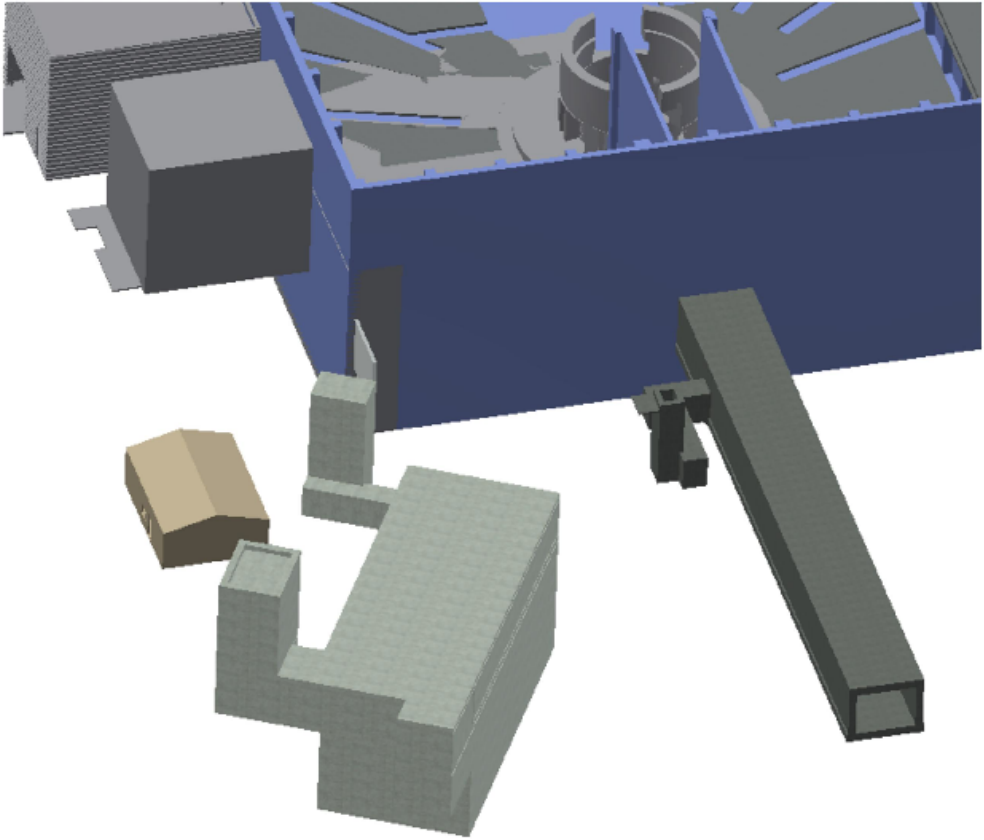}
   \caption{A schematic drawing of the detector location in relation to the SNS target hall.}
\label{sns_detector_schematic}
\end{figure*}

        The SNS target hall consists of thousands of various parts, 
many of them very complex shapes. Thus, for the preliminary 
simulation, a simplified target hall geometry was chosen.  The target hall model consists of 
the 400 mm x 104 mm x 500 mm liquid mercury target, surrounded in stainless steel casing.  We have 
 also included major bulk parts surrounding the target area, that are composed primarily
 of steel and concrete. 
The center of the detector is placed at $\sim 60$ m from the interaction point, 
at an angle of $\sim$150 degrees from the proton beam direction. Fig.~\ref{sns_detector_schematic} 
shows a schematic drawing of the detector location in relation to the SNS target hall.

\subsection{Description of Physics Processes}

In Geant4 terms, a physics list is a compilation of processes or interactions that particles are 
permitted to engage in, for varying energy ranges.  
Creating an acceptable physics list for a simulation is a challenging task 
(see e.g. \cite{tant}). The Geant4 collaboration provides a collection of pre-compiled physics 
lists for various applications.  For most physics applications, these physics lists are good 
only for a first order rate estimate.  Final experimental simulations require modifications to 
these pre-compiled lists.  Our rate estimates are based on the QGSP\_BERT list.   
It uses the quark-gluon string model for high energies ($>$ 20 GeV), 
low energy parameterized for medium energies (10 $<$ E $\leq$ 20 GeV), 
and Bertini-style cascade for low energies ($\leq$ 10 GeV). 

\begin{figure}
\centering
\includegraphics[scale=0.65]{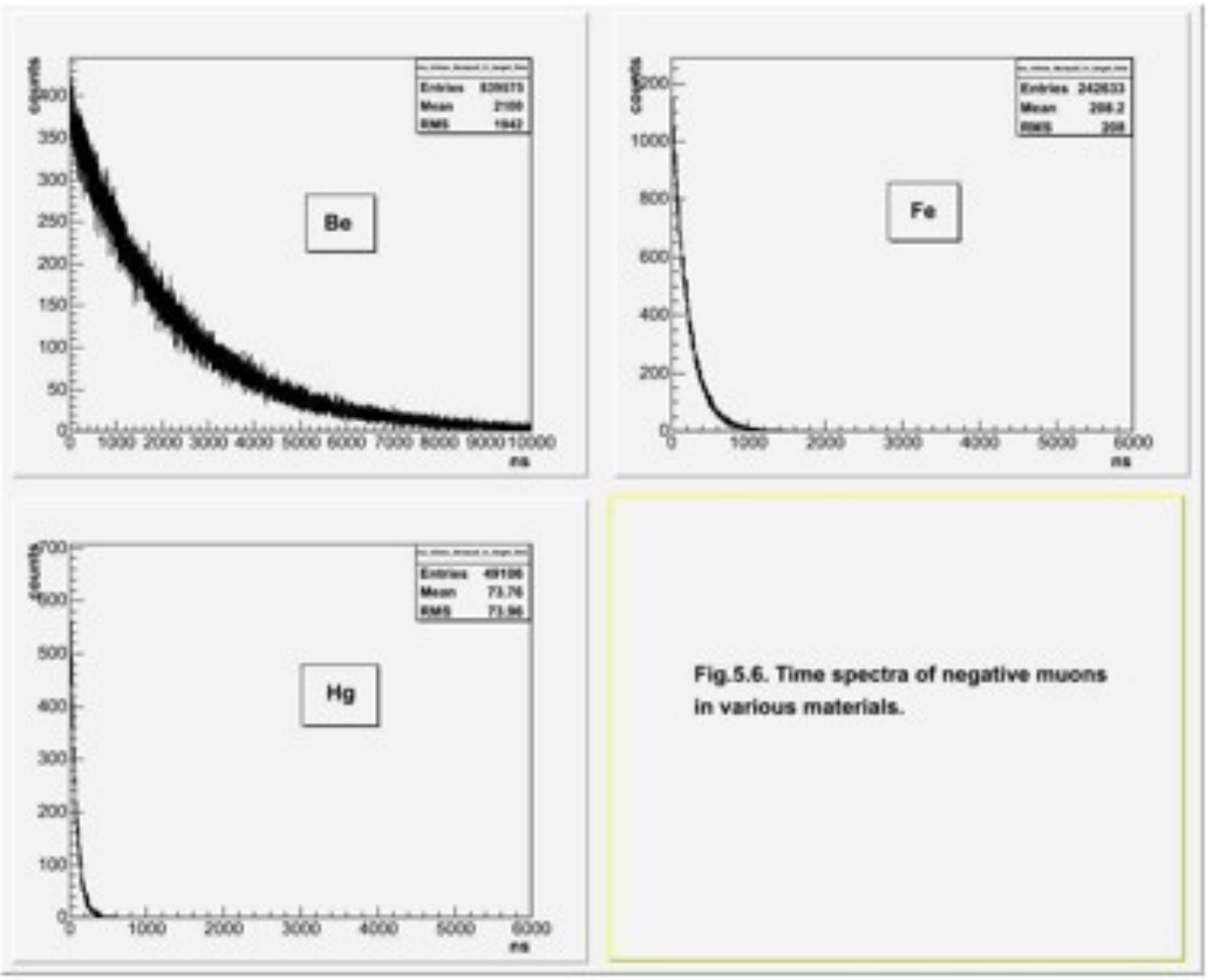}
\caption{Time spectra of negative muons in various materials.}
\label{fig_flux2}
\end{figure}

This list was validated by calculating the lifetime of negative 
muons in various materials. For better event statistics, a beam of 100 MeV $\mu^-$ was 
generated in the center of the target box, with an isotropic angular 
distribution. This list produces good agreement between the experimental data and the simulation 
results (see Table~\ref{tab1}).  The corresponding time spectra are shown in Figure~\ref{fig_flux2}. 
The lifetime values were deduced by fitting the time histograms with an exponent.

\begin{table}
\centering
\begin{tabular}{|c|c|c|}
\hline
Material & Lifetime, ns & Lifetime, ns \\
        & (experiment)
        & (calculation with QGSP\_BERT) \\
\hline
\hline
Be & 2162.1 $\pm$ 2.0 & 2161.0 \\
Fe & 206.0 $\pm$ 1.0 & 205.9 \\
Hg & 76.2 $\pm$ 1.5 & 72.3 \\
\hline
\end{tabular}
\caption{Lifetime values for negative muons calculated with Geant4, compared to experimental data.  Data is taken from \cite{suzuki}}
\label{tab1}
\end{table}

Using QGSP\_BERT, the yields (number of produced particles per proton-on-target) 
are 0.116 for $\pi^+$ and 0.059 for $\pi^-$. This results in the following yield values for 
various neutrino flavors: 0.09 for $\nu_e$, $\nu_{\mu}$, $\bar{\nu}_{\mu}$, and 1.15e$^{-4}$ 
for $\bar{\nu}_e$. We plan to further improve the Geant4 physics list for final rate estimates.

\subsection{Particle Counters and Histograms.} 
\begin{figure}[htbp]
\centering
\includegraphics[scale=0.50,angle=-90]{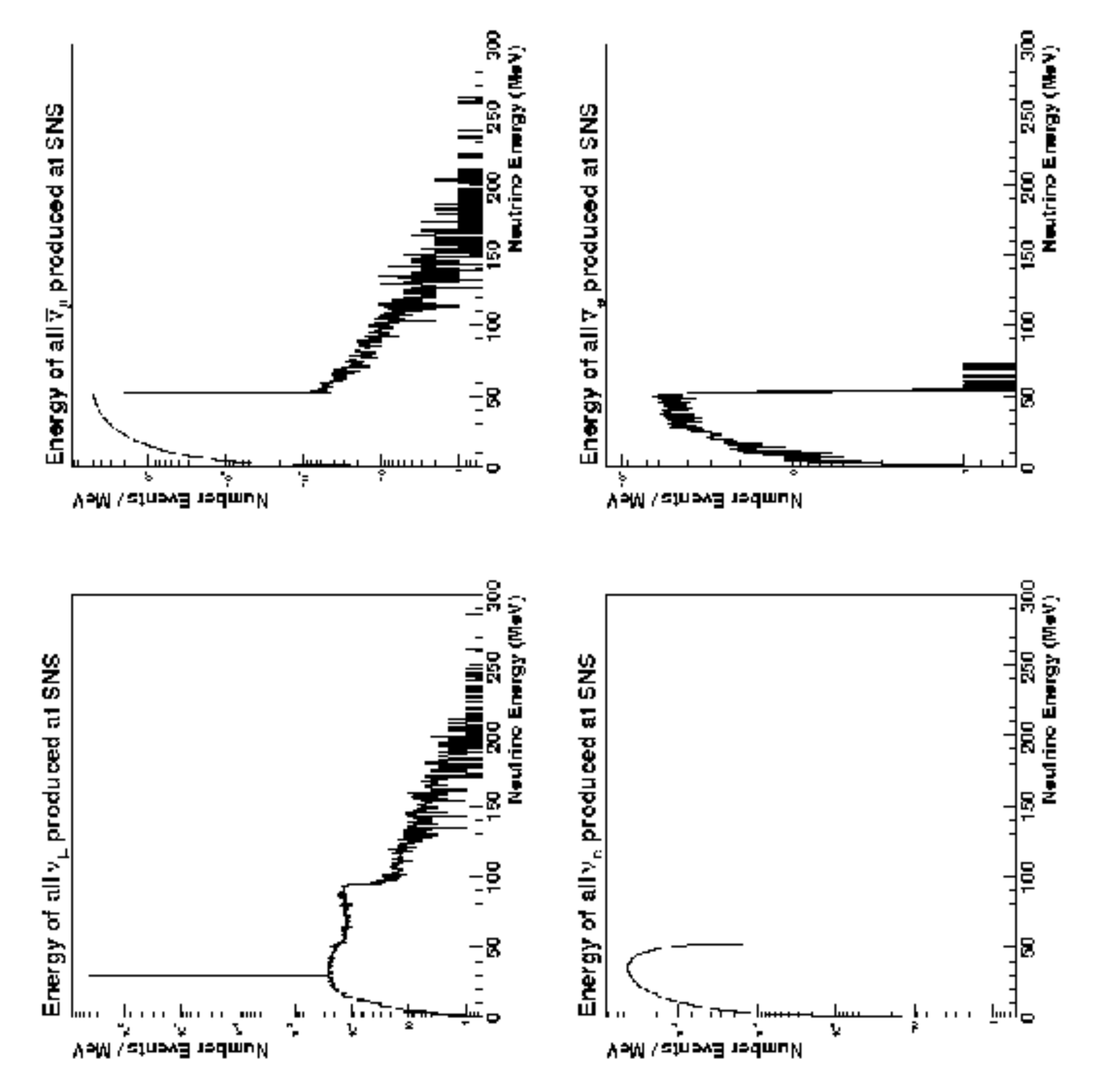}
   \caption{Energy spectra of neutrinos produced from the decay of pions and muons at the SNS.  From left to right, top: $\numu$, $\numubar$; bottom: $\nue$, $\nuebar$.  Rates are for 50M protons on target.  The prominent feature in the $\numu$ spectra around 90 MeV comes from the muMinusCaptureAtRest process in Geant4.}
\label{fig_flux3}
\end{figure}

\begin{figure}[htbp]
\centering
\includegraphics[scale=0.50,angle=-90]{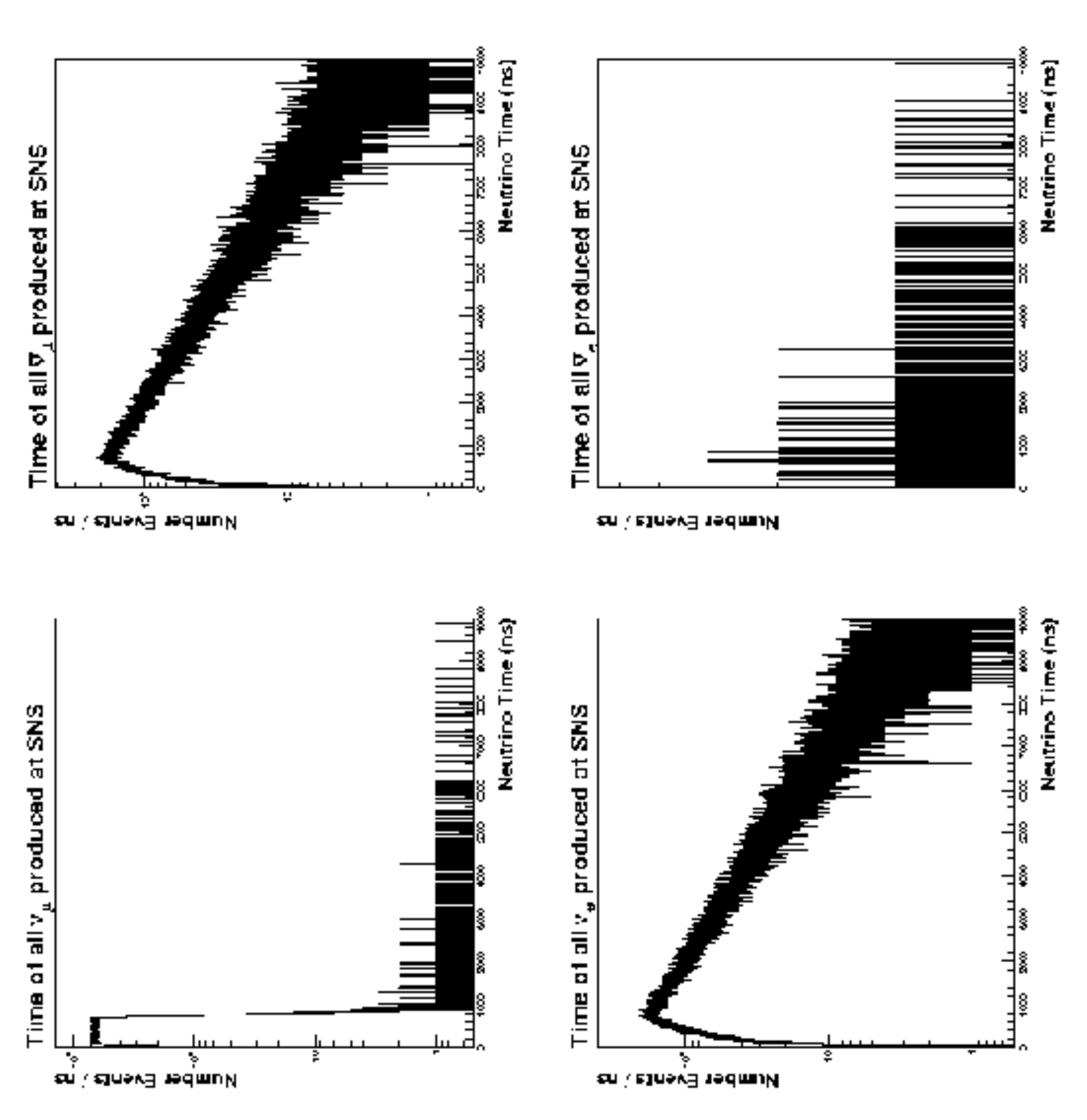}
   \caption{Time spectra corresponding to Figure~\ref{fig_flux3}.  From left to right, top: $\numu$, $\numubar$; bottom: $\nue$, $\nuebar$.  Rates are for 50M protons on target.}
\label{fig_flux4}
\end{figure}

Energy and time distributions of the charged pions, muons, and the corresponding neutrinos 
were written into ROOT \cite{root} histograms. The flux of neutrinos originating from
 the SNS target area (pion and muon decay, muon capture) are shown in the 
Figure~\ref{fig_flux3}.  The feature in the $\numu$ spectra around 
90 MeV comes from a Geant4 process, muMinusCaptureAtRest.  This process is built in to the 
Geant4 physics lists, and is model dependent.  There is no experimental data to guide the 
prediction of these interactions, and thus it 
has large uncertainties associated with it.  Future flux estimates will 
carefully study and consider the inclusion of this process into the final prediction.  
The corresponding time distributions of these 
neutrinos are found in Figure~\ref{fig_flux4}. 

\begin{figure}[htbp]
\centering
\includegraphics[scale=0.50,angle=-90]{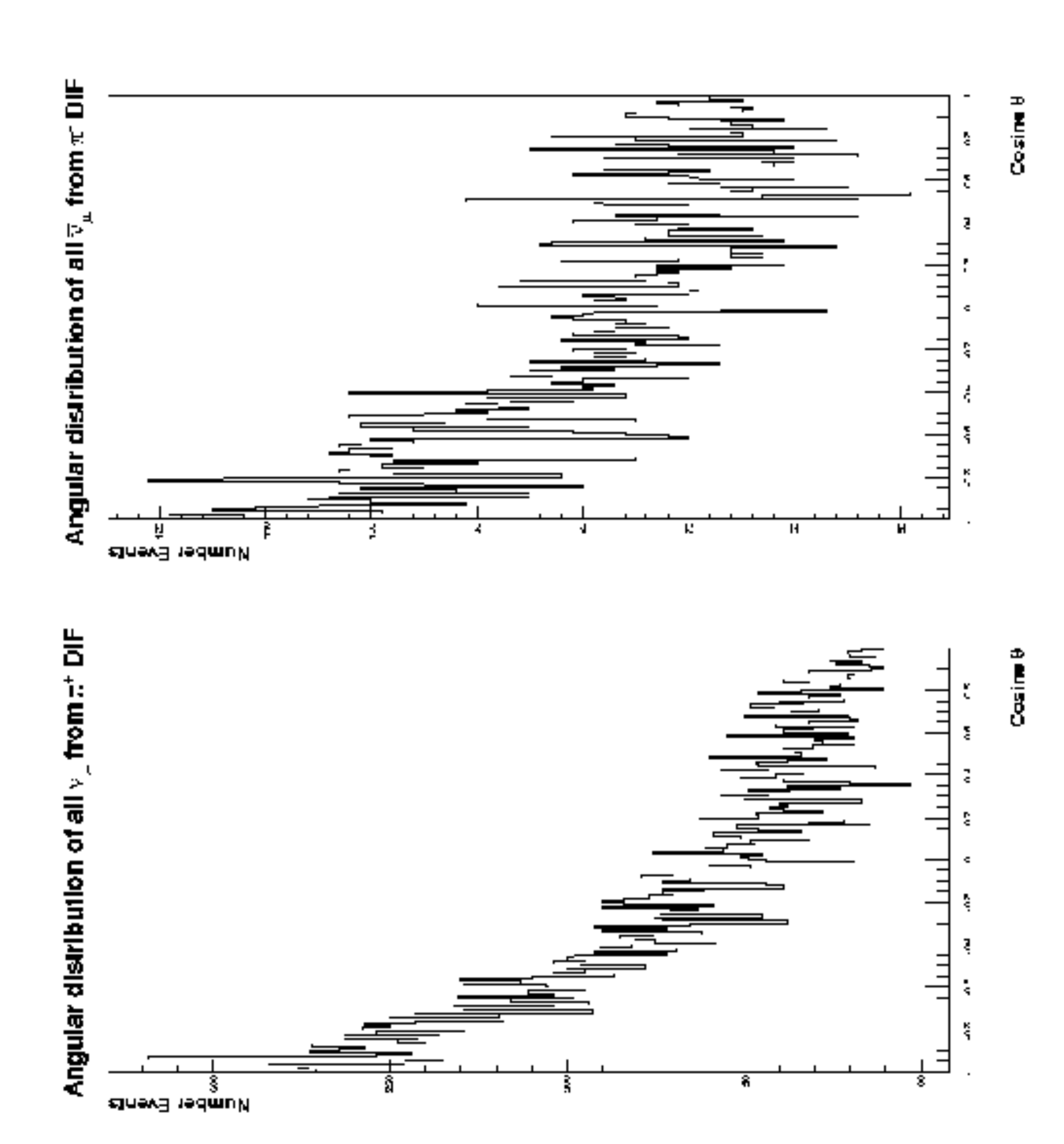}
   \caption{Angular distributions for muon neutrinos (left) and anti-neutrinos (right), from all decays in flight of positive and negative pions, respectively.  The angle is taken between the emitted neutrino direction and the direction to the detector.  Rates are for 50M protons on target.}
\label{fig_flux5}
\end{figure}

Almost all negative muons and pions are captured in the target prior to decaying. 
The small percent of negative mesons that decay in flight are boosted in the direction 
of the proton beam.  Neutrinos originating from decays in flight 
are primarily emitted in the forward direction (see Figure~\ref{fig_flux5}).  We have 
chosen to place our detector in the backward direction, relative to the proton beam, to 
reduce this background to a negligible level.  Almost no neutrinos 
from particles decaying in flight are seen in the detector.

The neutrinos produced from the decays at rest in the target are distributed isotropically.  This 
may be seen by comparing the ratios:

\begin{equation}
\frac{\mathrm{AREA \ SUBTENDED \ BY \ DETECTOR}}{4\pi \mathrm{R}^2}
\end{equation}
and
\begin{equation}
\frac{\mathrm{NUMER \ OF \ \nu \ IN \ DETECTOR}}{TOTAL \ NUMBER \ OF \ \nu},
\end{equation}

where R is the distance from the target to the detector cylinder. For 5 x 10$^7$ 
incident protons, these ratios for muon neutrinos are 0.0025 and 0.002548, respectively.  

It should be stressed that the $\bar{\nu}_e$ production is a factor of 10$^{-3}$ to 10$^{-4}$ 
lower than the other neutrino species. 
A large reduction in the $\bar{\nu}_e$ flux is due to the strong absorption of $\pi^-$ 
and $\mu^-$ in the SNS mercury target.  
Decay of $\mu^-$ is the primary source of $\bar{\nu}_e$; additional $\bar{\nu}_e$ are produced 
via the decay of $\pi^- \rightarrow e^- \bar{\nu}_e$, with a branching 
ratio of $\sim$1.2e$^{-4}$.  
Additional reductions to the $\bar{\nu}_e$ background will result from the use of the time 
structure in the SNS proton beam. The time spectrum of $\bar{\nu}_e$ from the decay at rest of $\mu^-$ 
consists of two basic components: (1) an exponential decay in low-Z materials, such as Be and Al, 
dominated by the 2.2 $\mu$s muon mean life, and (2) the much faster exponential decay in high-Z 
materials mercury, lead, and iron characterized by a fast absorption rate.  
Events are separated into two time groups, above and below 1000 ns from the start of the proton pulse.  
A time-cut at 1000 ns will eliminate the $\bar{\nu}_e$  background from $\mu^-$ decay in mercury and 
iron, thereby reducing the background by a factor of two. More importantly, comparing the rates of 
appearance of $\bar{\nu}_e$ before and after the 1000 ns cut provides for the first time an 
opportunity to experimentally evaluate the $\bar{\nu}_e$ background.

\section{Simulation of Neutrino Interactions}

At its present development stage, Geant4 does not simulate neutrino interactions.  Therefore, a 
separate program was written in Fortran77 for that purpose. The neutrino energy spectra histograms 
produced by Geant4 are used as inputs to this program.  

The Fortran code calculates cross-sections for quasi-elastic neutrino scattering 
with the nucleon bound inside a nucleus, and 
quasi-elastic neutrino-nucleus scattering using the Fermi gas model. 
The code can also be easily modified to calculate other reactions of interest.

The output file contains the coordinates of the outgoing lepton vertex, its energy, 
differential cross-section, and direction (as cos$\theta$ and $\phi$). Since the cross-section 
values are relative, the events have to be weighted. One way to achieve this is as
follows. For a given number of events, the maximum value of the cross-section is 
determined. Then, for each individual event, the ratio of its cross-section to the maximum 
value is calculated and is compared to the random number generated in the (0,1) interval. 
The event is passed to the output file if the ratio is greater than the random number, 
and is neglected otherwise.  The resulting file is supplied to another Geant4 package to 
simulate the propagation of produced leptons inside the detector.

\chapter{OscSNS Detector Simulation}

A Monte Carlo simulation was written to simulate events in the
detector using GEANT4 \cite{geant4}.  The simulated detector
consists of a cylindrical tank filled with mineral oil which has
been laced with butyl-PBD (butyl-phenyl-biphenyl-oxydiazole)
scintillator at a concentration of 0.031 g/l.  The inner region
of the detector has a radius of 3.65 m, and is surrounded by 3900
photomultiplier tubes.  The photomultiplier tubes were modeled as
hemispheres of radius 10.1 cm containing photocathodes of
cross-sectional radius 9.5 cm.  The photocathode surface was
located 15 cm inside the central tank wall.  This arrangement
yields a mean photocathode coverage of 25\%. 


Standard Geant4 packages were used to simulate particle
interactions within the detector.  Parameters used to simulate
the optical model were taken from studies conducted by previous
experiments.  In particular, the values for the refractive index
and absorption length spectra of the mineral oil were determined
using values measured by MiniBooNE \cite{mb_oil}.  The actual
detector will also contain a small concentration of butyl-PBD.  
The MiniBooNE values for these quantities are 
adequate for preliminary studies.

The parameters used to simulate scintillation light were taken
from studies conducted by the LSND collaboration at the Area A
test channel of the Los Alamos Meson Physics Facility (LAMPF)
\cite{LSND_om}.  Scintillation light typically has both fast and
slow components.  The ratio of slow to fast scintillation light
produced by an ionizing particle depends upon the dE/dx of the
particle, as the amount of fast scintillation light produced
saturates at higher dE/dx as described by Birks' law
\cite{birks}. The LSND study measured the properties of Cerenkov
and scintillation light for both positrons and protons.  While
the scintillation time constants should be the same for all types
of charged particles, the LSND measurements of the isotropic time
constants differed for positrons and protons.  This difference
was due to Cerenkov light produced by the positrons being
absorbed and re-emitted isotropically, thus contaminating the
isotropic light generated by pure scintillation light. For the OscSNS
optical model we are taking the scintillation time constants to
be those measured for protons, which were produced below the
Cerenkov threshold and should therefore have no such
contamination in their measured isotropic light. 

The LSND study also measured the relative amounts of slow and
fast scintillation light produced by both particle types, the
values of which are also used in the simulation. 
These values are listed in table \ref{table:om_scint}.  The
scintillation yield was set arbitrarily high so that studies
could be conducted to determine the optimum scintillation light
yield using the reconstruction code.

\begin{table}[h]
\centering

\vspace{0.2in}
\begin{tabular}{ c r l } \hline \hline
  Fast time constant & 2.99  & $\pm$ 0.06 ns \\
  Slow time constant & 34.34 & $\pm$ 1.51 ns \\ \hline
\end{tabular}
\vskip0.2in
\begin{tabular}{ c c c } \hline \hline
              & Fast           & Slow \\
              & Component (\%) & Component (\%) \\
  \hline
  e$^+$/e$^-$ & 56.5 $\pm$ 0.9 & 42.5 $\pm$ 0.9 \\
       p      & 43.1 $\pm$ 0.7 & 56.9 $\pm$ 0.7 \\ \hline \hline
\end{tabular}
\caption{Scintillation time constants, and relative yields of fast and 
slow scintillation light used in simulation.  Values are taken from~\cite{LSND_om}, see text for details.}
\label{table:om_scint}
\end{table}


\chapter{Event Reconstruction and Particle Identification}


The event reconstruction in OscSNS is performed using a maximum likelihood
algorithm which is based on the recorded charge and time information of all
3900 PMTs in the inner detector, regardless of whether they are hit or not.
For an event characterized by a set of parameters \mbox{\boldmath{$\alpha$}},
such as event vertex, time, energy, etc., the likelihood for an observed set
of PMT measurements is given by
\[ {\cal L}_{event} =
   \prod_{i=nohit}\left[\frac{}{}\! 1-P_{hit}(\mbox{\boldmath{$\alpha$}})
                 \right] \times
   \prod_{i=hit}P_{hit}(\mbox{\boldmath{$\alpha$}})
                f_q(q_i;\mbox{\boldmath{$\alpha$}})
                f_t(t_i;\mbox{\boldmath{$\alpha$}}), \]
where:
\begin{enumerate}
\item[(i)]   $P_{hit}(\mbox{\boldmath{$\alpha$}})$ is the probability of the
             {\em{i}}th PMT to be hit given the event parameters
             \mbox{\boldmath{$\alpha$}},
\item[(ii)]  $(q_i,t_i)$ are the measured charge and time on the
             \mbox{\em{i}}th hit PMT,
\item[(iii)] $f_q(q_i;\mbox{\boldmath{$\alpha$}})$ is the probability
             distribution function (PDF) for the measured charge on the
             \mbox{\em{i}}th hit PMT given the event parameters
             \mbox{\boldmath{$\alpha$}}, evaluated at the measured value $q_i$,
             and
\item[(iv)]  $f_t(q_i;\mbox{\boldmath{$\alpha$}})$ is the PDF for the measured
             time on the \mbox{\em{i}}th hit PMT given the event parameters
             \mbox{\boldmath{$\alpha$}}, evaluated at the measured value $t_i$.
\end{enumerate}
Generally, it is more convenient to work with the negative logarithm of the
event likelihood, and since the charge- and time-related portions of the
likelihood decouple naturally, we define
\[ F(\mbox{\boldmath{$\alpha$}}) \equiv
   - \ln {\cal L}_{event}(\mbox{\boldmath{$\alpha$}}) \equiv
   F_q(\mbox{\boldmath{$\alpha$}}) + F_t(\mbox{\boldmath{$\alpha$}}), \]
where
\begin{eqnarray*}
   F_q(\mbox{\boldmath{$\alpha$}}) & = &
   - \sum_{i=nohit}\ln\left[\frac{}{}\! 1-P_{hit}(\mbox{\boldmath{$\alpha$}})
                     \right]
   - \sum_{i=hit}\ln\left[\frac{}{}\! P_{hit}(\mbox{\boldmath{$\alpha$}})
                                      f_q(q_i;\mbox{\boldmath{$\alpha$}})
                   \right],\\
   F_t(\mbox{\boldmath{$\alpha$}}) & = &
   - \sum_{i=hit}\ln\left[\frac{}{}\! f_t(t_i;\mbox{\boldmath{$\alpha$}})
                   \right].
\end{eqnarray*}
For brevity, we shall refer to $F_q$ and $F_t$ as the charge and time
likelihoods, respectively, although strictly speaking they are the negative
logarithms of the likelihoods.

Both charge and time PDFs which enter the calculation of the likelihoods can
be obtained both from MC simulations and self-consistently from the data
(either calibration or cosmic-ray induced).
Furthermore, assuming that the PMT discriminator level is set low enough, the
hit probability for any given PMT can be assumed to be simply
$P_{hit}(\mbox{\boldmath{$\alpha$}})=1-\exp(-\mu)$,
where $\mu$ is the amount of predicted charge at the PMT given the set of
event parameters \mbox{\boldmath{$\alpha$}}.
The exact form of the PMT hit probability function can be measured
{\em in situ} from low-intensity laser calibration data.

For electrons, the event is fully characterized by a set of seven parameters:
the event vertex, time, direction and energy,
\[ \mbox{\boldmath{$\alpha$}}_e=(x_0,y_0,z_0,t_0,\varphi,\theta,E_e), \]
where $(\theta,\phi)$ give the event direction in the detector's coordinate
system, whereas a neutron is fully characterized by only five parameters,
namely the event 4-vertex and the visible energy:
\[ \mbox{\boldmath{$\alpha$}}_n=(x_0,y_0,z_0,t_0,E_n). \]
Each event is reconstructed under two different hypotheses, namely under the
electron model, which yields an event likelihood ${\cal L}_e$, and under the
neutron model, which yields an event likelihood ${\cal L}_n$.
The ratio of these two likelihoods, or equivalently, the difference in the
negative log-likelihoods,
\[ \ln\,\frac{{\cal L}_e}{{\cal L}_n} = F_n - F_e \]
serves as particle identification.

\begin{figure}
\vspace{5mm}
\centering
\includegraphics[width=12cm,angle=90,clip=true]{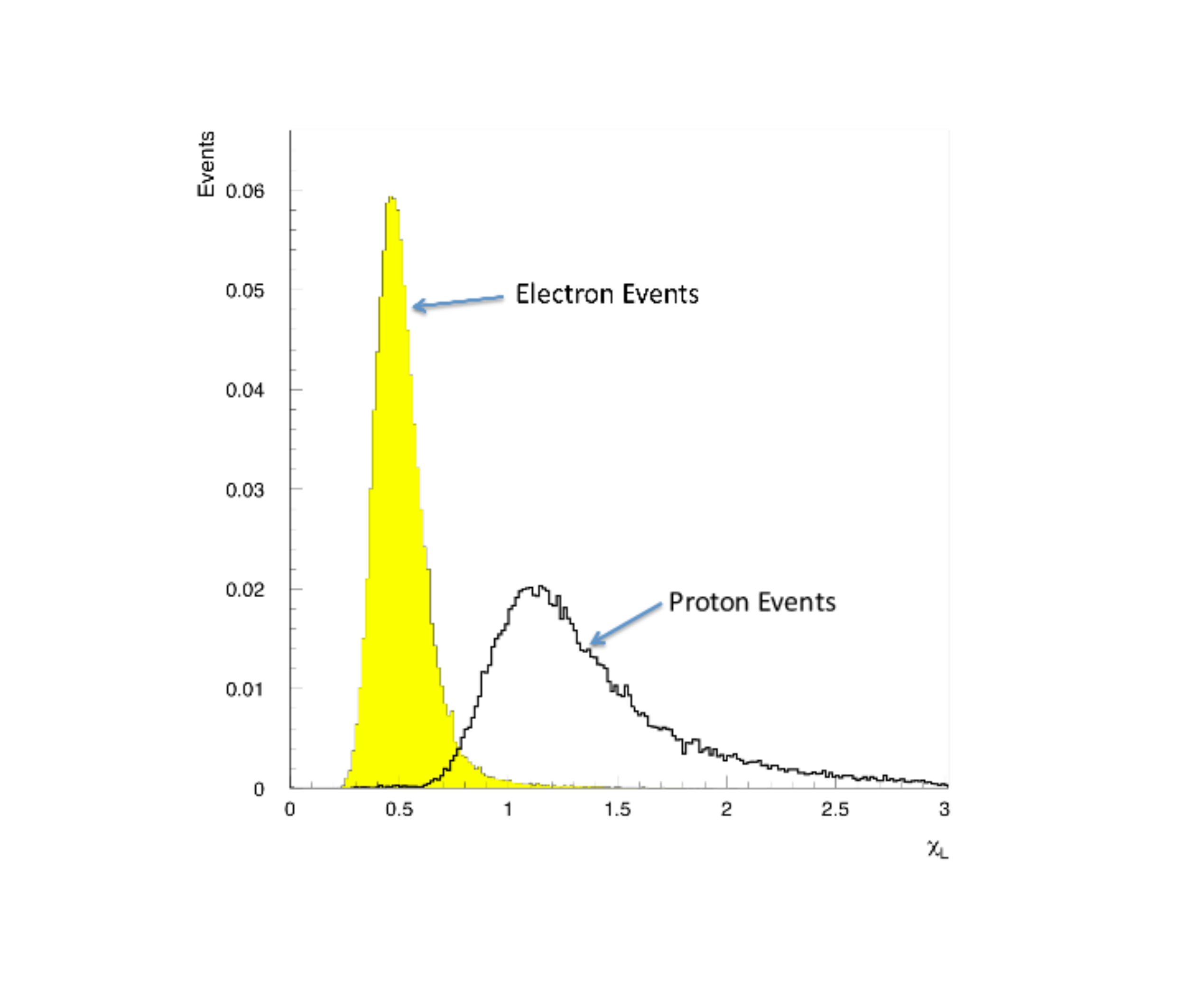}
\vspace{1mm}
\caption{
The separation of electrons and protons that was achieved with the
LSND experiment \cite{lsnd}.
}
\label{lsnd_pid}
\end{figure}

Fig. \ref{lsnd_pid} shows the separation of electrons and protons that was achieved with the
LSND experiment \cite{lsnd}. The parameter $\chi_L$ depended on three quantities: the fit to the Cerenkov
cone, the position and angle fit, and the fraction of hit PMTs that had times more than 10 ns later 
than the fitted times. Electron events could clearly be distinguished from proton and neutron events.

\chapter{Event Rates and Sensitivity Predictions}



The OscSNS experiment offers a complete physics program encompassing beyond the Standard Model searches 
as well as precision cross section measurements.  The following chapter details the event rate expectations 
for each of the physics goals, and provides sensitivity predictions for the oscillation searches.
These predictions are based on neutrino backgrounds only.  In addition, we expect a small contribution 
from cosmic-ray background, which is assumed to be negligible in this preliminary study.

The number of expected neutrino events of a given interaction type is
\begin{equation}
\Phi \ (\nu/year/cm^2) \cdot \sigma \ (cm^2) \cdot N_{targets},
\end{equation}
where $\Phi$ is the neutrino flux seen at the detector, $\sigma$ is the cross section
for the interaction, and N$_{targets}$ is the number of targets in the volume of the detector.  
Carbon is the target material 
for the sterile neutrino (disappearance) oscillation search, 
carbon and free protons are the targets for the appearance oscillation searches, and carbon only is the 
target for the $\nu_e$ C $\rightarrow$ e$^-$ N cross section measurement.    
Electrons are the target material for the elastic scattering cross section measurement.

\section{Cross Section and Target Information}

The following list describes the complete suite of measurements that OscSNS will be performing:

\begin{itemize}
\item Neutrino-electron elastic scattering used for cross section measurements:
        \begin{itemize}
        \item $\nue \ e^{-} \rightarrow \nue \ e^{-}$
        \item $\numu \ e^{-} \rightarrow \numu \ e^{-}$
        \item $\numubar \ e^{-} \rightarrow \numubar \ e^{-}$
        \end{itemize}

\item Neutral current (NC) interaction channels are used to search for oscillations into sterile neutrinos.  
        These analyses look for a deficit (disappearance) 
        of events over what is expected based on a non-oscillation hypothesis.  The relevant processes are: 
        \begin{itemize}
        \item $\nu \ ^{12}C \rightarrow \nu \ ^{12}C^*$ ($\numu$ disappearance, all disappearance)
        \item $\bar{\nu} \ ^{12}C \rightarrow \bar \nu \ ^{12}C^*$ (all disappearance)
        \end{itemize}

\item Charged current (CC) interaction channel are used to search for oscillations into sterile neutrinos.
        This analysis looks for a deficit (disappearance)
        of events over what is expected based on a non-oscillation hypothesis.  The relevant process is:
        \begin{itemize}
        \item $\nu_e \ ^{12}C \rightarrow e^- \ ^{12}N_{gs}$ ($\nu_e$ disappearance)
        \end{itemize}

\item Charged current (CC) interaction channels are employed in the appearance oscillation searches.  These analyses 
        look for an excess of events over those coming from neutrinos inherent, or intrinsic, to the beam (i.e. background 
        events).
        \begin{itemize}
        \item $\nuebar \ ^{12}C \rightarrow e^+ \ ^{12}B$ ($\nuebar$ appearance)
        \item $\nuebar \ p \rightarrow e^+ \ n$ ($\nuebar$ appearance)
        \item $\nue \ ^{12}C \rightarrow e^- \ ^{12}N_{gs}$ (mono-energetic $\nue$ appearance)
        \end{itemize}

\end{itemize}

\begin{figure}[htbp]
\centering
\includegraphics[scale=0.50,angle=-90]{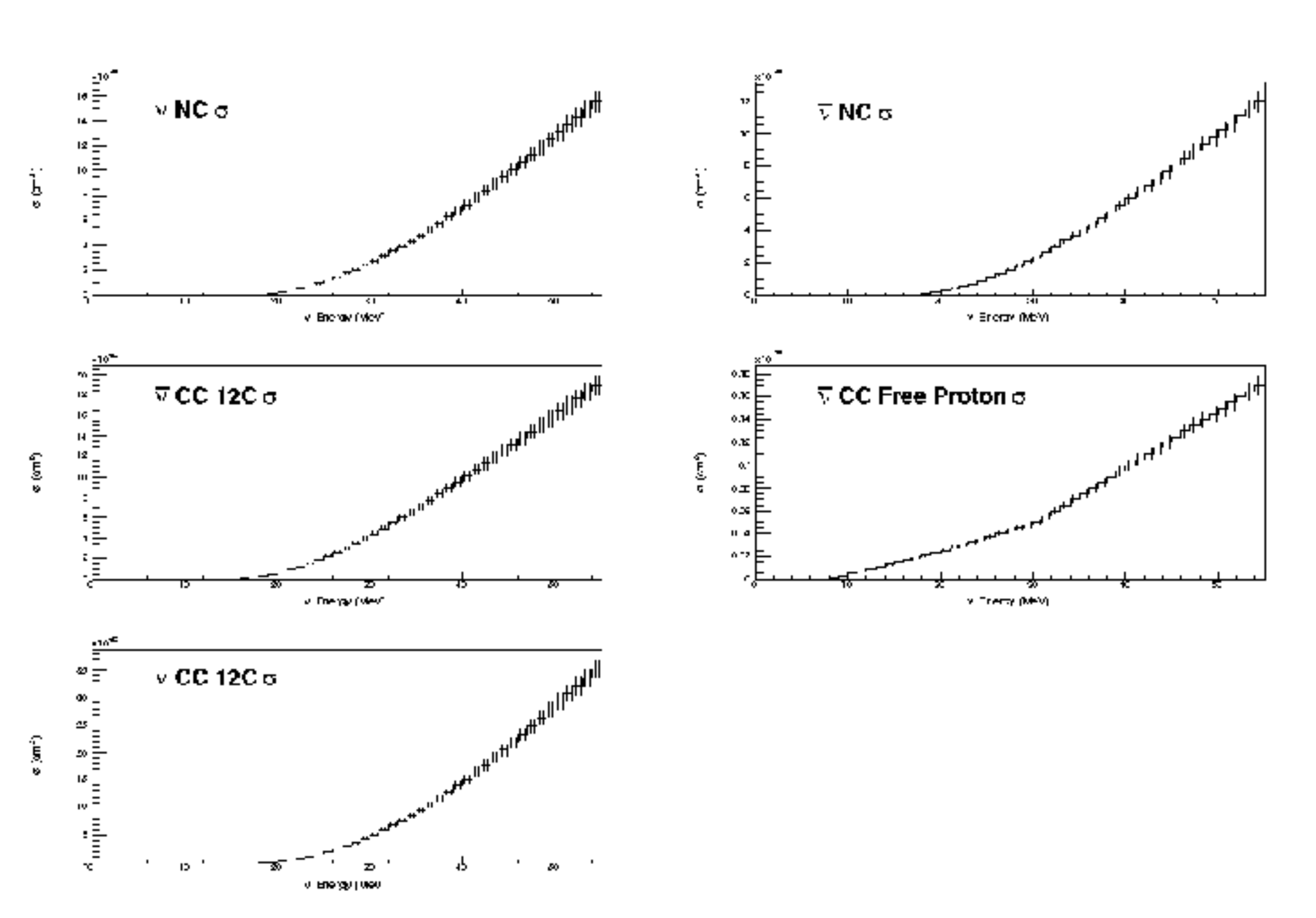}
\caption{Input cross sections as a function of incident neutrino energy.  Cross section plots include a 
5\% uncertainty.  
From left to right, top: $\nu \ ^{12}C \rightarrow ^{12}C^*$, $\bar{\nu} \ ^{12}C \rightarrow ^{12}C^*$; 
middle: $\nuebar \ ^{12}C \rightarrow e^+  \ ^{12}B$, $\nuebar \ p \rightarrow e^+ \ n$; bottom: $\nue \ ^{12}C \rightarrow 
e^- \ ^{12}N$.}
\label{fig_evtrate1}
\end{figure}

The cross section predictions used in the event rate estimates 
come from References \cite{fuku,kolbe,vogel}, and are provided 
as a function of incident neutrino energy.  Input cross section histograms are shown in 
Figure~\ref{fig_evtrate1}.

The density of mineral oil (CH$_2$) is 0.86 grams/cm$^3$.  The target mass is 
the density times the fiducial volume of the detector (523 m$^3$) or 450 tonnes.  The total mass divided 
by the mass of CH$_2$ (2.32 $\cdot \ \mathrm{10}^{-23}$ grams) 
gives the number of CH$_2$ molecules in the detector.  
For the above interactions on carbon, the number of molecules is the same as the 
number of $^{12}$C targets.  The total number of CH$_2$ targets is then
\begin{equation}
\frac{0.86 \ g/cm^3 \cdot 523 m^3}{2.32\cdot 10^{-23} \ g}, 
\end{equation}
or 1.94 $\cdot \ \mathrm{10}^{31}$ $^{12}$C targets.  
The total number of free proton targets (for the appearance search) 
is 2 times the number of carbon targets, or 3.9 $\cdot \ \mathrm{10}^{31}$ targets.
The total number of electron targets for the neutrino-electron scattering cross sections is 8 times the number of CH$_2$ molecules
or 15.5 $\cdot \ \mathrm{10}^{31}$ $^{12}$ electron targets.

\section{Total Expected Flux}

\begin{figure}[htbp]
\centering
\includegraphics[scale=0.50,angle=-90]{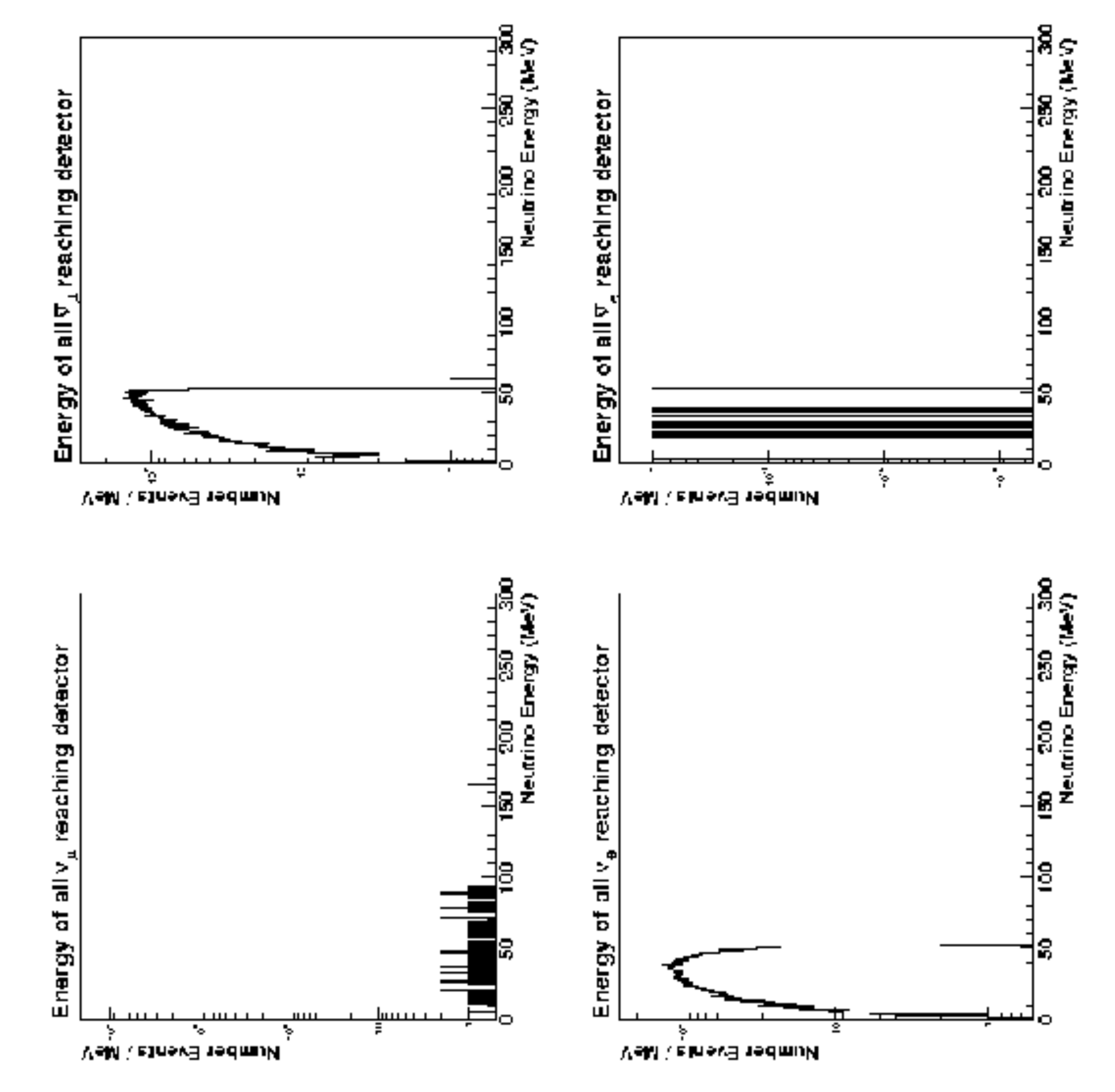}
   \caption{Input flux as a function of incident neutrino energy, for neutrinos reaching the detector.  These plots include neutrinos from DAR and DIF.  From left to right, top: $\numu$, $\numubar$; bottom: $\nue$, $\nuebar$.  Rates are for 50M simulated protons on target.}
\label{fig_evtrate2}
\end{figure}

The flux estimates come from a Geant4 simulation of the proton-mercury interaction, 
described in more detail in Chapter~\ref{chap:flux}.  
The flux rates are for a detector 
located 60 meters away from the interaction (creation of neutrinos) point, $\sim$150 degrees in the 
backward direction from the proton beam, and are for 50M simulated protons 
on target (POT), resulting in the following number of neutrinos passing through the detector:
$\numu$ = 11289, $\nue$ = 11224, $\numubar$ = 11357, $\nuebar$ = 15 
(See Figure~\ref{fig_evtrate2}).
The majority of these neutrinos come from particles decaying at rest (DAR).  Only 0.23\% of the 
neutrino beam seen by the detector originate from mesons decaying in flight (DIF).

The neutrino flux per year is found by assuming 8.7 $\cdot \ \mathrm{10}^{15}$ POT/second from the SNS (for $\sim$1.4 MW 
beam power), or 2.74 $\cdot \ \mathrm{10}^{23}$ POT/year.  Dividing this by 50M results in
a conversion factor of 5.5 $\cdot \ \mathrm{10}^{15}$ POT/year.  The expected neutrino fluxes are:
\begin{itemize}
\item $\numu$ = 2.48 $\pm$ 0.02 $\cdot \ \mathrm{10}^{22}$ $\nu$/year
\item $\nue$ =  2.46 $\pm$ 0.02 $\cdot \ \mathrm{10}^{22}$ $\nu$/year
\item $\numubar$ = 2.50 $\pm$ 0.02 $\cdot \ \mathrm{10}^{22}$ $\nu$/year
\item $\nuebar$ = 3.30 $\pm$ 0.85 $\cdot \ \mathrm{10}^{19}$ $\nu$/year, 
\end{itemize}

where errors are statistical.  The flux per cm$^2$ is 

\begin{equation}
\frac{\mathrm{neutrino \ flux}}{4 \pi \cdot (6000 \ cm)^2}, 
\end{equation}

which results in the following rates:

\begin{itemize}
\item $\numu$ = 5.48 $\pm$ 0.05 $\cdot \ \mathrm{10}^{13}$ $\nu$/year/cm$^2$
\item $\nue$ =  5.45 $\pm$ 0.05 $\cdot \ \mathrm{10}^{13}$ $\nu$/year/cm$^2$
\item $\numubar$ = 5.51 $\pm$ 0.05 $\cdot \ \mathrm{10}^{13}$ $\nu$/year/cm$^2$
\item $\nuebar$ = 7.28 $\pm$ 1.88 $\cdot \ \mathrm{10}^{10}$ $\nu$/year/cm$^2$
\end{itemize}

The total neutrino flux at the detector is $\sim$1.64 $\cdot \ \mathrm{10}^{14}$ $\nu$/year/cm$^2$.
The flux numbers quoted above are the final fluxes expected, integrated over the entire energy 
spectrum for each neutrino type.

\section{Disappearance Sensitivity}

For the disappearance sensitivity, we find the number of non-oscillated events
by multiplying the intrinsic neutrino flux by 
the 2-neutrino oscillation probability:

\begin{equation}
P \ = \ 1 - sin^2 2\theta \ sin^2(\frac{1.27 \Delta m^2 L}{E})
\end{equation}

In the above equation, $L$ is the distance traveled by the neutrino ($50<L<70$ m) and $E$ is the
neutrino energy ($E = 30$ MeV for $\nu_\mu$ disappearance and approximately $30<E<52.8$ MeV for
$\nu_e$ disappearance). 
Best fits to the world data \cite{sorel,karagiorgi,giunti,kopp} predict values of
$\sin^22\theta$ to be $\sim 10-15\%$ for both $\nu_\mu$ and $\nu_e$ disappearance.

\section{Appearance Sensitivity : $\numubar \rightarrow \nuebar$, $\numu \rightarrow \nue$}

For the appearance sensitivity we start with the $\numubar$($\numu$) neutrino flux, and assume 
100\% transmutation.  This number of transmuted $\nuebar$($\nue$) is multiplied by the 
2-neutrino oscillation 
probability:
\begin{equation}
P \ = \ sin^2 2\theta \ sin^2(\frac{1.27 \Delta m^2 L}{E})
\end{equation}

Best fits to the world data \cite{sorel,karagiorgi,giunti,kopp} predict values of
$\sin^22\theta$ to be $\sim 0.3-0.5\%$. (Note that $\sin^22\theta$ is different for appearance
and disappearance oscillations.)
The oscillated event spectrum is added to the intrinsic $\nuebar$($\nue$) spectrum, to produce the final 
spectrum of observed events.

\section{Event Rates and Background Estimates}

Event rate estimates are based on fluxes seen by the full volume of the detector.  
However, final estimates need to scale rates to the expected fiducial volume (6 m diameter by 18.5 m long)
and to include reductions due to detector efficiency (50\%) and beam-on efficiency (50\%). 

Errors on the rate estimates comprise statistical errors 
 and a 5\% systematic error on the neutrino fluxes and cross sections.

\subsection{$\nu_\mu$ Disappearance Analysis}

For the $\numu \rightarrow \nu_{sterile}$ search, we look for a deficit in the 
expected number of neutral current $\nu_\mu \ ^{12}C \rightarrow \nu_\mu \ ^{12}C^*$ events.  The total number of 
these interactions per year, in the absence of oscillations to a sterile neutrino, 
for a detector at 60 meters from the interaction point, is 
\begin{equation}
5.48\cdot 10^{13} \nu/year/cm^2 \cdot \sigma_{\nu_\mu ~ ^{12}C \rightarrow \nu_\mu ~ ^{12}C^*} \cdot 1.94\cdot 10^{31} \ ^{12}C \ targets,
\end{equation}
or 2977 events per year, where $\sigma_{\nu_\mu ~ ^{12}C \rightarrow \nu_\mu ~ ^{12}C^*} = 2.8 \times 10^{-42}$ cm$^2$ \cite{kolbe}.  
After reductions due to detector efficiency (50\%) and beam-on efficiency (50\%) this 
becomes {\bf 745 $\pm$ 42 expected events per year}.

\vspace{0.25in}

If we perform the search using the entire neutrino flux from $\nu_\mu$, $\nu_e$, and $\bar \nu_\mu$, 
where $\sigma_{\nu ^{12}C \rightarrow \nu ^{12}C^*} = 13.3 \times 10^{-42}$ cm$^2$ \cite{kolbe}, we expect 14,140
events per year, in the absence of oscillations.  This reduces to {\bf 3535 $\pm$ 182 events per year} after reductions
due to detector efficiency (50\%) and beam-on efficiency (50\%).

\vspace{0.5in}

\begin{figure}[hbp]
\centering
\includegraphics[scale=0.5,angle=90]{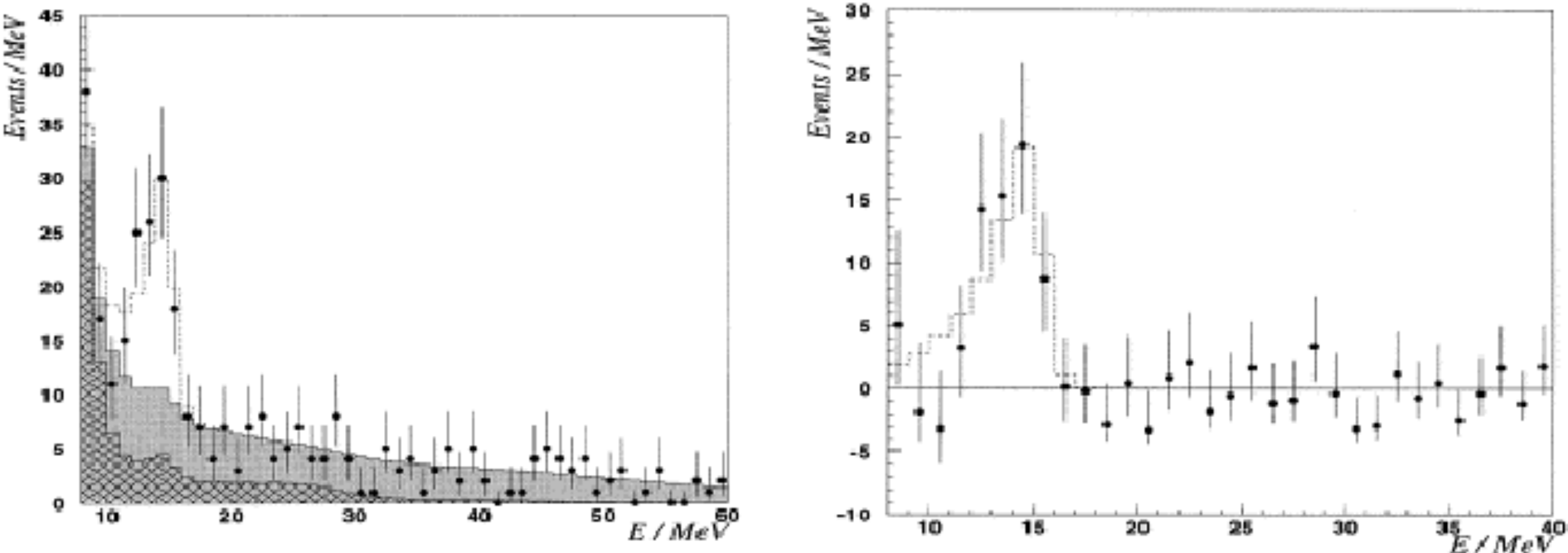}
\caption{The KARMEN measurement of $\nu_\mu ~ ^{12}C \rightarrow \nu_\mu ~ ^{12}C^*$, where
the shaded region corresponds to the fast neutron background and the hatched region corresponds to the
neutrino and cosmic ray backgrounds. The right plot shows the background subtracted signal.}
\label{numuC}
\end{figure}

\begin{figure}[htbp]
\hspace{-0.75cm}
\begin{minipage}[h]{0.46\linewidth}
\includegraphics[scale=0.40,angle=0]{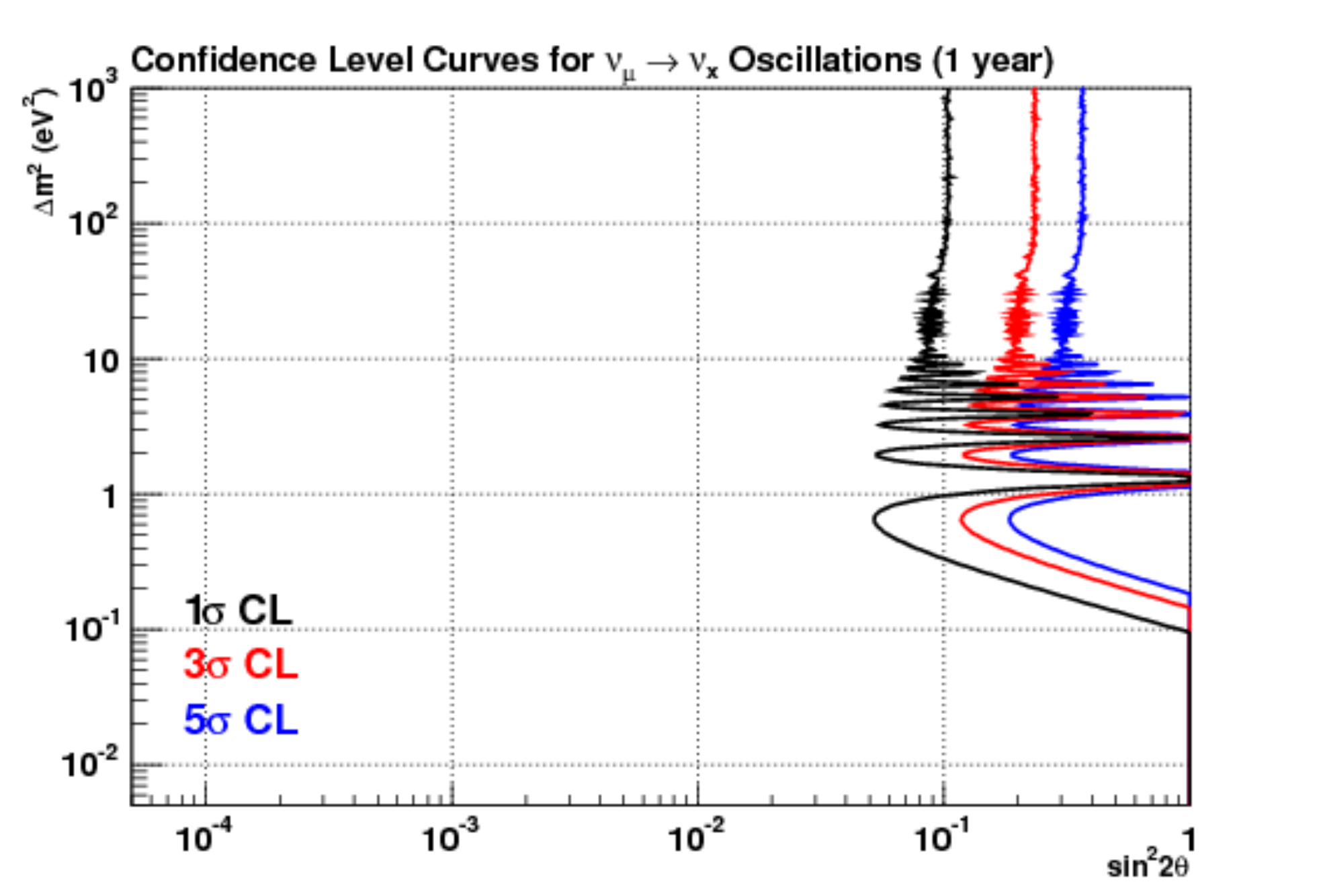}
\end{minipage}
\hfill
\begin{minipage}[h]{0.46\linewidth}
\includegraphics[scale=0.40,angle=0]{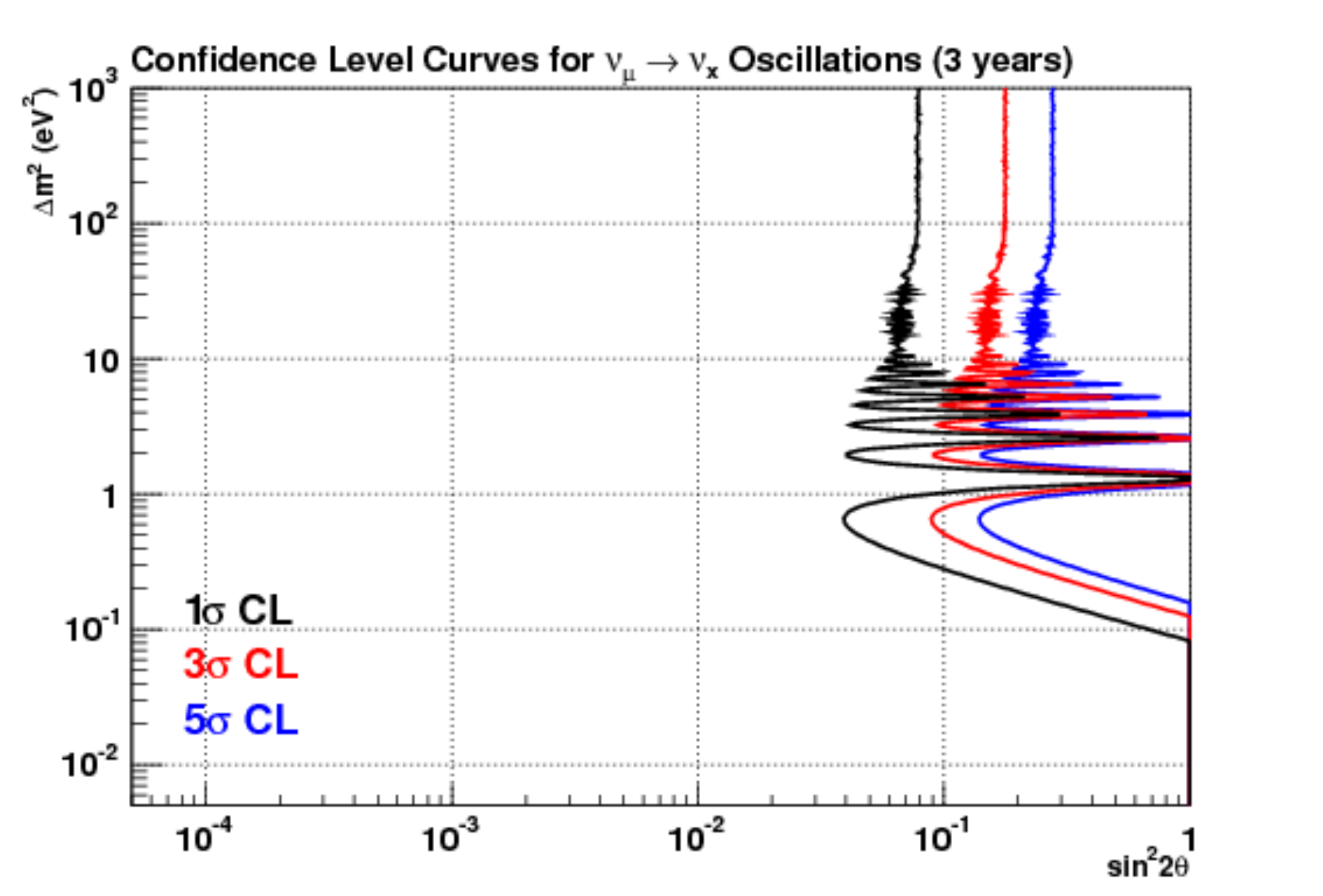}
\end{minipage}
   \caption{OscSNS sensitivity for $\numu$ disappearance, for two calendar years of run time (left), and six 
calendar years of run time (right).}
\label{fig_dis_sen}
\end{figure}

\begin{figure}[htbp]
\hspace{-0.75cm}
\begin{minipage}[h]{0.46\linewidth}
\includegraphics[scale=0.40,angle=0]{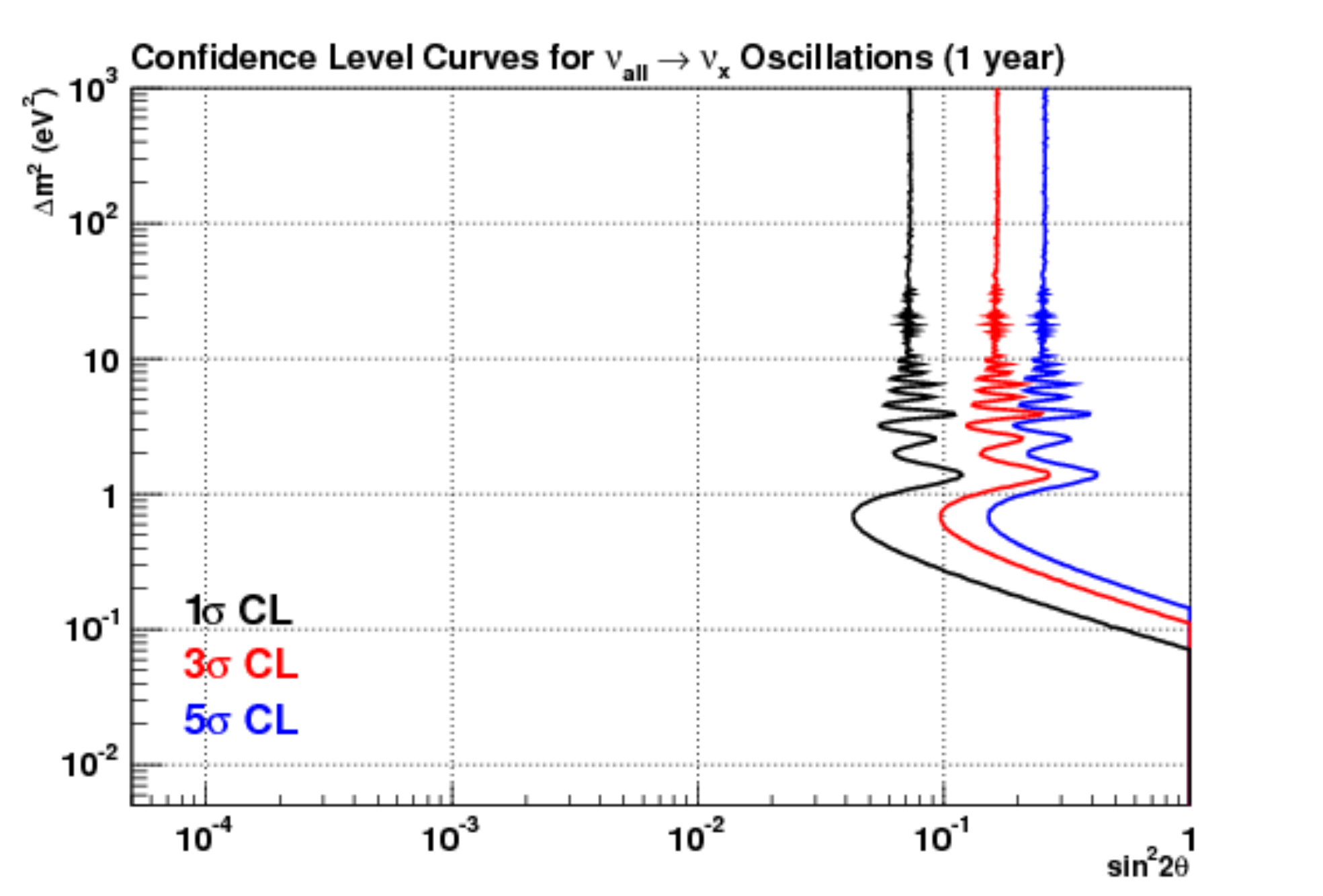}
\end{minipage}
\hfill
\begin{minipage}[h]{0.46\linewidth}
\includegraphics[scale=0.40,angle=0]{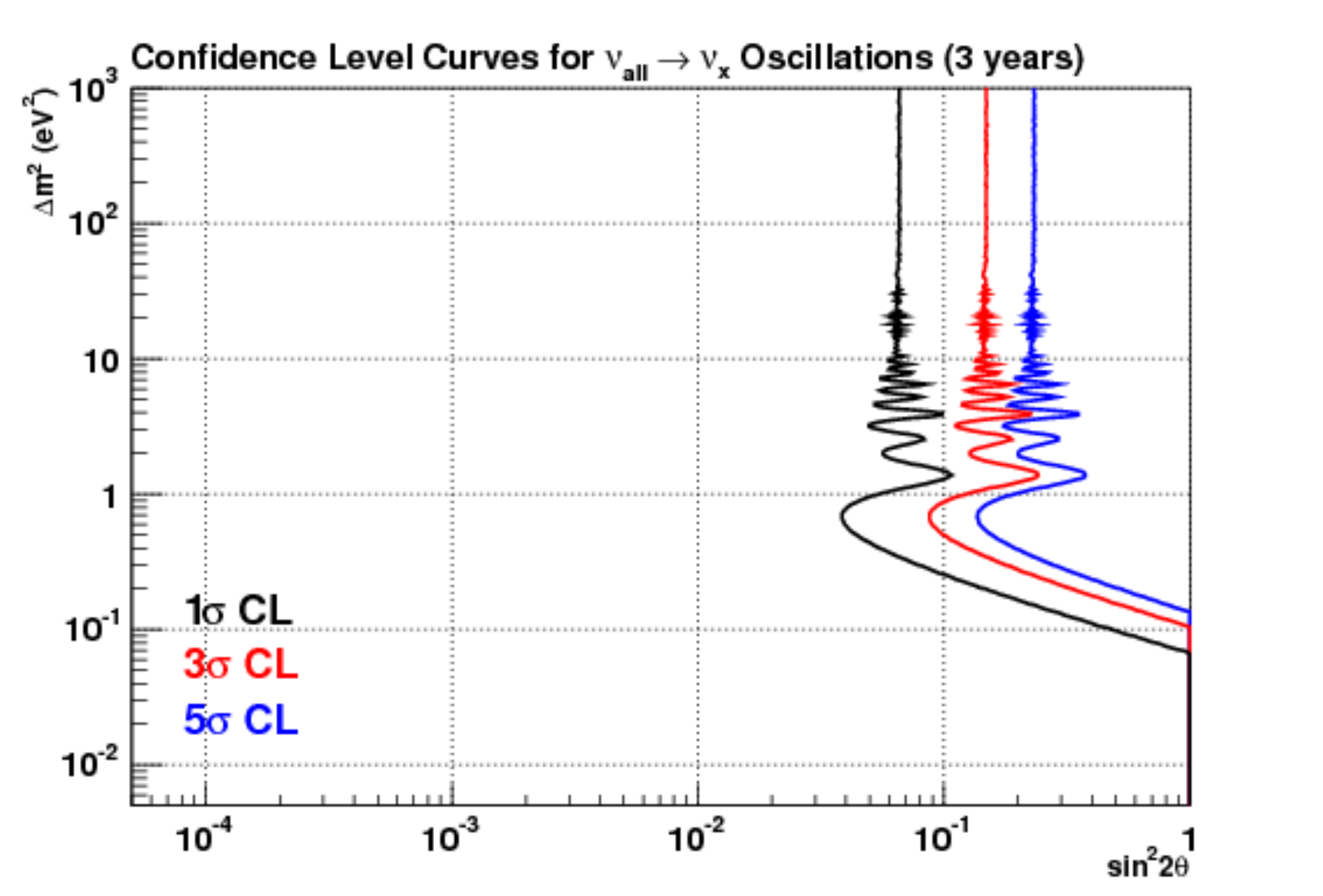}
\end{minipage}
   \caption{OscSNS sensitivity for the total flux disappearance, for two calendar years of run time (left), 
and six calendar years of run time (right).}
\label{fig_dis_all}
\end{figure}

The estimated background under the 15.11 MeV $\gamma$ peak is approximately 30\%, which is obtained 
from the KARMEN experiment \cite{karmen} after subtracting the fast neutron background.  This should be much lower 
for OscSNS due to the increased amount of passive shielding between the beam dump and the detector. 
Fig. \ref{numuC} shows the KARMEN measurement of $\nu_\mu ~ ^{12}C \rightarrow \nu_\mu ~ ^{12}C^*$, where
the shaded region corresponds to the fast neutron background and the hatched region corresponds to the
neutrino and cosmic ray backgrounds. The right plot shows the background subtracted signal.
Sensitivity curves in $\sinsqtheta$-$\dmsq$ space are shown in Figure~\ref{fig_dis_sen} and 
Figure~\ref{fig_dis_all} for $\numu$ and the complete neutrino flux, respectively.  Sensitivity 
curves are given for two calendar years and for six calendar years of running. 

\subsection{$\nu_e$ Disappearance Analysis}

For the $\nu_e \rightarrow \nu_{sterile}$ search, we look for a deficit in the
expected number of charged current $\nu_e \ ^{12}C \rightarrow e^- \ ^{12}N_{gs}$ events. The signature for
this reaction is an electron correlated with a positron for $^{12}N_{gs}$ beta decay. 
The number of events per year is:
\begin{equation}
5.45\cdot 10^{13} \nu/year/cm^2 \cdot \sigma_{\nue \ ^{12}C \rightarrow e^- \ ^{12}N} \cdot 1.94\cdot 10^{31} \ ^{12}C \
targets
\end{equation}
or 9,410 events per year, where the $\sigma_{\nue \ ^{12}C \rightarrow e^- \ ^{12}N_{gs}} = 8.9 \times 10^{-42}$ cm$^2$ \cite{kolbe}.
Folding in the detector efficiency (50\%) and beam-on efficiency (50\%),
we obtain {\bf 2353 $\pm$ 123 expected events per year}.

\vspace{0.25in}

\begin{figure}[htbp]
\centering
\includegraphics[scale=0.5,angle=90]{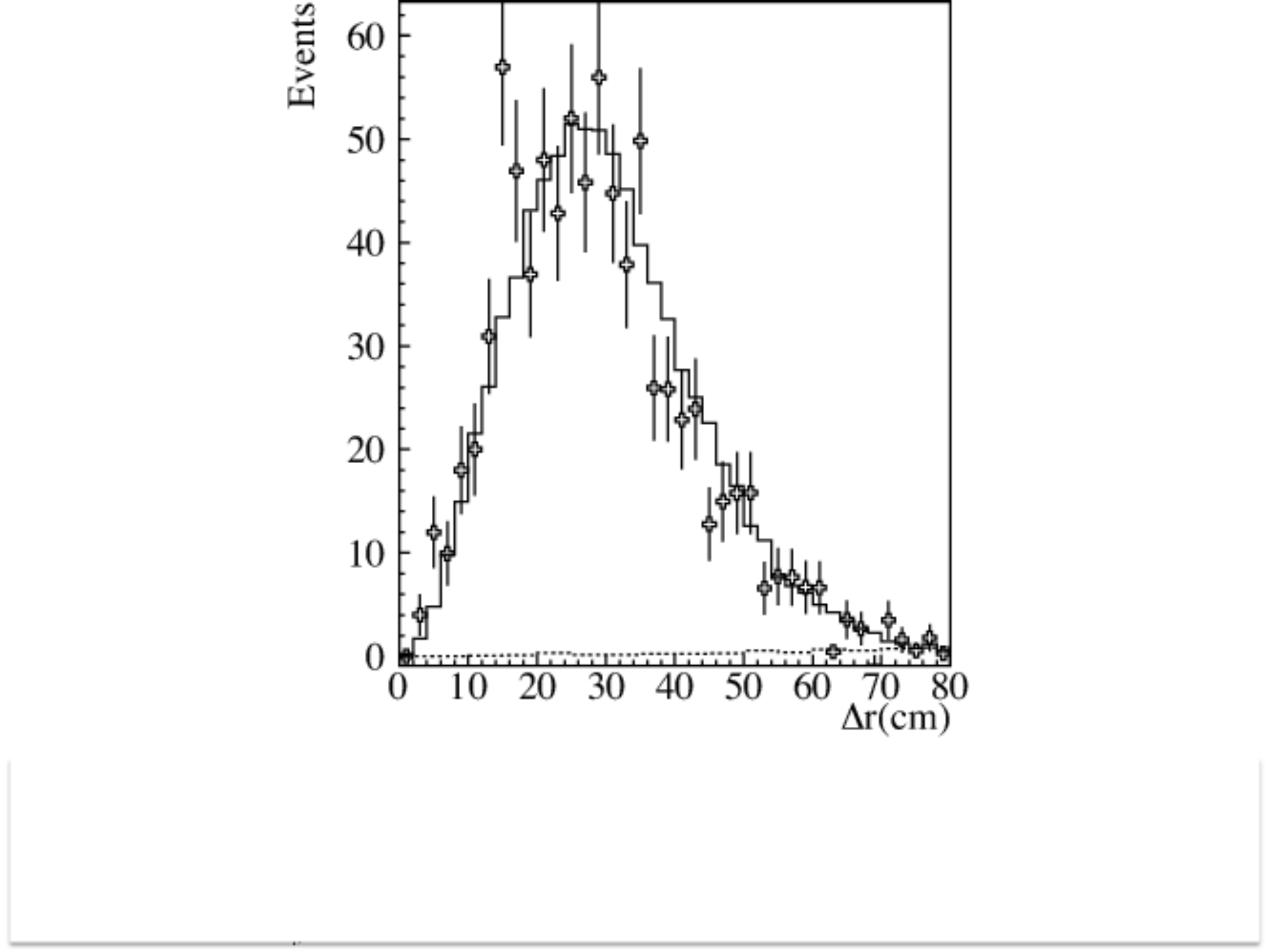}
\caption{The measurement of the reaction $\nu_e \ ^{12}C \rightarrow e^- \ ^{12}N_{gs}$ by
the LSND experiment. The background, which is estimated to be $\sim 1\%$, is shown as the 
dotted curve.}
\label{nueCgs}
\end{figure}

The estimated background for the $\nu_e \ ^{12}C \rightarrow e^- \ ^{12}N_{gs}$ reaction is 1\%, which
is obtained from the measurement by the LSND experiment \cite{lsnd_nuec}, as shown in Fig. \ref{nueCgs}.

\vspace{0.5in}

\subsection{$\numubar \rightarrow \nuebar$ Appearance Analysis}

The signature for these events is $\nuebar$ interactions in the detector, in the form of 
$\nuebar \ ^{12}C \rightarrow e^+ \ ^{11}B \ n$ and $\nuebar \ p \rightarrow e^+ \ n$ followed by
$n p \rightarrow D \gamma(2.2)$ and the emission of a 2.2 MeV $\gamma$.

The number of intrinsic (background) $\nuebar$ events per year is a sum of the contributions:
\begin{equation}
7.28\cdot 10^{10} \nu/year/cm^2 \cdot \sigma_{\nuebar \ p \rightarrow e^+ \ n} \cdot 3.9\cdot 10^{31} \ free \ p \ targets,
\end{equation}
and 
\begin{equation}
7.28\cdot 10^{10} \nu/year/cm^2 \cdot \sigma_{\nuebar \ ^{12}C \rightarrow e^+ \ ^{11}B\ n} \cdot 1.94\cdot 10^{31} \ ^{12}C \ 
targets,
\end{equation}
or 207 total events per year, where $\sigma_{\nuebar \ p \rightarrow e^+ \ n} = 0.72 \times 10^{-40}$ cm$^2$
and $\sigma_{\nuebar \ ^{12}C \rightarrow e^+ \ ^{11}B \ n} = 2 \times 10^{-42}$ cm$^2$ \cite{vogel}.  
We expect {\bf 42 $\pm$ 5 final background events per year}, after 
reductions due to detector efficiency (50\%), beam-on efficiency (50\%), and the $E_e >$ 20 MeV selection (81\%) are applied.

\vspace{0.2in}

The total number of $\numubar$ that can oscillate into $\nuebar$ is 5.51 $\cdot \ \mathrm{10}^{13}$ $\nu$/year/cm$^2$.  
If we assume a 100\% transmutation rate, this full amount becomes $\nuebar$.
The total number of oscillated $\nuebar$ events per year is: 
\begin{equation}
5.51\cdot 10^{13} \nu/year/cm^2 \cdot \sigma_{\nuebar \ p \rightarrow e^+ \ n} \cdot 3.9\cdot 10^{31} \ free \ p \ targets,
\end{equation}
plus 
\begin{equation}
5.51\cdot 10^{13} \nu/year/cm^2 \cdot \sigma_{\nuebar \ ^{12}C \rightarrow e^+  \ ^{11}B \ n} \cdot 1.94\cdot 10^{31} \ ^{12}C 
\ targets,
\end{equation}
or 207,350 events per year, where $\sigma_{\nuebar \ p \rightarrow e^+ \ n} = 0.95 \times 10^{-40}$ cm$^2$
and $\sigma_{\nuebar \ ^{12}C \rightarrow e^+ \ ^{11}B \ n} = 3 \times 10^{-42}$ cm$^2$ \cite{vogel}.

\vspace{0.2in}

\begin{figure}[htbp]
\hspace{-0.75cm}
\begin{minipage}[h]{0.46\linewidth}
\includegraphics[scale=0.40,angle=0]{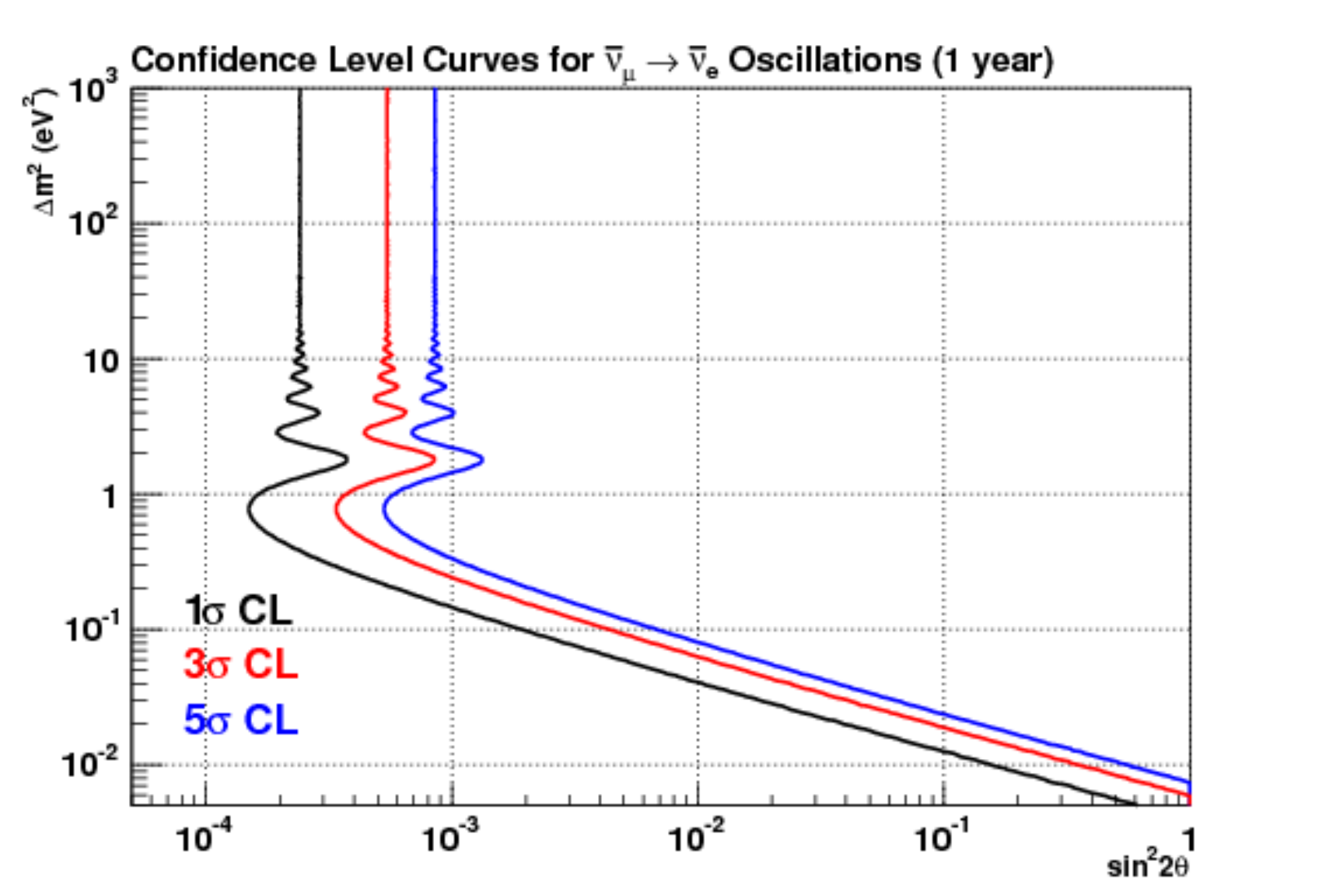}
\end{minipage}
\hfill
\begin{minipage}[h]{0.46\linewidth}
\includegraphics[scale=0.40,angle=0]{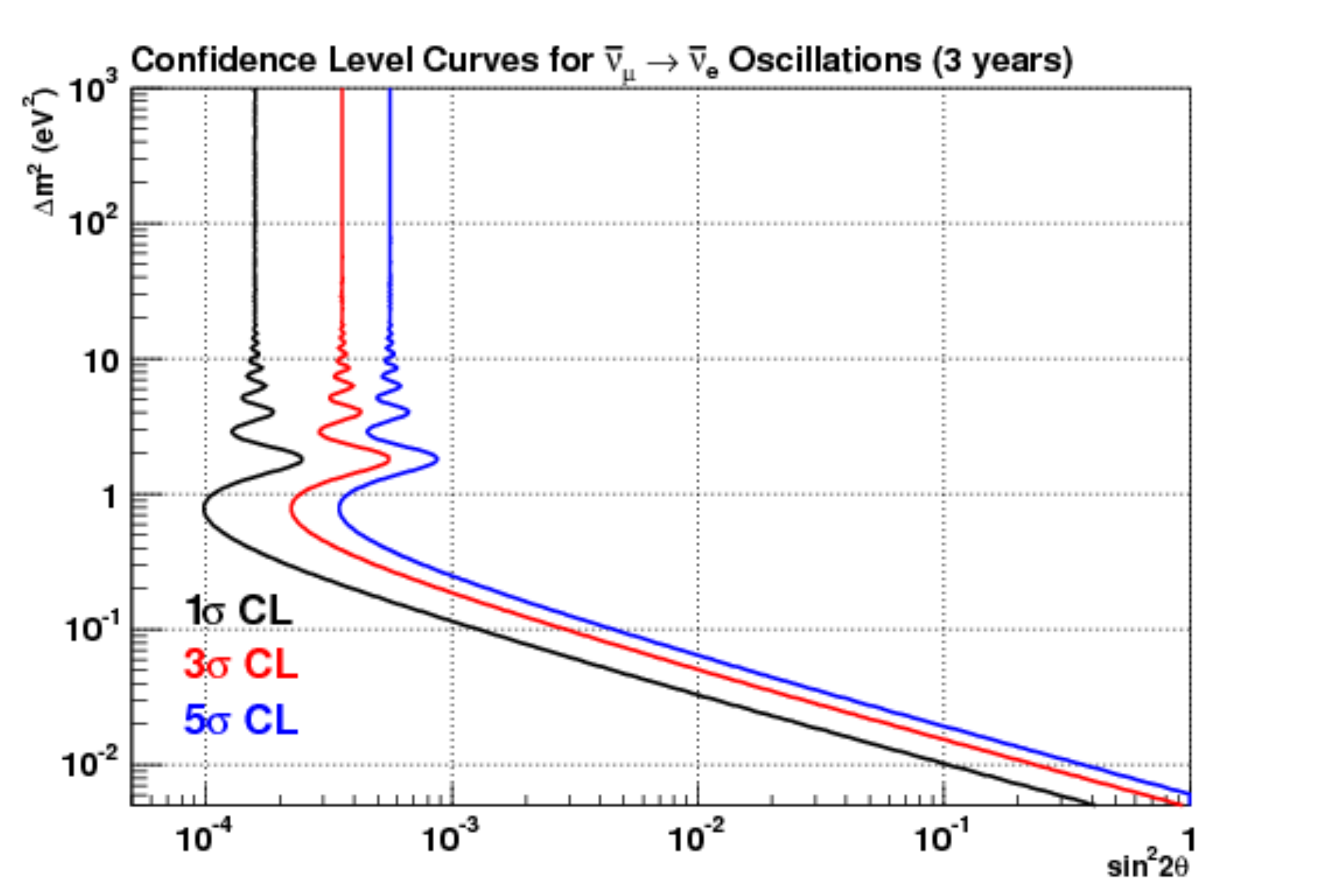}
\end{minipage}
   \caption{OscSNS sensitivity for $\numubar \rightarrow \nuebar$ oscillations, for two calendar years of 
run time (left), and six calendar years of run time (right).}
\label{fig_nuebar_app}
\end{figure}

We expect {\bf 120 $\pm$ 10 events} after the oscillation probability (0.26\%), detector 
efficiency (50\%), beam-on efficiency (50\%), and $E_e >$ 20 MeV selection (89\%) have been applied.  The final expected sensitivity for two calendar years and six calendar years run time is 
shown in Figure~\ref{fig_nuebar_app}.  
A detector at 60 meters will give a signal to background of $\sim$3 for this search channel.

\subsection{$\numu \rightarrow \nue$ Appearance Analysis}

The signature of these events are $\nue$ interactions in the detector, in the form of 
$\nue \ ^{12}C \rightarrow e^- \ ^{12}N_{gs}$, followed by $^{12}N_{gs}$ beta decay.
Therefore, there is a 2-fold signature of a $12.5$ MeV mono-energetic electron and a
correlated positron from beta decay.

The number of intrinsic $\nue$ background events per year is:
\begin{equation}
5.45\cdot 10^{13} \nu/year/cm^2 \cdot \sigma_{\nue \ ^{12}C \rightarrow e^- \ ^{12}N} \cdot 1.94\cdot 10^{31} \ ^{12}C \ 
targets
\end{equation}
or 9410 events per year, where the $\sigma_{\nue \ ^{12}C \rightarrow e^- \ ^{12}N_{gs}} = 8.9 \times 10^{-42}$ cm$^2$ \cite{kolbe}.
Folding in the detector efficiency (50\%) and beam-on efficiency (50\%),
we expect 2353 background events per year, spread over the entire event collection window.

However, this calculation does not take into account the timing structure of the beam.  The $\numu$ 
come primarily from the $\pip$ decay, and occur during the beam spill window.  Accepting only events 
occurring in the first 500 ns with respect to the beam on target signal will result in a $\sim$pure 
$\numu$ beam for use in a $\numu \rightarrow \nue$ oscillation search.  
Only $\sim$8.6\% of all $\nue$ reaching the detector occur in the first 500 ns.  The background 
expectation is reduced from 2353 to 203 events.
An additional cut may be applied to this analysis to further reduce background events.
The $\numu$ undergoing oscillations come from the $\pi^+$ DAR, and are mono-energetic.  
The energy of the produced electron will appear between 11 and 15 MeV, after smearing due 
to the energy resolution of the detector.  Applying a requirement that the electron be in this 
energy range reduces the background by a factor of $\sim$0.06.  Approximately {\bf 12 $\pm$ 3 
intrinsic background events} are expected per calendar year in this analysis.

\vspace{0.2in}

\begin{figure}[htbp]
\hspace{-0.75cm}
\begin{minipage}[h]{0.46\linewidth}
\includegraphics[scale=0.40,angle=0]{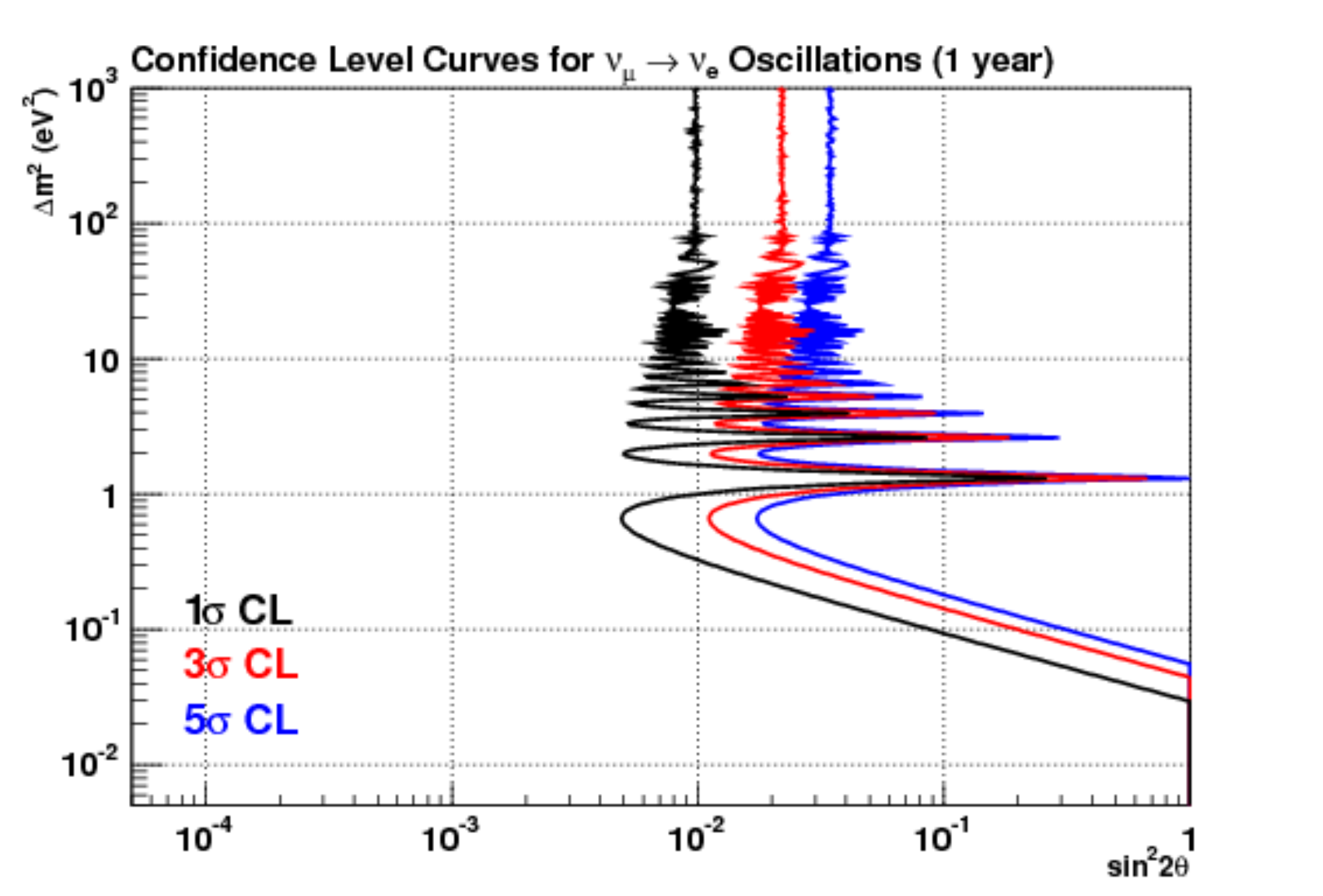}
\end{minipage}
\hfill
\begin{minipage}[h]{0.46\linewidth}
\includegraphics[scale=0.40,angle=0]{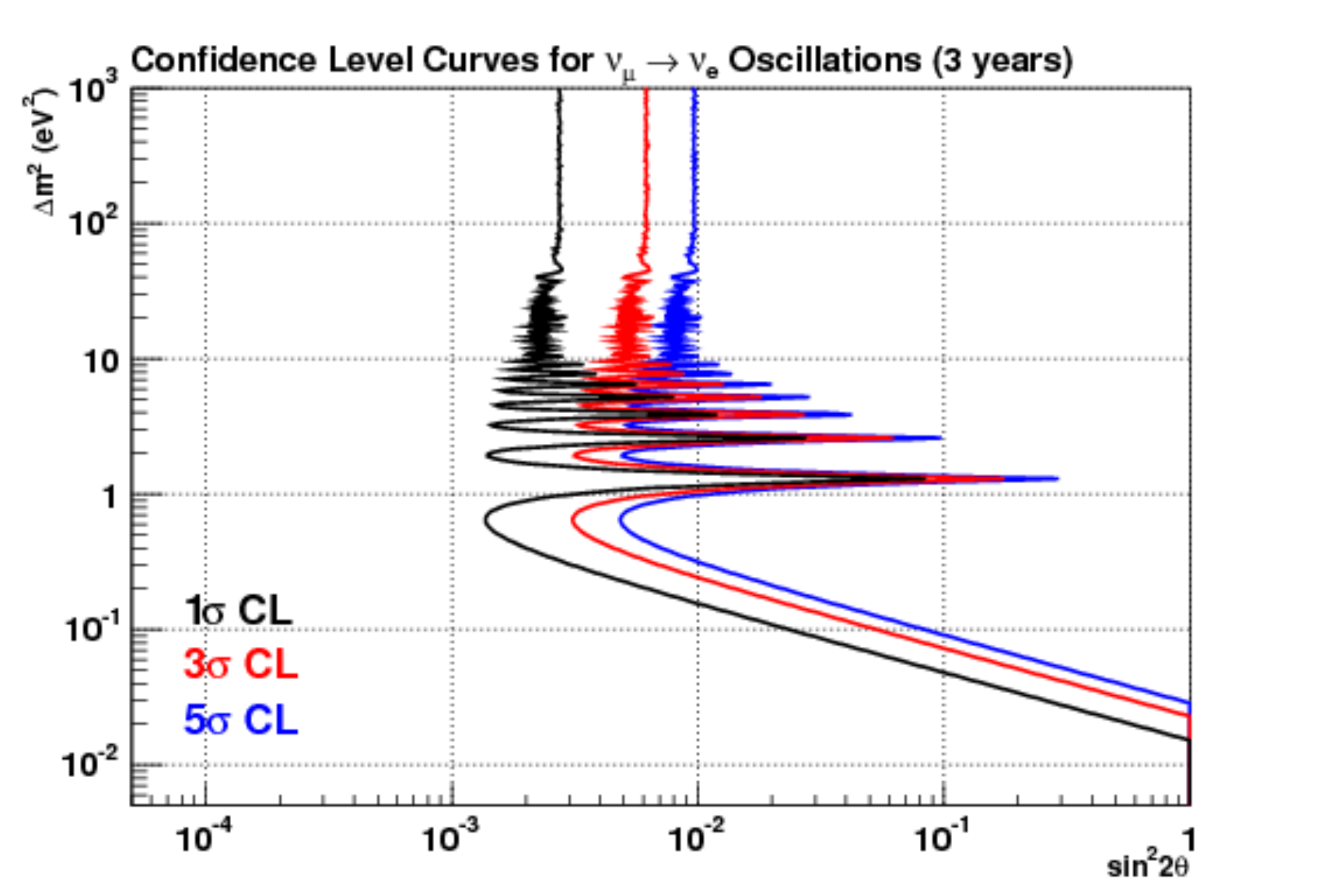}
\end{minipage}
   \caption{OscSNS sensitivity for $\numu \rightarrow \nue$ oscillations, for two calendar years of 
run time (left), and six calendar years of run time (right).}
\label{fig_nue_app}
\end{figure}

The total number of $\numu$s that can oscillate into $\nue$ is 5.48 $\cdot \ \mathrm{10}^{13}$ $\nu$/year/cm$^2$.  
If we assume a 100\% transmutation rate, this full amount becomes $\nue$.
The total number of oscillated $\nue$ events per year is: 
\begin{equation}
5.48\cdot 10^{13} \nu/year/cm^2 \cdot \sigma_{\nue \ ^{12}C \rightarrow e^- \ ^{12}N} \cdot 1.94\cdot 10^{31} \ ^{12}C \ targets,
\end{equation}
or 5316 events per year, where the $\sigma_{\nue \ ^{12}C \rightarrow e^- \ ^{12}N_{gs}} = 5 \times 10^{-42}$ cm$^2$ \cite{kolbe}.
After including factors for detector efficiency (50\%), beam-on efficiency (50\%), and oscillation probability
(0.26\%), we expect {\bf 3.5 $\pm$ 1.5 total oscillated events} per calendar year.  
Figure~\ref{fig_nue_app} shows the expected reach 
for this analysis channel, for two and six calendar years run time.

\subsection{Oscillation Analysis Summary}

Table~\ref{table:nums} summarizes the expected event sample sizes for the disappearance and appearance 
oscillation searches, per calendar year.  Figures~\ref{fig_dis_sen}, ~\ref{fig_dis_all}, 
~\ref{fig_nuebar_app}, and ~\ref{fig_nue_app} present the expected sensitivity for each of these analyses.  
Figures show the expected reach for each analysis for two calendar years and for six calendar years run time.

\begin{table}[ht]
\begin{center}
\begin{tabular}{|c|c|c|} \hline
 \multicolumn{1}{c}{Channel} &  \multicolumn{1}{c}{Background} & \multicolumn{1}{c}{Signal}  \\ \hline \hline
 \multicolumn{3}{c}{Disappearance Search}  \\ \hline
$\nu_\mu \ ^{12}C \rightarrow \nu_\mu \ ^{12}C^*$    &      & \\ 
$\nu_e \ ^{12}C \rightarrow \nu_e \ ^{12}C^*$    &      & \\ 
$\bar{\nu_\mu} \ ^{12}C \rightarrow \bar \nu_\mu \ ^{12}C^*$ & 1060 $\pm$ 36 & 3535 $\pm$ 182  \\ \hline
$\numu \ ^{12}C \rightarrow \numu \ ^{12}C^*$  & 224 $\pm$ 75 & 745 $\pm$ 42  \\ \hline

$\nu_e \ ^{12}C \rightarrow e^- \ ^{12}N_{gs}$  & 24 $\pm$ 13 & 2353 $\pm$ 123  \\ \hline

\multicolumn{3}{c}{Appearance Search} \\ \hline
$\numubar \rightarrow \nuebar$: $\nuebar \ ^{12}C \rightarrow e^+  \ ^{11}B \ n$  &  & \\ 
$\numubar \rightarrow \nuebar$: $\nuebar \ p \rightarrow e^+  \ n$          & 42 $\pm$ 5 & 120 $\pm$ 10 \\ \hline

$\numu \rightarrow \nue$: $\nue \ ^{12}C \rightarrow e^- \ ^{12}N_{gs}$  & 12 $\pm$ 3 & 3.5 $\pm$ 1.5 \\ 

\hline
\hline 

\end{tabular}
\normalsize
\caption{Summary of per calendar year event rate predictions for a detector located at the SNS a distance of 60 meters from the interaction point, at $\sim$150 
degrees in the backward direction from the proton beam.  The first column is oscillation channel, the second column is the 
expected intrinsic background, and the third column is the expected signal for appearance searches and the total
number of events for disappearance searches.  All event rates account for a 50\% detector efficiency, a 50\% 
beam-on efficiency, and a fiducial volume of 523 m$^3$, and are in units of expected events per calendar 
year.  Appearance signal estimates assume a 0.26\% oscillation probability.}
\label{table:nums}
\end{center}
\end{table}

\subsection{Cross Section Analyses}

The three flagship cross section analyses of OscSNS are the measurement of neutrino-electron elastic scattering 
$\nue  \ e^{-} \rightarrow \nue  \ e^{-}$, $\numu C \rightarrow \numu C^*$ NC scattering, 
and $\nue  \ ^{12}C \rightarrow e^- \ ^{12}N_{gs}$ CC scattering.

The number of $\nue  \ e^{-} \rightarrow \nue  \ e^{-}$ events expected per year is:
\begin{equation}
5.45\cdot 10^{13} \nu/year/cm^2 \cdot \sigma_{\nue \ e^{-} \rightarrow \nue e^{-}} \cdot 8 \cdot 1.94\cdot 10^{31} \ e^{-} \ targets,
\end{equation}
or 2707 events per year, where $\sigma_{\nue \ e^{-} \rightarrow \nue e^{-}} = 3.2 \times 10^{-43}$ cm$^2$ \cite{lampf}.

If we assume 50\% detector efficiency and 50\% beam-on efficiency, this becomes 677 $\pm$ 39 events per
calendar year.  
The best measurement of this interaction so far
arises from the LSND experiment in a sample of only 191 events \cite{lampf}, and has a 17\% total error.  OscSNS will far 
surpass this measurement in statistics and total uncertainty, with only one year of run time.

The NC $\numu C$ interaction has been measured by the KARMEN experiment to be 3.2 $\pm$ 0.5 $\pm$ 0.4  
$\cdot \ \mathrm{10}^{-42}$ cm$^2$ \cite{karmen}.  This 
measurement was performed using only 86 $\numu$ events.  While this result is consistent with theory, 
it has a 20\% total error, half of which 
is due to statistics.  OscSNS will collect 745 $\pm$ 42 of these events in only one calendar year of run time 
(Table~\ref{table:xsecnums}), and is expected 
to have smaller systematic errors.  This will allow for the world's most precise 
cross section measurement; any deviations from theory could 
indicate the presence of sterile neutrinos.

For the CC $\nue  \ ^{12}C \rightarrow e^-  \ ^{12}N_{gs}$ measurement, we are able to use the entire spectrum of intrinsic 
$\nue$, 
in contrast to the oscillation search, where we only consider the neutrino flux in the first 500 ns of the beam spill.
We expect 2353 $\pm$ 123 of these events per calendar year, in the absence of oscillations.

\begin{table}[ht]
\begin{center}
\begin{tabular}{|c|c|} \hline
 \multicolumn{1}{c}{Channel} &  \multicolumn{1}{c}{Event Rate}  \\ \hline \hline

$\nu_e  e^- \rightarrow \nu_e  e^-$ & 677 $\pm$ 39 \\ \hline
NC $\numu \ ^{12}C \rightarrow \numu \ ^{12}C^*$ & 745 $\pm$ 42 \\ \hline
CC $\nue  \ ^{12}C \rightarrow e^-  \ ^{12}N_{gs}$ & 2353 $\pm$ 123\\ 

\hline
\hline 
\end{tabular}
\normalsize
\caption{Summary of per calendar year event rate predictions for a detector located at the SNS a distance of 60 meters from the neutrino source, at $\sim$150 
degrees in the backward direction from the proton beam.  The first column is cross section channel, the second column is the 
expected event rate. All event rates account for a 50\% detector efficiency and a 50\% beam-on efficiency,
and are in units of expected events per year.}
\label{table:xsecnums}
\end{center}
\end{table}

\chapter{L/E Distributions}

\begin{figure}[htbp]
\centering
\includegraphics[width=12cm,angle=90]{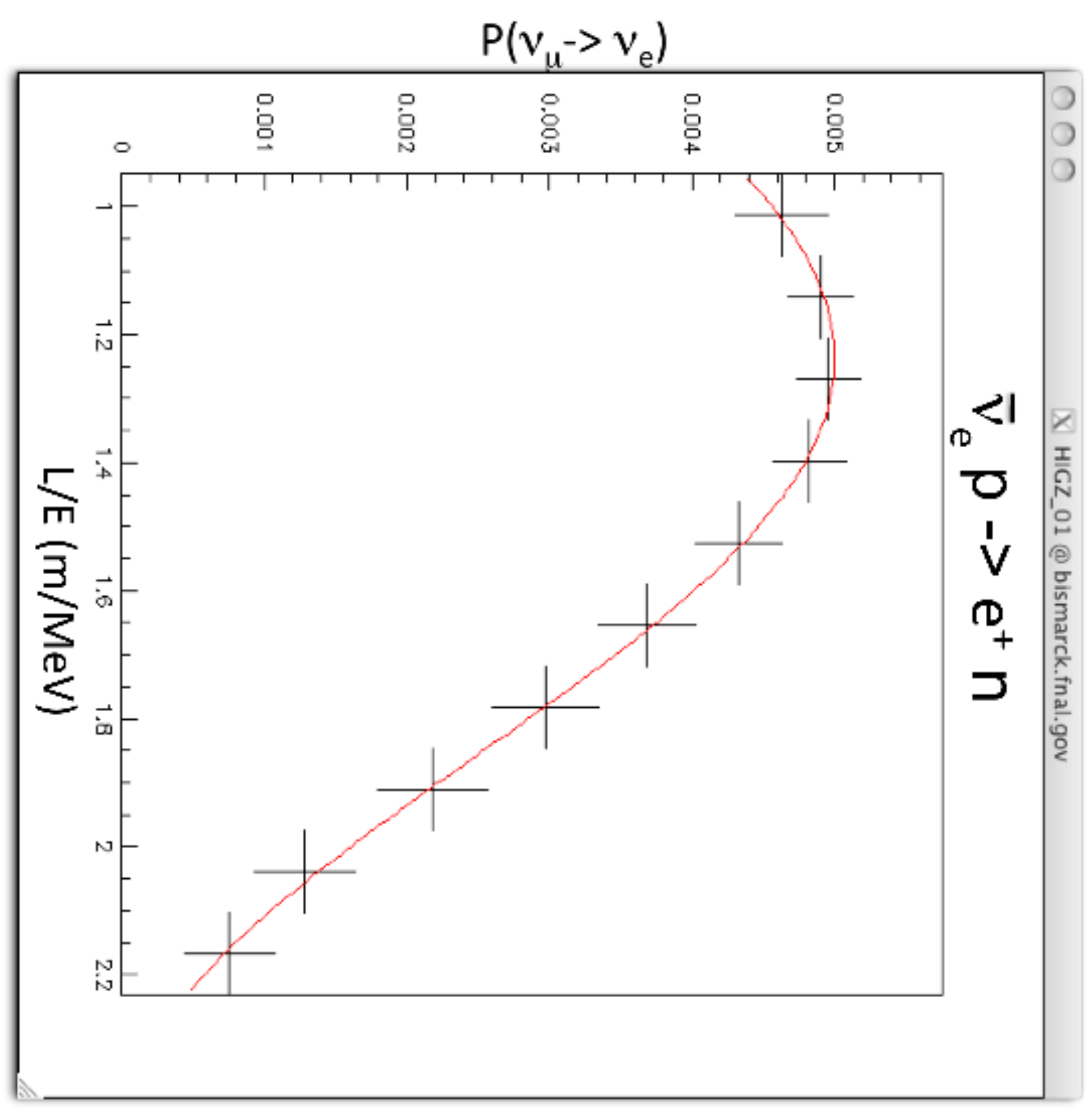}
\caption{The expected oscillation probability from $\bar \nu_e$ appearance as a
function of $L/E$ for
$\sin^22\theta = 0.005$ and $\Delta m^2 = 1$ eV$^2$, for ten {\it calendar} years of data collection at 50\% beam live-time.}
\label{loe_nuebar1}
\end{figure}

\begin{figure}[htbp]
\centering
\includegraphics[width=12cm,angle=90]{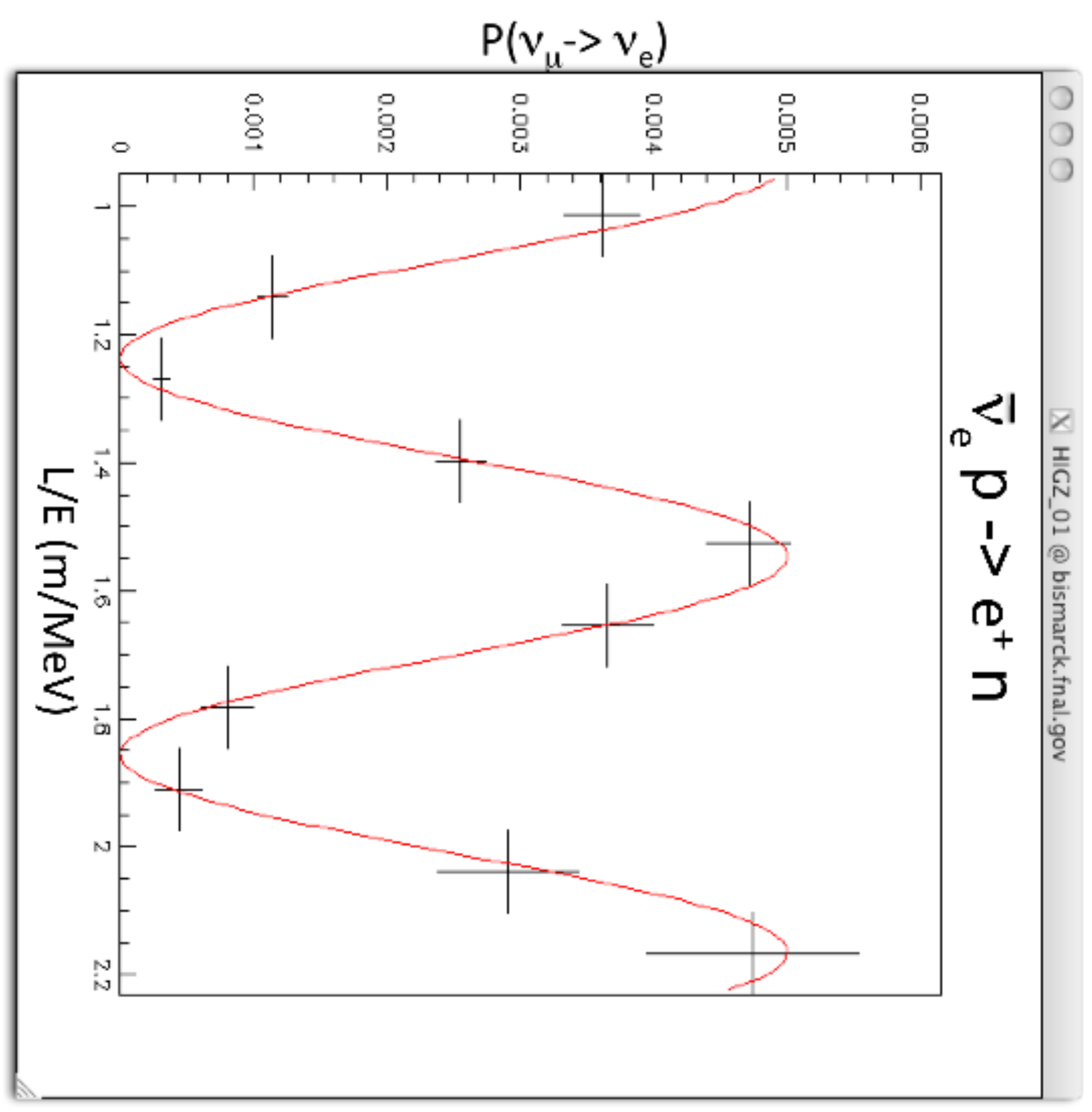}
\caption{The expected oscillation probability from $\bar \nu_e$ appearance as a
function of $L/E$ for
$\sin^22\theta = 0.005$ and $\Delta m^2 = 4$ eV$^2$, for ten {\it calendar} years of data collection at 50\% beam live-time.}
\label{loe_nuebar4}
\end{figure}

Due to the low neutrino energies from $\pi^+$ and $\mu^+$ decay
at rest, it may be possible to actually observe oscillations in
the detector if there is a $\Delta m^2$ of approximately 1 eV$^2$
or greater. Figs. \ref{loe_nuebar1} and \ref{loe_nuebar4} show
the expected oscillation probability from $\bar \nu_e$ appearance as a
function of $L/E$ for 
$\sin^22\theta = 0.005$ and $\Delta m^2 = 1$ eV$^2$ and 4 eV$^2$,
respectively, after 10 calendar years of data collection. This corresponds to 5 years of data collection 
with continuous beam.  The oscillation probability is defined as the
event excess divided by the number of events expected for 100\%
$\bar \nu_\mu \rightarrow \bar \nu_e$ transmutation, while $L$
is the reconstructed distance traveled by the antineutrino from the
mean neutrino production point to the interaction vertex and
$E$ is the reconstructed antineutrino energy. The systematic error associated
with the $L/E$ distribution is assumed to be negligible; however, the estimated background
is approximately 20\% of the signal. The $L/E$ resolution is estimated to be approximately
5\%. Taking into account the background,
an approximate $5\sigma$
($10\sigma$) shape distortion is observed for $\Delta m^2 = 1$ eV$^2$
(4 eV$^2$).

\begin{figure}[htbp]
\centering
\includegraphics[width=12cm,angle=90]{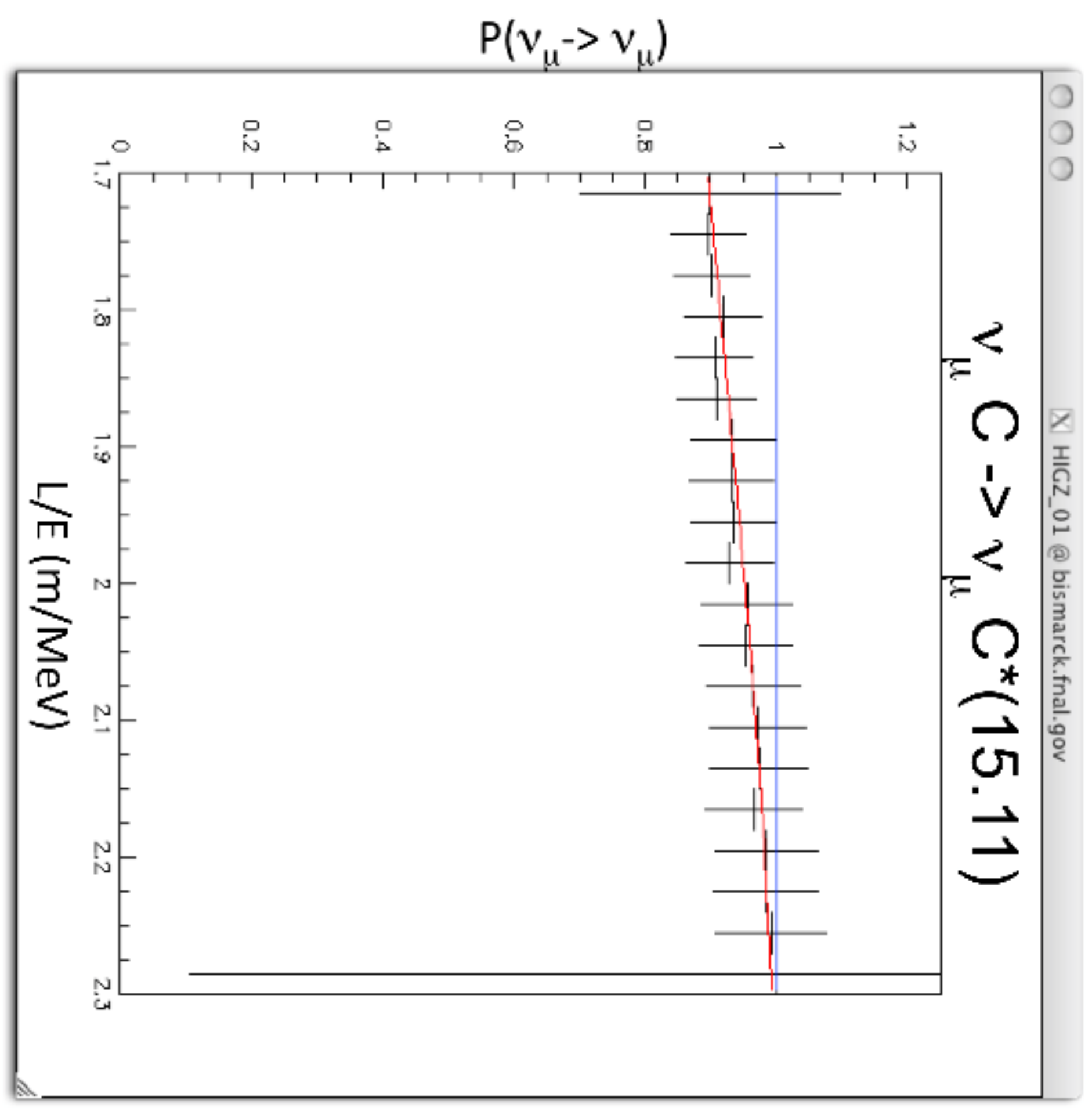}
\caption{The expected probability for $\nu_\mu$ to remain a $\nu_\mu$ as a
function of $L/E$ for
$\sin^22\theta = 0.15$ and $\Delta m^2 = 1$ eV$^2$, for ten {\it calendar} years of data collection at 50\% beam live-time.}
\label{loe_numu1}
\end{figure}

\begin{figure}[htbp]
\centering
\includegraphics[width=12cm,angle=90]{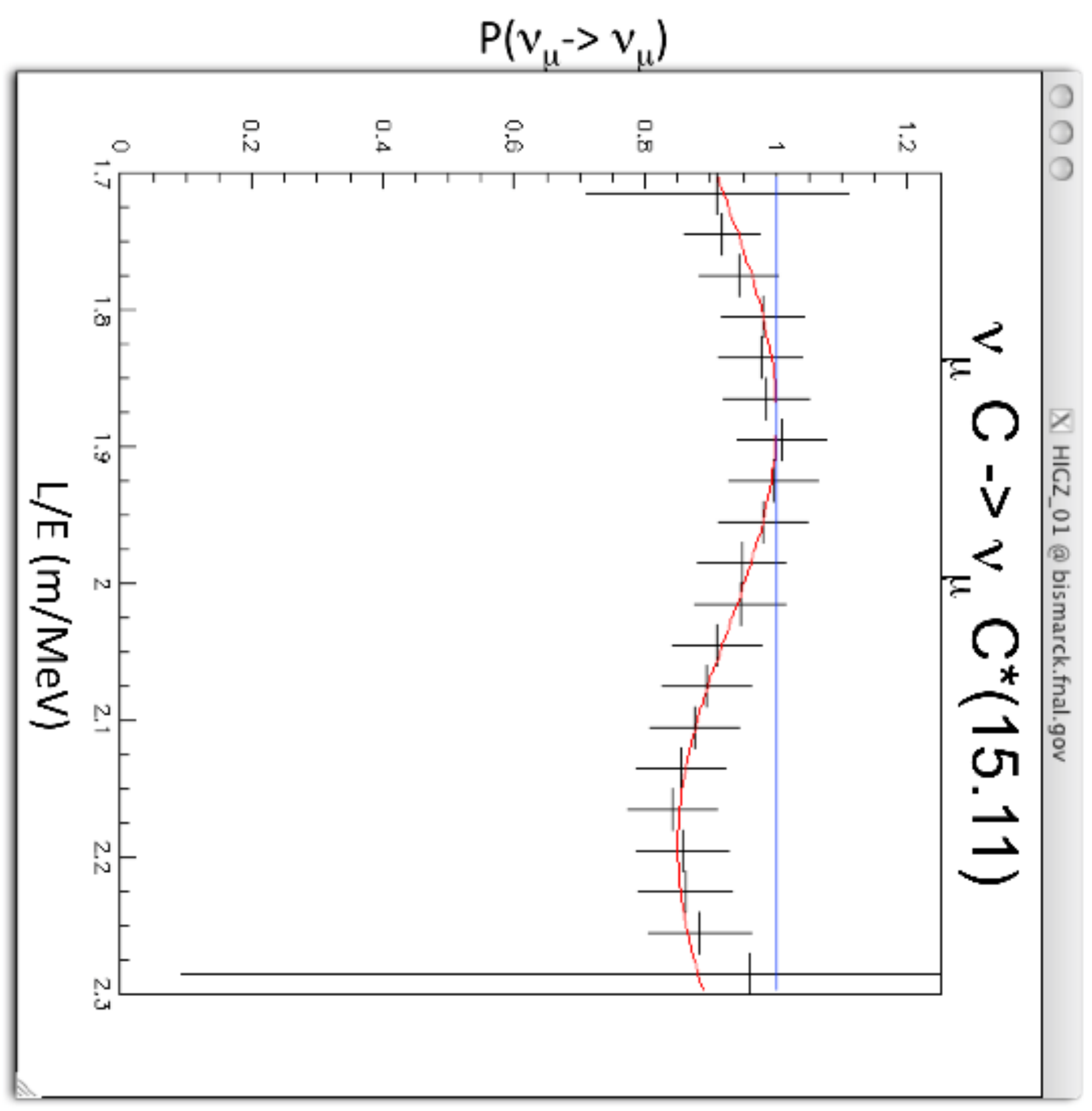}
\caption{The expected probability for $\nu_\mu$ to remain a $\nu_\mu$ as a
function of $L/E$ for
$\sin^22\theta = 0.15$ and $\Delta m^2 = 4$ eV$^2$, for ten {\it calendar} years of data collection at 50\% beam live-time.}
\label{loe_numu4}
\end{figure}

Figs. \ref{loe_numu1} and \ref{loe_numu4} show
the expected probability for $\nu_\mu$ to remain a $\nu_\mu$ as a
function of $L/E$ for
$\sin^22\theta = 0.15$ and $\Delta m^2 = 1$ eV$^2$ and 4 eV$^2$,
respectively, after 10 calendar years of data collection. The probability is defined as the
number of events observed divided by the number of events expected for 
no oscillations. $L$
is the reconstructed distance traveled by the neutrino from the
mean neutrino production point to the interaction vertex and
$E$ is the reconstructed neutrino energy. The $L/E$ resolution is estimated to be 
approximately 1\%. Taking into account the estimated
background of 30\% that is obtained from the KARMEN experiment after
subtracting the fast neutron background \cite{karmen}, 
an approximate $2\sigma$
($3\sigma$) shape distortion is observed for $\Delta m^2 = 1$ eV$^2$
(4 eV$^2$).

\begin{figure}[htbp]
\centering
\includegraphics[width=12cm,angle=90]{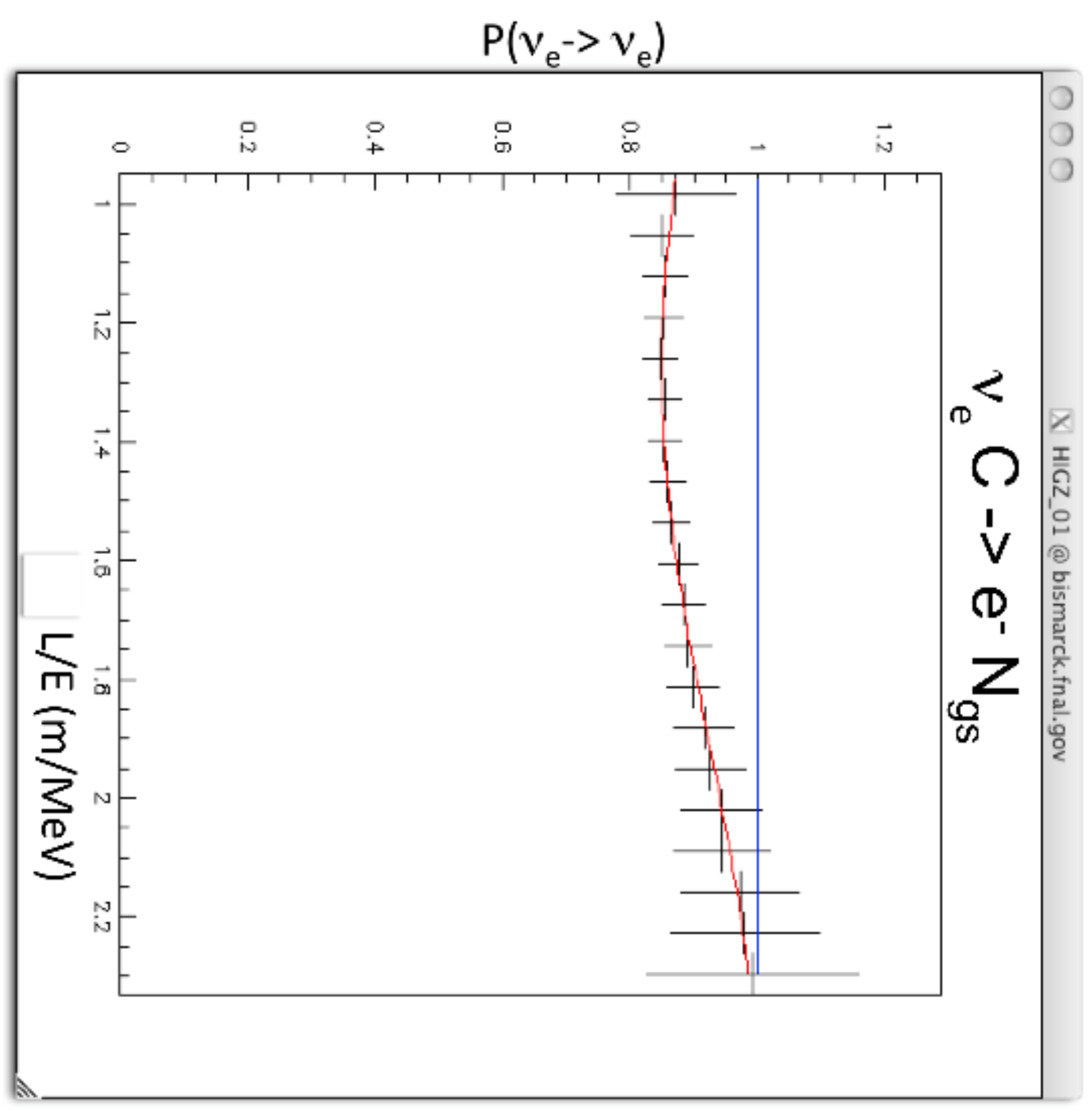}
\caption{The expected probability for $\nu_e$ to remain a $\nu_e$ as a
function of $L/E$ for
$\sin^22\theta = 0.15$ and $\Delta m^2 = 1$ eV$^2$, for ten {\it calendar} years of data collection at 50\% beam live-time.}
\label{loe_nue1}
\end{figure}

\begin{figure}[htbp]
\centering
\includegraphics[width=12cm,angle=90]{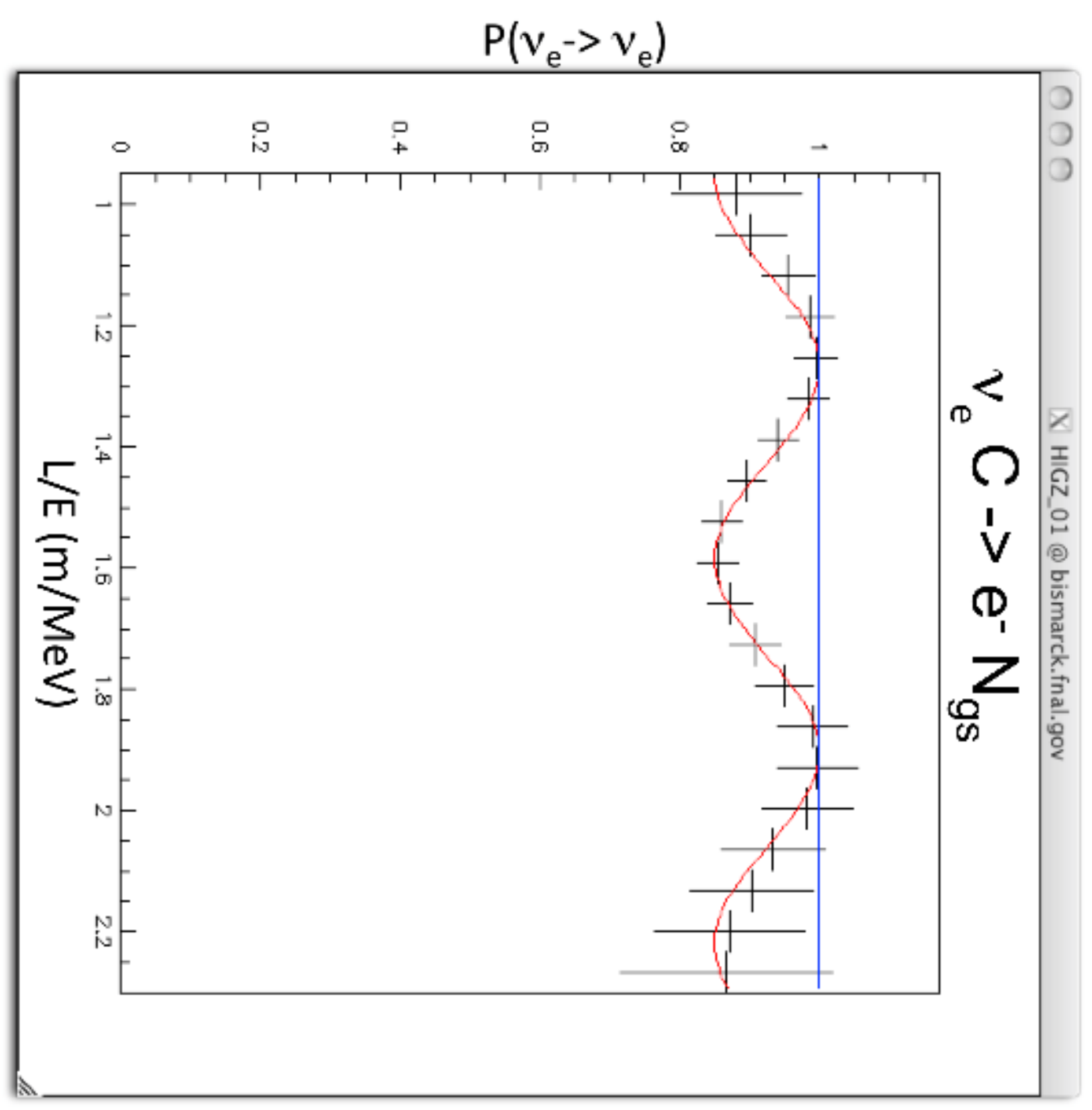}
\caption{The expected probability for $\nu_e$ to remain a $\nu_e$ as a
function of $L/E$ for
$\sin^22\theta = 0.15$ and $\Delta m^2 = 4$ eV$^2$, for ten {\it calendar} years of data collection at 50\% beam live-time.}
\label{loe_nue4}
\end{figure}

Figs. \ref{loe_nue1} and \ref{loe_nue4} show
the expected probability for $\nu_e$ to remain a $\nu_e$ as a
function of $L/E$ for
$\sin^22\theta = 0.15$ and $\Delta m^2 = 1$ eV$^2$ and 4 eV$^2$,
respectively, after 10 calendar years of data collection. The probability is defined as the
number of events observed divided by the number of events expected for 
no oscillations. $L$
is the reconstructed distance traveled by the neutrino from the
mean neutrino production point to the interaction vertex and
$E$ is the reconstructed neutrino energy. The $L/E$ resolution is estimated to be 
approximately 5\%. Taking into account the estimated
background of 1\% that was obtained in LSND \cite{lsnd_nuec}, an approximate $3\sigma$
($5\sigma$) shape distortion is observed for $\Delta m^2 = 1$ eV$^2$
(4 eV$^2$).

Table \ref{chisquares} shows the average shape-only $\chi^2$ values
(for 10 data bins assuming no oscillations) after 10 calendar years of data taking 
for $\bar \nu_e$ appearance, $\nu_e$ disappearance, and $\nu_\mu$
disappearance, taking into account the expected backgrounds. Also shown are the 
corresponding probabilities for no oscillations. Two
sets of oscillation parameters are assumed for each case: 
$(\Delta m^2, \sin^22\theta) = (0.91 eV^2, 0.0053)$ and $(3.6 eV^2, 0.0053)$ 
for $\bar \nu_e$ appearance, and $(\Delta m^2, \sin^22\theta) = 
(0.91 eV^2, 0.169)$ and $(3.6 eV^2, 0.169)$ for $\nu_e$ and $\nu_\mu$
disappearance. Significant evidence for neutrino oscillations can be
obtained by performing a simple shape-only fit to the $L/E$ distributions.

If the absolute neutrino flux normalization can be obtained by fitting the data,
then the significance of the fits to the $\chi^2$ distributions improves greatly.
Table \ref{chisquares2} shows the average normalized $\chi^2$ values
(for 10 data bins assuming no oscillations) after 10 calendar years of data taking
for $\bar \nu_e$ appearance, $\nu_e$ disappearance, and $\nu_\mu$
disappearance, taking into account the expected backgrounds. Also shown are the 
corresponding probabilities for no oscillations. 
Very significant evidence for neutrino oscillations can be
obtained by performing a normalized fit to the $L/E$ distributions.

\begin{table}[htbp]
\centering
\begin{tabular}{|c|c|c|c|c|c|}
\hline
Oscillation&$\Delta m^2$&$\sin^22\theta$&$\chi^2$&No Oscillation Probability&Significance \\
\hline
$\bar \nu_e$ Appearance&0.91 eV$^2$&0.0053&123.3&$<10^{-10}$&$>6 \sigma$ \\
$\bar \nu_e$ Appearance&3.6 eV$^2$&0.0053&166.3&$<10^{-10}$&$> 6 \sigma$ \\
\hline
$\nu_e$ Disappearance&0.91 eV$^2$&0.169&25.4&0.0046&$2.8 \sigma$ \\
$\nu_e$ Disappearance&3.6 eV$^2$&0.169&49.8&$2.9 \times 10^{-7}$&$5 \sigma$ \\
\hline
$\nu_\mu$ Disappearance&0.91 eV$^2$&0.169&15.7&0.1085&$1.6 \sigma$ \\
$\nu_\mu$ Disappearance&3.6 eV$^2$&0.169&25.1&0.0051&$2.7 \sigma$ \\
\hline
\end{tabular}
\caption{The average shape-only $\chi^2$ values
(for 10 data bins assuming no oscillations) after 10 calendar years of data taking
for $\bar \nu_e$ appearance, $\nu_e$ disappearance, and $\nu_\mu$
disappearance, taking into account the expected backgrounds. Also shown are the 
corresponding probabilities for no oscillations.}
\vspace{0.2in}
\label{chisquares}
\end{table}

\begin{table}[htbp]
\centering
\begin{tabular}{|c|c|c|c|c|c|}
\hline
Oscillation&$\Delta m^2$&$\sin^22\theta$&$\chi^2$&No Oscillation Probability&Significance \\
\hline
$\bar \nu_e$ Appearance&0.91 eV$^2$&0.0053&11,930&$<10^{-10}$&$>6 \sigma$ \\
$\bar \nu_e$ Appearance&3.6 eV$^2$&0.0053&2992&$<10^{-10}$&$>6 \sigma$ \\
\hline
$\nu_e$ Disappearance&0.91 eV$^2$&0.169&468.2&$<10^{-10}$&$>6 \sigma$ \\
$\nu_e$ Disappearance&3.6 eV$^2$&0.169&126.1&$<10^{-10}$&$>6 \sigma$ \\
\hline
$\nu_\mu$ Disappearance&0.91 eV$^2$&0.169&33.7&0.0002&$3.6 \sigma$ \\
$\nu_\mu$ Disappearance&3.6 eV$^2$&0.169&25.9&0.0038&$2.9 \sigma$ \\
\hline
\end{tabular}
\caption{The average normalized $\chi^2$ values
(for 10 data bins assuming no oscillations) after 10 calendar years of data taking
for $\bar \nu_e$ appearance, $\nu_e$ disappearance, and $\nu_\mu$
disappearance, taking into account the expected backgrounds. Also shown are the 
corresponding probabilities for no oscillations.}
\vspace{0.2in}
\label{chisquares2}
\end{table}

\chapter{Cost \& Schedule}

Table \ref{tab_hywel} shows a breakdown of the OscSNS cost estimate, which
is based on the MiniBooNE construction costs. The largest component is the
civil construction cost of \$5.461M, which has been been estimated by the 
BWSC company of Knoxville, TN. 
The
total estimated cost of the OscSNS experiment is \$21.852M, including
contingency ($\sim 54\%$) and escalation ($\sim 5\%$). The OscSNS
construction is assumed to start in the beginning of
FY15 and last for 3 years. Expenditures
per year are estimated to be
\$5896K, \$9048K, and \$6908K in FY15, FY16, and FY17, 
respectively. 

\begin{table}[h]
\centering
\begin{tabular}{|c|c|c|c|c|}
\hline
Item & Cost (\$K) & Contingency & Escalation & Total Cost (\$K) \\
\hline
\hline
Phototubes&5219&30\%&4\%&7056 \\
Preamps&53&20\%&12\%&71 \\
Electronics&665&30\%&12\%&969 \\
DAQ&50&20\%&12\%&67 \\
Civil Constr.&5462&100\%&0\%&10,924 \\
Plumbing&20&20\%&8\%&26 \\
Oil&1034&20\%&2\%&1265 \\
Detector Tank&1030&30\%&8\%&1446 \\
Support&20&30\%&8\%&28 \\
\hline
Total&13,553&54\%&5\%&21,852 \\
\hline
\end{tabular}
\caption{
A breakdown of the OscSNS cost estimate, including contingency and
escalation. The OscSNS
construction is assumed to start in the beginning of
FY15 and last for 3 years.
}
\label{tab_hywel}
\end{table}

The OscSNS cost estimate can be reduced by reusing the MiniBooNE oil
and PMTs. The use of additional parts from MiniBooNE 
may also be considered.

\chapter{The OscSNS Near Detector}

\subsection{The OscSNS Near Detector}
%

We would ideally like to install two neutrino detectors at the SNS.  
A near detector for OscSNS will provide a cross check for the neutrino flux
normalization.
A space adjacent to BL-18 has been identified as a possible location for
such a detector in terms of proximity to the neutrino source, available space,
and maximum floor loading.
Given the current constraints, the detector would have a footprint of
$3.7 \times 5.1 \times 3.4\,\mbox{m}$.
The outer $40\,\mbox{cm}$ consists of passive shielding, followed by
a 25-cm-thick active veto, which is optically isolated from the inner volume
of the detector.
The active veto is painted white to enhance reflections, and is instrumented
with about 120 5-inch PMTs.
The inner detector is instrumented with 808 5-inch PMTs, which correspond to
the same photocathode surface coverage as the main detector, namely 25\%.
These PMTs are embedded within the active veto, and are mounted such that their
equator is flush with the optical barrier.
Additional flat photodetection devices, currently under development at the
Argonne National Laboratory, could also be interspersed with the standard
5-inch PMTs, assuming that these devices are fully operational by construction
time.

The inner detector is painted black (to avoid reflections) and is filled with
exactly the same active medium as the main detector.
Assuming a neutrino production rate of $2.8 \times 10^{22}$ per year at 100\%
SNS efficiency, the neutrino flux at the location of the near detector yields
$\Phi = 5.57 \times 10^{14}\,\mbox{cm}^{-2}\,\mbox{yr}^{-1}$, which in turn
will yield an event rate of $183.4$ per year per cubic meter of fiducial
volume for the $\nu_e\,C \to e^-\,\mbox{}^{12}N_{gs}$ reaction.
This reaction is well-known theoretically, while its two-fold signature
(i.e., the space-time correlation between the electron and the positron from
the subsequent $\mbox{}^{12}N_{gs}$ $\beta$-decay) gives a strong discrimination
from backgrounds (in particular the relatively abundant beam-related neutrons
given the proximity of the neutron beam lines).

Assuming a fiducial volume of $7.4\,\mbox{m}^3$, which corresponds to a minimum
distance of $30\,\mbox{cm}$ from the edge of the optical barrier, the near
detector could record a total of 1620 $\nu_e\,C \to e^-\,\mbox{}^{12}N_{gs}$
events (assuming a combined efficiency of 40\% for the SNS beam and the near
detector), corresponding to a 2.5\% statistical error.

\chapter{Appendix A - Letters of Support}

This space is reserved for letters of support.

\chapter{Appendix B - BWSC Cost Estimate}

This space is reserved for a complete civil engineering study, performed by Barge Waggoner Sumner and Cannon, Inc., of Knoxville, TN.  The survey is in the final stages, and will be completed and appended shortly.



\clearpage


\begin{thebibliography}{99}




\bibitem{fogli}
G.~L.~Fogli, E.~Lisi, A.~Marrone, D.~Montanino, A.~Palazzo, and A.~M.~Rotunno,
arXiv:1205.5254 (2012).

\bibitem{lsnd}
  C.~Athanassopoulos {\em et~al.},
  Phys.\ Rev.\ Lett. 75, 2650 (1995);
  77, 3082 (1996); 81, 1774 (1998);
  Phys. Rev. C. {\bf 58}, 2489 (1998);
  A.~Aguilar {\em et~al.},
  Phys.\ Rev.\ D 64, 112007 (2001).

\bibitem{mb_osc_anti}
  A.~A.~Aguilar-Arevalo {\em et~al.},
  Phys.\ Rev.\ Lett.\  105, 181801 (2010);
  Phys.\ Rev.\ Lett.\ 110, 161801 (2013).

\bibitem{radioactive}
C.~Giunti and M.~Laveder, Phys.\ Rev.\ C 83, 065504 (2011).

\bibitem{reactor}
G.~Mention, M.~Fechner, T.~Lasserre, T.~A.~Mueller, D.~Lhuillier, M.~Cribier, 
and A.~Letourneau, Phys.\ Rev.\ D 83, 073006 (2011).

\bibitem{3+N}
J.~M.~Conrad, C.~M.~Ignarra, G.~Karagiorgi, M.~H.~Shaevitz, and J.~Spitz,
arXiv:1207.4765 [hep-ex] (2012).

\bibitem{kopp}
Joachim~Kopp, Pedro~A.~N.~Machado, Michele~Maltoni, and Thomas~Schwetz, 
arXiv:1303.3011 [hep-ph] (2013).

\bibitem{ornl} 
The Spallation Neutron Source (SNS) is an accelerator-based 
source built in Oak Ridge, Tennessee, by the U.S.\ DOE, http://sns.gov/.
Also see http://www.phy.ornl.gov/workshops/sns2/ for details on the neutrino 
source and cross section detector $\nu$-SNS.










\bibitem{mb_lowe}
  A.~A.~Aguilar-Arevalo {\em et~al.},
  Phys.\ Rev.\ Lett.\ 102, 101802 (2009).

\bibitem{karmen}
  B.~Armbruster {\em et~al.},
  Phys.\ Rev.\  D 65, 112001 (2002).

\bibitem{nustorm}
D.~Adey {\it et al.}, arXiv:1305.1419 [physics.acc-ph].


\bibitem{bigpaper1}
C.\ Athanassopoulos {\it et\ al.}, Nuclear Instruments and Methods in Physics Research
A {\bf 388}, 149 (1997).
            

\bibitem{sno}
B. A. Moffat, {\it et al.}, Nuclear Instruments and Methods in Physics Research A {\bf 554}, 255 (2005) 
[nucl-ex/0507026].


            
\bibitem{geant4}
S. Agostinelli, {\it et al.}, Nuclear Instruments and Methods in Physics Research A {\bf 506}, 250 (2003); http://geant4.cern.ch.
                              
\bibitem{gps}
http://reat.space.qinetiq.com/gps/.

\bibitem{tant} K. A. Walaron, UKNF Note 30: Simulations of Pion Production in a Tantalum Rod Target using GEANT4 with comparison to MARS; http://hepunx.rl.ac.uk/uknf/wp3/uknfnote\_30.pdf.

\bibitem{suzuki} T. Suzuki, D. F. Measday, J. P. Roaisvig, Phys. Rev. C, {\bf 35}, 2212 (1987).

\bibitem{root} R. Brun, F. Rademakers, Nuclear Instruments and Methods in Physics Research A {\bf 389}, 81 (1997); http://root.cern.ch.



\bibitem{mb_oil} http://www.physics.uc.edu/\verb|~|johnson/Boone/oil\_page/Index.html.

\bibitem{LSND_om} R. A. Reeder {\it et al.}, Nuclear Instruments and Methods in Physics Research A {\bf 334}, 353 (1993).

\bibitem{birks} J. B. Birks, Theory and Practice of Scintillation Counting (Pergamon. Oxford, 1964).




\bibitem{fuku} M. Fukugita, Y. Kohyama, K. Kubodera, Phys. Lett. B {\bf 212}, 139 (1988).

\bibitem{kolbe}
E. Kolbe, K. Langanke, and P. Vogel, Nucl. Phys. A {\bf 652}, 91 (1999). 

\bibitem{vogel}
P. Vogel and J. F. Beacom. Phys. Rev. D {\bf 60}, 053003 (1999).

\bibitem{sorel}
M.~Sorel, J.~M.~Conrad and M.~Shaevitz, Phys.\ Rev.\ D 70, 073004 (2004).

\bibitem{karagiorgi}
G.~Karagiorgi, Z.~Djurcic, J.~M.~Conrad, M.~H.~Shaevitz and M.~Sorel,
Phys.\ Rev.\ D 80, 073001 (2009) [Erratum- ibid. D 81, 039902 (2010)].

\bibitem{giunti}
C.~Giunti and M.~Laveder, Phys.\ Lett.\ B 706, 200 (2011).
C.~Giunti and M.~Laveder, Phys.\ Rev.\ D84, 073008, (2011).

\bibitem{lsnd_nuec} L. B. Auerbach {\it et al.}, Phys. Rev. D {\bf 64}, 065501 (2001).

\bibitem{lampf} L. B. Auerbach {\it et al.}, Phys. Rev. D {\bf 63}, 112001 (2001).



 
                   
\end{thebibliography}
\end{document}